\documentclass[11pt]{article}

\usepackage{bigfoot}
\usepackage{varioref}
\usepackage{amsmath,amsfonts,amsthm,mathrsfs}

\usepackage[normalem]{ulem}
\usepackage{multirow,color,graphics}
\usepackage{amsmath}
\usepackage{amsfonts}
\usepackage{amssymb}
\usepackage{jheppub}
\usepackage{enumerate}
\usepackage{fancyvrb}
\usepackage{verbatim}
\usepackage{wrapfig}
\usepackage{appendix}
\usepackage{amstext}
\usepackage{amssymb}
\usepackage{graphicx}
\usepackage{color}
\usepackage{varioref}
\usepackage{multirow,graphics}
\usepackage{epstopdf}
\usepackage{simplewick}

\usepackage[dvipsnames]{xcolor}
 
\definecolor{Green}{RGB}{0, 142, 0}

\newcommand{\nn}{\nonumber}

\numberwithin{equation}{section}

\def\[{\left[}
\def\]{\right]}
\def\({\left(}
\def\){\right)}
\def\d{\partial}

    \newcommand{\beq}{\begin{equation}}
    \newcommand{\eeq}{\end{equation}}

    \newcommand\beqa{\begin{eqnarray}}
    \newcommand\eeqa{\end{eqnarray}}
\newcommand\bea{\begin{array}}
\newcommand\eea{\end{array}}

\newcommand{\Blue}[1]{{\color{blue}#1\color{black}}}

\newcommand{\bQ}{{\bf Q}}

\newcommand{\bP}{{\bf P}}

\newcommand{\la}[1]{\label{#1}}
\newcommand{\eq}[1]{(\ref{#1})}

\usepackage[dvipsnames]{xcolor}

    \def\CO{{\cal O}}

    \def\bQ{{\bf Q}}
    \def\bP{{\bf P}}

\makeatletter
     \@ifundefined{usebibtex}{} {}
\makeatother

    \def\bQ{{\bf Q}}

        \def\bP{{\bf P}}

     \def\Tr{\text{Tr}}

\newcommand{\cO}{\mathcal{O}}

\newcommand{\tr}{\text{tr}}

\renewcommand{\d}{\partial}

\newcommand{\<}{{\langle}}
\renewcommand{\>}{{\rangle}}

\newcommand{\qc}{  q^{\uparrow}  }

\newcommand{\cB}{\mathcal{B}}

\newcommand{\bx}{\mathbf{x}}
\newcommand{\by}{\mathbf{y}}

\title{
Colour-Twist Operators I: Spectrum and Wave Functions}

\author[a]{~~Andrea Cavagli\`a,}
\author[a]{~~David Grabner,}
\author[a,b]{~~Nikolay Gromov,}
\author[c,d]{~~Amit Sever}

\affiliation[a]{Mathematics Department, King's College London,
The Strand, London WC2R 2LS, UK}
\affiliation[b]{St.Petersburg INP, Gatchina, 188 300, St.Petersburg,
  Russia}
\affiliation[c]{School of Physics and Astronomy, Tel Aviv University, Ramat Aviv 69978, Israel}
\affiliation[d]{CERN, Theoretical Physics Department, 1211 Geneva 23, Switzerland}

\emailAdd{andrea.cavaglia$\bullet$kcl.ac.uk, david.grabner$\bullet$kcl.ac.uk, nikgromov$\bullet$gmail.com, amit.sever$\bullet$gmail.com}

\abstract{

We introduce a new class of operators in any theory with a 't Hooft large-$N$ limit that we call colour-twist operators. They are defined by twisting the colour-trace with a global symmetry transformation and are continuously linked to standard, un-twisted single-trace operators. In particular, correlation functions between operators that are twisted by an R-symmetry of ${\cal N}=4$ SYM extend those in the $\gamma$-deformed theory.
The most general deformation also breaks the Lorentz symmetry but preserves integrability in the examples we consider.
In this paper, we focus on colour-twist operators in the fishnet model.
We exemplify our approach for the simplest colour-twist operators with one and two scalar fields, which we study non-perturbatively using field-theoretical as well as integrability methods, finding a perfect match. We also propose the  
quantisation condition for the Baxter equation appearing in the integrability calculation in the fishnet model.
The results of this paper constitute a crucial step towards building the separation of variable construction for the correlation
functions by means of the Quantum Spectral Curve approach.

}

\begin{document}

\begin{flushright}
CERN-TH-2020-012
\end{flushright}
\maketitle

\newpage
\section{Introduction}\la{intro}

Twist operators play an important role in two-dimensional conformal field theories. They are defined by the action of a {\it symmetry} as one goes {\it around} the operator \cite{Dixon:1986qv,Calabrese:2004eu}. Twist operators also exist in quantum field theories in higher dimensions. As in the two-dimensional case, these co-dimension two surface-operators are defined by the action of a symmetry transformation as one goes around the surface in the two-dimensional transverse space \cite{Calabrese:2004eu,Casini:2011kv,Hung:2014npa,Balakrishnan:2017bjg}.\footnote{Such operators were studied in the context of entanglement entropy. In that case, the relevant symmetry acts in the replicated theory by interchanging between the replica copies.} In this paper, we study a new type of twist operators in the 't Hooft large-$N$ limit that we call ``colour-twist operators". They may be defined as a generalisation of single-trace operators, where the colour-trace is accompanied by a symmetry transformation. 
In this picture, going around the operator takes place in colour space instead of spacetime.\footnote{In general, colour-twist operators are different from the large-$N$ limit of the twist operators mentioned above. However, in some special cases they turn out to be the same \cite{Brigante:2005bq}.} In holographic theories, we expect our field theory definition to coincide with twisted vertex operators in the two-dimensional worldsheet CFT of the dual string. 

Our motivation for considering these somewhat exotic operators comes from studying correlation functions in ${\cal N}=4$ SYM. The aim of this program is to compute planar correlation functions of single-trace operators at finite 't Hooft coupling. The most efficient way for computing their conformal dimensions is called Quantum Spectral Curve (QSC). This integrability based method yields not only the quantum spectrum of operators, but also their wave functions in the so-called separation of variables (SoV) basis. Hence, one expects that this method can be further developed for computing correlation functions.\footnote{This expectation was partly confirmed for the case of the cusp correlation function in the ladders limit \cite{Cavaglia:2018lxi,McGovern:2019sdd}
and in the near-BPS limit \cite{Giombi:2018qox,Giombi:2018hsx}.} The twisting procedure we study in this paper turns out to be essential for building such a coordinate system where the degrees of freedom become independent~\cite{Gromov:2016itr}. Our strategy then is to first compute correlation functions between colour-twist operators in terms of the SoV wave functions and, in the end, send the twist parameters to zero.

To make progress in this hard problem, it is often useful to first study simplified limits. One such limit is the strongly $\gamma$-deformed limit of \cite{Gurdogan:2015csr}. It leads to a much simpler integrable planar CFT called the fishnet model. Correspondingly, in the bulk of this paper, we will focus on colour-twist operators in the fishnet model. In \cite{colortwist2} we will study the correlation functions between these operators using the corresponding SoV wave functions.

This paper is organised as follows. In section \ref{sec_ct}, we start with a perturbative definition of colour-twist operators in any large-$N$ QFT. In section \ref{sec:sec3}, we focus on colour-twist operators in ${\cal N}=4$ SYM theory and the fishnet model. 
In particular, the $\gamma$-deformation can be obtained as a particular case of the colour-twist and so the  fishnet model can be obtained from ${\cal N}=4$ SYM using colour-twist operators. 
In section \ref{secOneScalar}, we study twist operators in the fishnet model that consist of one scalar. We map the computation of their scaling dimensions and wave functions to a Schr\"odinger problem of one degree 
of freedom and use it to analyse the spectrum. In section \ref{sec:onemagnon}, we consider colour-twist operators with two orthogonal scalars (i.e. the one-magnon case). We study their spectrum at one-loop order and also at finite coupling in a certain case. 
In section \ref{sec:integrability}, we study the spectrum using integrability. We explicitly construct the Baxter equation and generalised quantisation condition for the Q-functions of colour-twist operators. 
We then reproduce and extend the direct field theory results of sections \ref{secOneScalar} and \ref{sec:onemagnon}. 
In section \ref{secmap}, we give an explicit map between the Q-function and the CFT wave function for colour-twist operators with one scalar. Under this map, the Baxter equation of section \ref{sec:integrability} becomes the Schr\"odinger equation of section \ref{secOneScalar}. We end in section \ref{secdis} with a short discussion.

\section{Colour-Twist Operators}\la{sec_ct}
In this section, we define a new type of 
{\it colour-twist operators}, which are continuously linked to the operators in the theory without the twist. 
These colour-twist operators are a certain generalisation of single-trace operators in which the cyclic permutation of the fields in the trace is accompanied by a symmetry transformation. To give a more precise definition of these operators, we will use the perturbative expansion. 
The perturbation theory in planar limits is usually convergent, and thus this definition extends to finite coupling, too. Below, we describe the prescription for computing correlation functions between twisted operators, by directly twisting the  Feynman diagrams contributing to the correlator. Throughout this paper we will sometimes refer to these operators in short as {\it twisted} operators.

\subsection{Perturbative definition}\la{pertdef}

\paragraph{Twist symmetry.} To twist an operator, one may use any global symmetry of the theory. This symmetry transformation, which will be denoted by $R$, can be an internal symmetry acting on fields, a spacetime rotation, translation, or even a conformal transformation in a CFT. In this paper, we will only consider global symmetries that have fixed points and will place the corresponding twisted operator at one of these points. More generally, one may also consider non-local twist operators, see for instance \cite{Ben-Israel:2018ckc}. A useful example to keep in mind is when $R$ is a spacetime rotation around the insertion point of a local operator.

\paragraph{Twisted field.} 

A fundamental field twisted by a symmetry transformation $R$ is simply defined to be the transformed field. For example, 
a scalar field in four dimensions, twisted by a conformal transformation $R$, is given by
\beq\la{scalartransform}
R:\quad\phi(x)\quad\rightarrow\quad\phi_\circlearrowleft(x)=\left|{\d x_\circlearrowleft\over\d x}\right|^{\frac{1}{4}} \phi\(x_\circlearrowleft\)\ ,
\eeq
where $
x_{\circlearrowleft} \equiv R(x)$ denotes the image of a space-time point under the transformation $R$, and where
the factor
$| 
{\d x_\circlearrowleft\over\d x}
|$ is the determinant of the Jacobian.
We will call $\phi_\circlearrowleft$ in (\ref{scalartransform}) a {\it twisted scalar}. Similarly, twisted fermions and gauge fields are fields that have been transformed covariantly under the symmetry transformation $R$. 
In the case when $R$ is a rotation,  the Jacobian is trivial and we  simply have $\phi_\circlearrowleft(x)=\phi(x_\circlearrowleft)$.

\paragraph{Twisted propagator.} A twisted propagator is a propagator between a twisted and an un-twisted field. For scalar fields it takes the form
\beq\la{twistedprop}
\contraction{}{\phi}{(y)}{\phi}\phi(y)\,\phi_\circlearrowleft(x)=\left|{\d x_\circlearrowleft\over\d x}\right|^{\frac{1}{4} }\contraction{}{\phi}{(y)}{\phi}\phi(y)\,\phi(x_\circlearrowleft)\ .
\eeq
In the case where the twist is by a rotation and the scalars are $N\times N$ matrices
, this twisted propagator is simply 
\beq
\includegraphics[width=0.5 \textwidth]{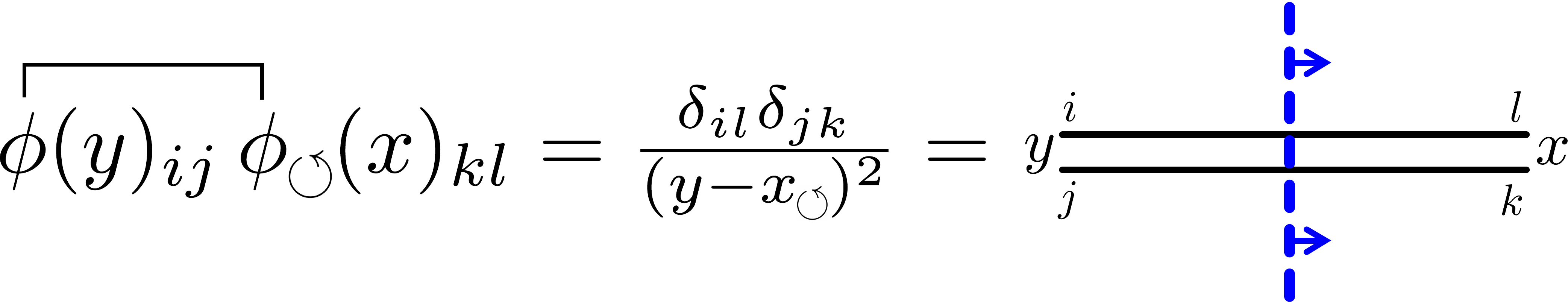}\ .
\eeq
Here, we have drawn the propagator in the double line notation. The twist is represented by the dashed blue line and the arrows indicate the direction upon which it acts.

\begin{figure}[h]
\centering
\def\svgwidth{16cm}
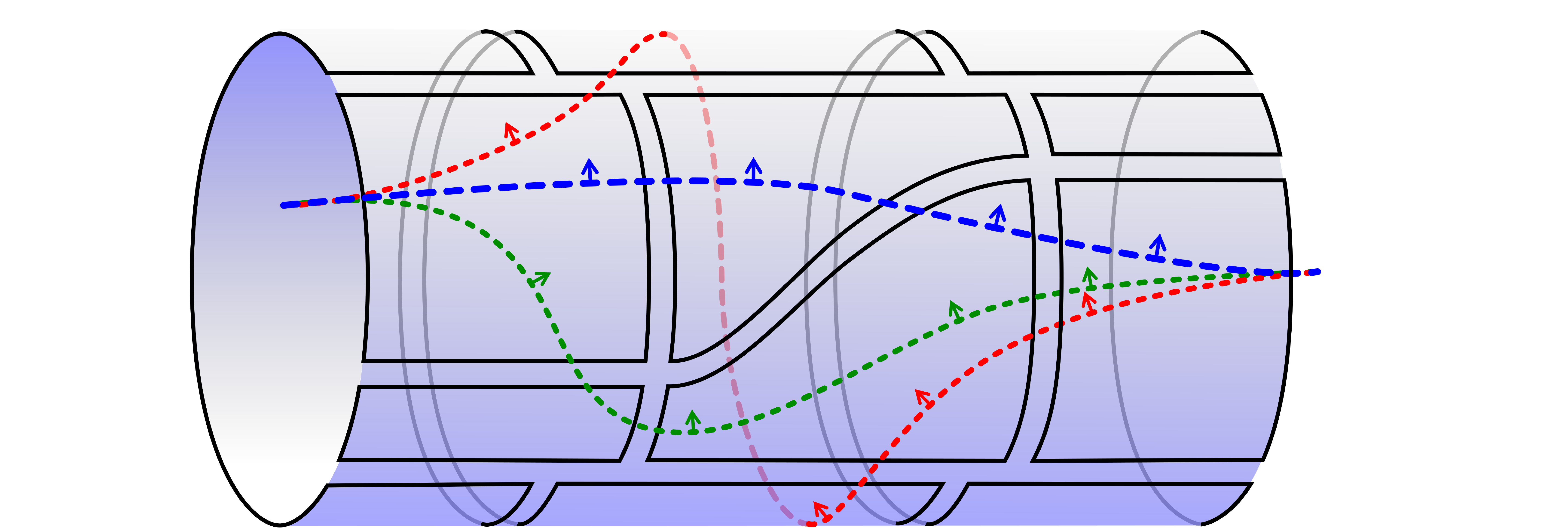
\caption{\small A Feynman diagram that contributes to the two-point functions of twisted operators of length $L=3$. The dashed lines indicate different possible choices for the cut used in the perturbative definition. Every propagator crossing the cut is replaced by a twisted propagator as explained in the text. The red dashed line wraps around the cylinder once more than the blue one does, and correspondingly the result differs by the action of the symmetry transformation, $R$, on the external operator $\text{Tr}( \phi^3)^\dagger(y) $.  }\label{twistcut}
\end{figure}

\paragraph{Twisted Feynman diagram.} 
We now show how to construct twisted Feynman diagrams  and define correlators of twisted operators. 
 Consider the two-point function of two single-trace operators, $\<\cO(x )\,\cO^\dagger(y)\>$. Working in double line notation, every   planar Feynman diagram 
 that contributes to this correlator
can be drawn on a cylinder. To twist the operators, we add {an oriented} non self-intersecting cut on the cylinder starting from $\cO(x)$ and ending at $\cO^\dagger(y)$; see figure \ref{twistcut}. On the way, the cut crosses a set of propagators. 
We then replace every propagator that crosses the cut by a twisted propagator connecting the twisted and un-twisted fields as is indicated by the arrow on the cut. Propagators that do not cross the cut are left unchanged.\footnote{Note that here we assume that the theory is orientable. Namely, the matrix fields are Hermitian, the colour-traces have a distinct orientation along which $R$ is acting.
This definition can be extended for symmetric groups. In such cases we would have to symmetrise (automatically) between $R$ and $R^{-1}$.}
By definition, we say the diagram contributes to the correlator involving  $\mathcal{O}(x)$ \emph{twisted by the transformation} $R$, and  the operator $\mathcal{O}^\dagger(y)$ \emph{twisted by the transformation} $R^{-1}$.

Importantly, for the consistency of this definition we can verify that the result for an integral represented by a diagram is unchanged if we reverse the orientation of the cut (i.e., flip the direction of the arrows), and simultaneously replace $R \rightarrow R^{-1}$. 
This follows from a simple identity for the twisted propagators
\beq\label{eq:twisted inv}
\contraction{}{\phi}{(y)}{\phi}\phi(y)\,\phi_R(x) = \contraction{}{\phi}{(x)}{\phi}\phi_{ }(x)\,\phi_{R^{-1}} (y)\ ,
\eeq
where we are denoting the field twisted by the transformation $R$ ($R^{-1}$) as $\phi_R$ ($\phi_{R^{-1}}$), respectively.  
The relation (\ref{eq:twisted inv}) is a consequence of the covariance of the propagator (which is a two-point function) under a conformal transformation. This useful relation can be written as
\beq\label{eq:covariance}
\frac{1}{(x-y)^2} = \frac{  \left|{\d x_\circlearrowleft\over\d x}\right|^{\frac{1}{4}}  \left|{\d y_\circlearrowleft\over\d y}\right|^{\frac{1}{4} }  }{(x_{\circlearrowleft} - y_{\circlearrowleft})^2 }\ . 
\eeq
Changing $y \rightarrow R^{-1}(y)$, (\ref{eq:covariance}) implies
\beq
\frac{ \left|{\d R^{-1}(y) \over\d y }\right|^{\frac{1}{4} } }{(x-R^{-1}(y))^2} = \frac{  \left|{\d R (x) \over\d x}\right|^{\frac{1}{4}}  }{(R(x) - y )^2 }\ , 
\eeq
which is precisely the property (\ref{eq:twisted inv}). 
\begin{figure}[h]
\centering
\def\svgwidth{6cm}
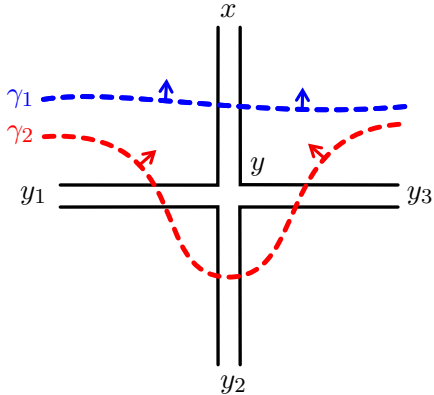
\caption{Local cut deformation. Corresponding to the cuts $\gamma_1$ and $\gamma_2$, the Feynman integral is modified as in (\ref{eq:expr1}) and (\ref{eq:expr2}), respectively. The two choices are equivalent. }\label{pic:local}
\end{figure}

\paragraph{Local cut deformations.} So far, our definition involves the choice of a cut for any diagram. 
For consistency, the resulting correlator must be independent of these arbitrary choices. For simplicity, we assume that each interaction term in the Lagrangian is independently invariant under the transformation.\footnote{This is not always the case, for instance in the case of the conformal symmetry. However, provided that the Lagrangian is invariant under the transformation $R$, one can use the Lagrangian insertion technique  \cite{Eden:2010zz} which only generates symmetric loop integrals (e.g. conformal integrals in a CFT). Namely, one should group the diagrams into integrated correlators with the Lagrangian, symmetrise over the insertion points, and take the twist cut in between the Lagrangian insertions. }
In this case, consider a local deformation of the twist cut across an interaction vertex, such as the one from the blue to the green dashed lines in figure \ref{twistcut}. Such a deformation can be recast as the action of the symmetry transformation on that integrated vertex and hence, modulo possible subtleties due to the regularisation scheme, does not change the result for the diagram. 

Let us consider a simple example of this for a conformal scalar integral. Suppose we have a diagram with a four-scalar interaction vertex that is connected by four propagators to the rest of the diagram. We start from a twist cut that is crossing only one of these four propagators, see ${\color{blue} \gamma_1}$ in figure \ref{pic:local}. This part of the diagram will result in the following expression
\beq\label{eq:expr1}
{\cal M}_{\color{blue} \gamma_1}=\dots\times\int d^4 y \frac{ \left|{\d x_\circlearrowleft\over\d x}\right|^{\frac{1}{4} } }{(x_{\circlearrowleft}-y)^2(y-y_1)^2(y-y_2)^2(y-y_3)^2}\;,
\eeq
where the dots stand for the rest of the diagram that is independent of $y$. For the same diagram but with the deformed cut ${\color{red} \gamma_2}$ we get
\beq\label{eq:expr2}
{\cal M}_{\color{red}\gamma_2}=\dots\times\int d^4 y \frac{ \left|{\d y_{ \circlearrowleft}\over\d y}\right|^{\frac{3}{4} }  }{(x-y)^2(y_{\circlearrowleft}-y_{1{}})^2(y_{\circlearrowleft}-y_{2{}})^2(y_{\circlearrowleft} -y_{3} )^2}\;.
\eeq
 The equivalence of the two expressions can be proven 
 using the identity (\ref{eq:covariance}) for the propagator.  
 Plugging (\ref{eq:covariance})  into (\ref{eq:expr2}), and 
  using $ d^4 y \, \left|{\d y_{ \circlearrowleft}\over\d y}\right| = d^4 y_{\circlearrowleft} $ to change the integration variable from $y$ to $y_{ \circlearrowleft}=R(y)$, the expression (\ref{eq:expr2}) becomes
\beq
{\cal M}_{\color{red}\gamma_2}=\dots\times \int d^4 y_{\circlearrowleft}\frac{ \left|{\d x_\circlearrowleft\over\d x}\right|^{\frac{1}{4}}  }{(x_{\circlearrowleft}-y_{\circlearrowleft} )^2(y_{\circlearrowleft}-y_{1{}})^2(y_{\circlearrowleft}-y_{2{}})^2(y_{\circlearrowleft} -y_{3} )^2}\ ,
\eeq
which is exactly the expression we got for the initial contour ${\cal M}_{\color{blue}\gamma_1}$ in (\ref{eq:expr1}) after renaming the variable of integration $y_\circlearrowleft$ to $y$. This demonstrates that any local 
deformation of the cut does not affect the resulting correlator.

\paragraph{Non-local cut deformations.}
In order to ensure self-consistency of the twisting procedure, we also have to require that the result for the deformed diagram (or sum of diagrams) stays the same even under topologically non-trivial deformations of the contour. 
This requirement, however, imposes a constraint on the combination of a symmetry transformation and a local operator. Namely, the operator ${\cal O}$ should be invariant under the action of $R$. Indeed, the difference between two topologically inequivalent cuts, such as the blue and the red cuts in figure \ref{twistcut}, is a closed loop which has a non-trivial winding number around the operator ${\cal O}$. Each winding has the effect of acting on the operator with the twist transformation, $\cO(x)\to \cO_\circlearrowleft(x)$. Therefore, for consistency of the definition, the twisted operators $\cO(x)$ must be invariant under $R$.\footnote{An analogous projection is familiar from the construction of twist operators in 2d CFTs.} This requirement implies that local operators can be twisted only by spacetime transformations that admit \emph{fixed points}. {\it Local} twisted operators are constrained to sit at a fixed point of $R$. In general, one may also consider non-local twist operators that are supported on a region of spacetime that is mapped to itself under $R$, but not necessarily point by point (for instance a line, invariant under a translation \cite{Ben-Israel:2018ckc}).

 A particularly important class of twists is when $R$ is an internal symmetry, such as the R-symmetry in ${\cal N}=4$ SYM. In this case, the whole ${\mathbb R}^4$ is invariant and there is no constraint on the positions of the twisted local operators. However, the quantum numbers of the operators should be such that they are invariant under this internal symmetry transformation. In section \ref{sec:sec3} we will consider such a case explicitly.

\paragraph{Marked point.} Every single-trace twist operator comes with a marked point where the twist cut begins/ends. The dependence on this point drops out of any physical computation, once the operator is properly normalised. For example, for the operator in figure \ref{twistcut} we choose that point to lay between $\phi_1$ and $\phi_2$. Deforming that point to start between $\phi_2$ and $\phi_3$ differs by the action of the twist transformation on $\phi_2(x)\to\phi_{2\,\circlearrowleft}(x)$. Since $x_\circlearrowleft=x$, 
this difference is just a linear transformation of the basis of operators.

\paragraph{Possible issues and ambiguities.} 
Even though the construction of the colour-twist described above is very general, potential ambiguities could arise in particular theories. 
The first source of potential problems is the regularisation of the Feynman graphs, which may break the contour deformation symmetry under the local deformations.
Similarly, local operators need to be regularised and renormalised. One should make sure that the regularisation is compatible with the invariance, or it could result in additional subtleties. 

Finally, in gauge theories we have to worry about gauge invariance of our prescription and potential new anomalies. At the technical level, one should make a gauge fixing choice before proceeding to the diagrams, and we have to ensure the result does not depend on that choice. In particular, some gauge choices could break explicitly the global twist symmetry and may lead to a restriction on the allowed gauges.

All these cases require further study of our procedure.\footnote{The twisted diagrams in a gauge theory have already appeared in the literature before in~\cite{Brigante:2005bq} for the case of planar ${\cal N}=4$ SYM theory on $S^3\times S^1$ in the confined phase
under the name {\it Inheritance principle}.} In this paper, we will mostly concentrate on the simple bi-scalar fishnet theory, but we believe that this definition can be applied in a much wider context. In particular, in the case of ${\cal N}=4$ SYM, we believe that our definition should lead to the spectrum described by the deformed Quantum Spectral Curve of~\cite{Gromov:2013pga,Kazakov:2015efa}. 
In this paper, we will give support to this claim, by matching the resummation of twisted Feynman diagrams in the fishnet model with the corresponding QSC prediction. Another piece of evidence comes from considering a Wilson loop with a cusp instead of the single-trace operators. In that case, the same prescription reduces to a simple shift in the cusp angles and therefore agrees with the QSC of~\cite{Gromov:2015dfa} by construction.

\paragraph{Higher point functions.} Similar to the way we compute two-point functions of twisted single-trace operators by twisting the corresponding Feynman diagrams, we may also consider higher point functions. Planar diagrams that contribute to an $n$-point correlation function have the topology of a sphere with $n$ punctures, one for each operator. For twisted operators, we have a set of oriented non-intersecting cuts on the Feynman diagram that emerge from the operators and meet at one of the faces of the diagram. As for the two-point function, the propagators along the cuts are twisted by the corresponding symmetry transformation. When two twist cuts meet, they join to a new twist that is given by the ordered product of the two symmetry transformations.
At the face where all the twist cuts meet, they must satisfy the monodromy condition:
\begin{figure}[h]
\centering
\includegraphics[width=1 \textwidth]{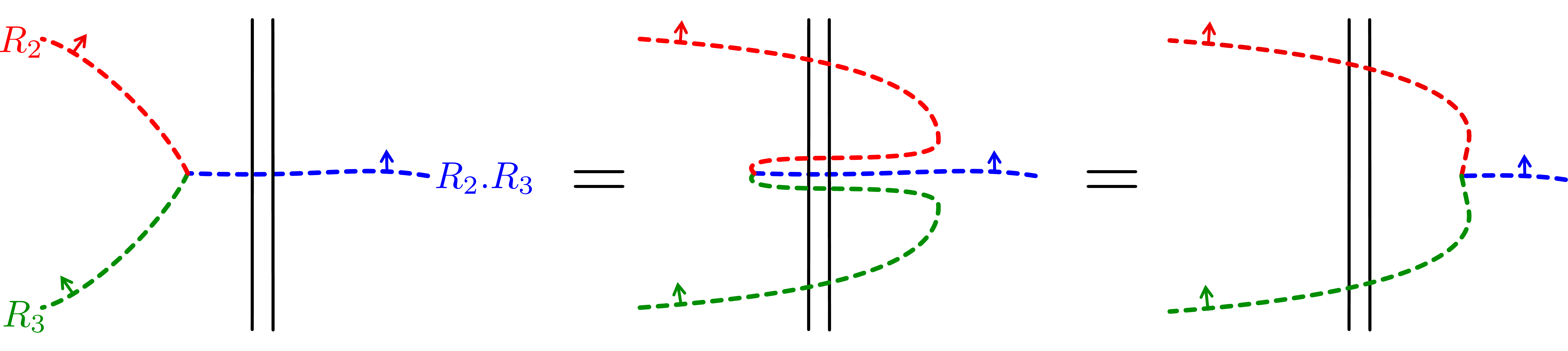}
\caption{\small The monodrony condition (\ref{monodromy}) ensures that the cuts can cancel each other at a face of the diagram. It allows us to drag the meeting point freely across the propagators.}\label{twistcut_example3}
\end{figure}
\beq\la{monodromy}
R_{1}\cdot R_{2}\cdot\dots\cdot R_{n}=1\ .
\eeq
This condition ensures that the twist cuts can  cancel each other at the face of the diagram. The same condition also allows us to deform the cuts so that they meet at any face of the diagram without affecting the result, see figure \ref{twistcut_example3}.

Notice that for more than two operators, this definition requires a particular ordering of the twist transformations in (\ref{monodromy}). This ordering determines the order in which the corresponding twist cuts meet at a face. While this definition is self-consistent for any fixed ordering, 
for some applications one may need to sum over contributions  corresponding to all  possible permutations.  

\begin{figure}[h]
\centering
\def\svgwidth{16cm}
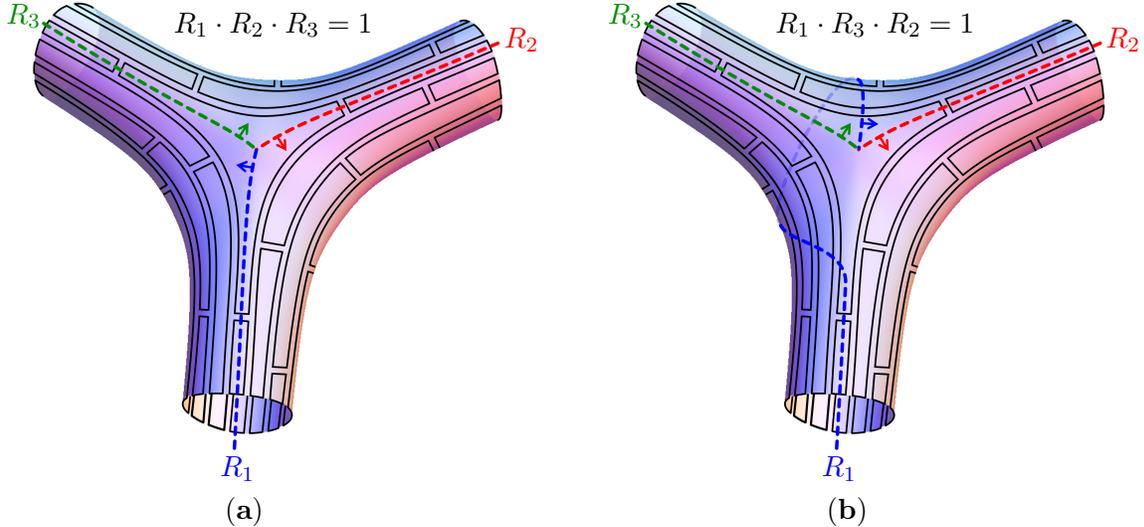
\caption{\small Two different ordering of twist transformations that can contribute to the correlator of three twisted operators with twists $R_1$, $R_2$ and $R_3$. For a given ordering $\sigma$ to contribute, the three twist transformations have to satisfy the monodromy condition $R_{\sigma(1)}.R_{\sigma(2)}.R_{\sigma(3)}=1$.}\label{3D3P}
\end{figure}

The condition \eq{monodromy} ensures that the result does not depend on where we choose the cuts to meet. It can be thought of as a sort of conservation of twist. 
For example, in the case of the two-point function this condition simply becomes $R_2=R_1^{-1}$, which ensures that the arrow along the cut goes from one operator to another. 

In figure \ref{3D3P}.a we draw a twisted Feynman diagram that contributes to the three-point function of three twisted operators. In figure \ref{3D3P}.b we have a different ordering of the twists for the same three-point function. Each ordering gives a separate different deformation to the planar correlator and may be considered in its own right.\footnote{There is a certain analogy with the colour decomposition of a planar scattering amplitude in terms of colour-ordered partial amplitudes. There, one has to sum together all possible orderings to find the physical result.}

Our definition of correlation functions of twist operators can, in principle, be extended order by order in the $1/N$ expansion. However, we will not study this type of corrections in this paper. In the non-planar case, one needs to deal also with possible splittings of the cuts along the diagram surface, making the structure more involved. This question definitely deserves to be addressed in  future studies. 

Note that for the case of the two-point function, one does not need to require the invariance under the twist transformation of both operators.
In this case, the general definition of a correlation function  still remains independent of the  choice of the twist cuts ambiguity. 
The reason is that a twist cut that winds around the cylinder 
can always be unwound through the invariant operator in the correlator. In other words, if one of the two operators is not invariant under the twist, the two-point function is only sensitive to the projection of that operator to the set of invariant twist operators. In section \ref{secOneScalar}, we will use this fact to define the so-called ``CFT wave function".

\begin{figure}[h]
\centering
\includegraphics[width=1 \textwidth]{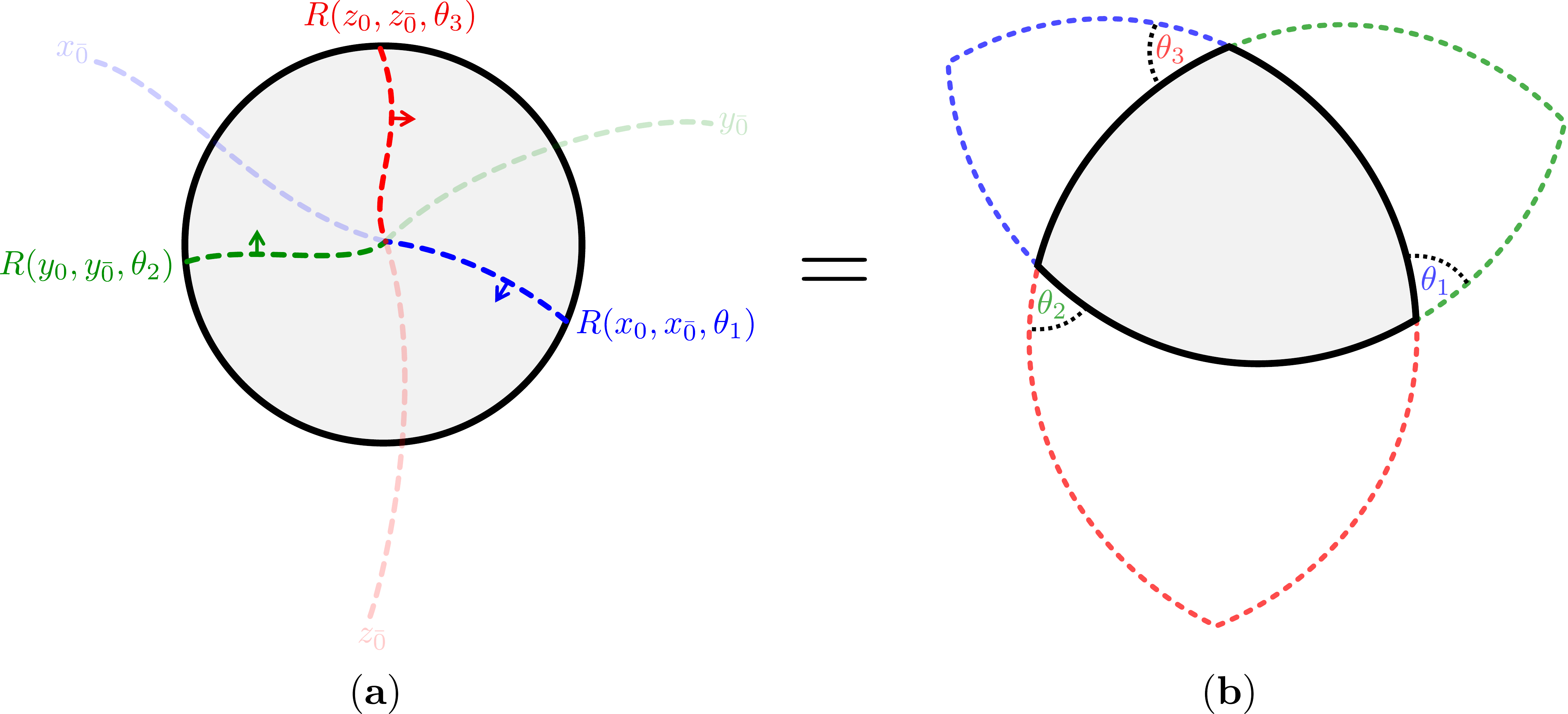}
\caption{\small ({\bf a}) A circular Wilson loop with three colour-twist insertions at three points along the loop $x_0$, $y_0$, and $z_0$. The three twist transformations map the plane of the loop to itself. They have two fixed points each, given by $(x_0,x_{\bar 0})$, $(y_0,y_{\bar 0})$, and $(z_0,z_{\bar0})$. Near these points, they act as rotations by three angles $\theta_1$, $\theta_2$, and $\theta_3$, respectively. The points along the loop and the angles are arbitrary. The points $x_{\bar 0}$, $y_{\bar 0}$, and $z_{\bar 0}$ are fixed by the monodromy condition $R(z_0,z_{\bar 0},\theta_3).R(y_0,y_{\bar 0},\theta_2).R(z_0,z_{\bar 0},\theta_1)=1$. This correlator is equivalent to a Wilson loop with three cusps of angles $\theta_1$, $\theta_2$, $\theta_3$ and no colour-twists insertions that is plotted in ({\bf b}). It is built from three circular arcs.
One arc {(say, the arc  connecting $z_0$ and $x_0$)} can be chosen to be a part of the initial circle, the other two are obtained as a result of the action of $R(z_0,z_{\bar 0},\theta_1)$ 
and
$R(z_0,z_{\bar 0},\theta_1).R(y_0,y_{\bar 0},\theta_2)$ on arcs of the initial circle {(the ones connecting $x_0$ with $y_0$ and $y_0$ with $z_0$, respectively)}. The set-up on the figure ({\bf b}) was studied in \cite{Cavaglia:2018lxi,McGovern:2019sdd} in the ladders limit.}\label{twistcusp}
\end{figure}

 {Finally, the case of correlation function of cusp operators on a piecewise circular Wilson loop that was recently studied in \cite{Cavaglia:2018lxi,McGovern:2019sdd,Giombi:2018qox, Giombi:2018hsx} has an equivalent description in terms of a circular Wilson loop without the cusps. In this description, the effect of the cusp angles is accounted for by including colour-twist operators inserted along the loop, see figure \ref{twistcusp}.} 

\subsection{Kinematics of twisted correlators}\label{sec:kinematics}
Twist operators transform covariantly under a global symmetry transformation $K$. In particular, the twist map $R$ transforms as
\beq\la{Rtrans}
R\rightarrow \widetilde R= K \cdot R\cdot K^{-1}\ .
 \eeq
 
 In this paper, we will focus on the case where using this transformation law, $R$ can be mapped to a simple rotation matrix times an internal symmetry transformation. Such a rotation matrix commutes with dilatations and some other rotations. Correspondingly, local operators in a conformal theory that are twisted by such a transformation $R$ can be characterised by a scaling dimension $\Delta$, some spins $\vec S$, and internal charges $\vec J$, $\CO_i=\CO_{R_i,\Delta_i,\vec S_i,\vec J_i}(x_i)$. The correlation function between $n$ of them transforms covariantly. For example, in the case of scalar operators ($\vec S=\vec J=0$), the correlator transforms as
 \beq\label{eq:kinematics}
\langle \mathcal{O}_{R_1, \Delta_1}(x_{1} )  \dots  \mathcal{O}_{R_n, \Delta_n} (x_n) \rangle =   \left| \frac{\partial \tilde x_1 }{\partial x_1 } \right|^{\frac{\Delta_1}{4} } \dots \left| \frac{\partial \tilde x_n }{\partial x_n } \right|^{\frac{\Delta_n}{4} } \langle \mathcal{O}_{\widetilde R_1, \Delta_1}(\tilde x_{1} )  \dots  \mathcal{O}_{ \widetilde R_n, \Delta_n} (\tilde x_n) \rangle\ ,
\eeq
where $\tilde x_i \equiv K(x_i)$.  
 In the case where $R_i$ are all internal symmetries and $K$ is a conformal transformation, $\widetilde R_i=R_i$, and this transformation law reduces to the standard one of un-twisted local scalar operators.
 
A special case is that of the two-point function $\langle \mathcal{O}_{R_1 , \Delta_1}(x_1) \mathcal{O}_{R_2 , \Delta_2}(x_2)\rangle$ of such scalar twist operators.
Due to the constraint \eq{monodromy}, $R_1=R_2^{-1}\equiv R$, and thus both points $x_1=x_0$ and $x_2=x_{\bar 0}$ are fixed points of $R$. The two-point function is invariant under conformal transformations that leave $x_0$, $x_{\bar{0}}$ fixed. Therefore it can be written purely as a function of these points.  Further, it is invariant under translations and rotations which act on $R$ as in (\ref{Rtrans}). 
Hence, it is only a function of the distance $|x_0-x_{\bar 0}|$. For operators with fixed scaling dimensions $\Delta$ and $\Delta'$, such a correlator must take the standard form\footnote{The orthogonality follows from \eq{eq:kinematics}, when acting with the conjugate dilatation operator, which keeps $x_0$ 
and $x_{\bar 0}$ invariant,
whereas the coordinate dependence follows from \eq{eq:kinematics}
for $K(x)=\frac{x-\eta x_0}{|x-\eta x_0|^2}-\frac{x_{\bar 0}-\eta x_0}{|x_{\bar 0}-\eta x_0|^2}$ and taking $\eta\to 1$.}
\beq
\langle \mathcal{O}_{R , \Delta}(x_0) \mathcal{O}_{R^{-1} , \Delta'}(x_{\bar{0}} )\rangle \propto \frac{\delta_{\Delta, \Delta'} }{|x_0 - x_{\bar{0}} |^{2 \Delta} }\;.
\eeq

The functional form of three and higher point functions is not fixed by the conformal symmetry. The reason is that in these cases there exist conformally invariant functions of the twist maps and positions.
We will see some examples below in section~\ref{secOneScalar}.

\subsection{The holographic dual of twist operators}\la{holographicsection}

We end this section with a short discussion about the holographic dual of twisted operators. 
We postpone a more detailed investigation and explicit checks to future study. We consider some examples below in section~\ref{sec:dual}.

Consider a single-trace operator in the planar limit of ${\cal N}=4$ SYM theory. Such an operator is dual to a single closed string state in $AdS_5\times S^5$. In particular, this duality maps the periodicity of the trace to the periodicity of the closed string. 

Recall that an operator can only be twisted by a symmetry transformation that leaves it invariant. Global symmetries of ${\cal N}=4$ SYM theory are dual to isometries of $AdS_5\times S^5$. Hence, in the holographic dual picture, we are considering a string state with zero charge or momentum along this isometry direction in the bulk.
\begin{figure}[h]
\centering
\def\svgwidth{12cm}
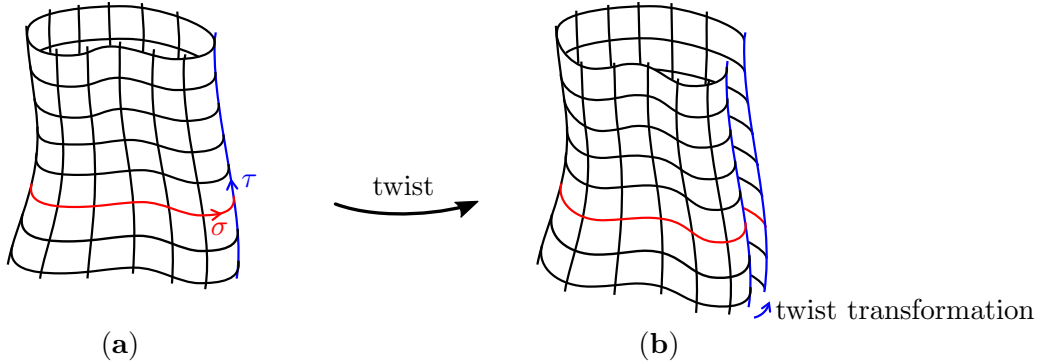
\caption{\small {\bf a}) A closed string in $AdS_5\times S^5$ that is dual to a single-trace operator in the boundary ${\cal N}=4$ SYM theory. {\bf b}) After twisting of the single-trace operator, the dual string 
is no longer closed. Instead, it is described by a map from the worldsheet to $AdS_5\times S^5$ that is not single valued. As we go around the worldsheet, we end at two different points in spacetime that are related to each other by the twist isometry transformation.}\label{twisted_string}
\end{figure}

As we discussed above, twisting such a single-trace operator amounts to dressing the periodicity of the trace with a symmetry transformation. Hence, it is natural to expect that the periodicity of the dual closed string is modified in an analogous way. Namely, we expect the string dual of a twisted operator to be described by a map from a cylindrical worldsheet to $AdS_5\times S^5$ that is no longer single valued. As we go around the cylinder, we end at two different points in $AdS_5\times S^5$ that are related to each other by the twist transformation. In the target space the string will not be closed, but will have an extension in the twist-isometry direction.
Essentially, it becomes periodic with period given by the twist transformation, so that both coordinates and all derivatives are matched by the twist transformation, see figure \ref{twisted_string}.\footnote{Such strings were considered in \cite{Ben-Israel:2018ckc} in the context of non-planar scattering amplitudes.} For example, 
for the $\gamma$-deformation, we end up with a string that is extended along the equator of $S^5$.\footnote{One way of studying such twisted string states is using T-duality. The reason is that T-duality along the twist isometry direction maps a twisted string to a normal un-twisted closed string. At the same time, this closed string propagates in the T-dual background and carries non-zero momentum. For the case of the $\gamma$-deformation one should get the closed string moving in the Lunin-Maldacena~\cite{Lunin:2005jy} background.}

\section{Twisted operators in planar ${\cal N}=4$ SYM theory and the fishnet model}\label{sec:sec3}

In this section, we study twisted operators in ${\cal N}=4$ SYM theory. This theory enjoys a $PSU(2,2|4)$ symmetry, and one may study operators twisted by any element of that group.\footnote{
The twisted ABA equations that correspond to such twisted operators were studied in \cite{Beisert:2005if}.}
We  focus on two types of such twist transformations, which are of particular relevance for the rest of the paper. 
The first type is a twist by an element of the $SU(4)$ R-symmetry. 
{As we discuss in more detail below, a subset of correlators between operators twisted by this symmetry coincides with the ones arising in the $\gamma$-deformed theory \cite{Frolov:2005iq,Frolov:2005dj}.} In particular, the correlators in the strongly $\gamma$-deformed fishnet model of \cite{Gurdogan:2015csr} can be interpreted as a double scaling limit of correlation functions of colour-twist operators. 
Our alternative interpretation, where we deform the operators, rather than the theory, affords us more freedom in choosing the twist parameters for different operators, and results in a more general class of correlation functions. 
{For example, in the fishnet limit, we can assign an independent coupling to each of the operators.} The second type of twists is by spacetime symmetries with two fixed points. Operators twisted by this type of symmetry  still have a well-defined scaling dimension; however, the degeneracy between primaries and descendants is lifted.

\subsection{R-symmetry twist and relation to $\gamma$-deformed ${\cal N}=4$ SYM theory}\label{gamma-deformetion}

In general, we may consider the correlation functions between operators twisted by any set of $SU(4)$ transformations, subject to the monodromy condition (\ref{monodromy}). We will now prove that correlation functions between operators in ${\cal N}=4$ SYM theory that are twisted by a certain family of such transformations coincide with the correlation functions of the $\gamma$-deformed ${\cal N}=4$ SYM theory \cite{Lunin:2005jy,Frolov:2005iq,Frolov:2005dj}. 

\paragraph{The $\gamma$-deformation twist.} Consider a general operator of $\mathcal{N}=4$ SYM, $\cO_{\vec J}$, carrying charge $\vec \Phi \cdot \vec J$ under the three-parameter family of commuting transformations
\beq\label{eq:gammatwist}
G(\vec\Phi)\circ(Z,X,Y) = (e^{i\Phi_1 }Z,e^{i \Phi_2}X, e^{i \Phi_3}Y )\ .
\eeq
Such an operator is left invariant under the transformations  $G(\vec\gamma\times\vec J)$, where $(\vec\gamma\times\vec J)_i=\epsilon_{ijk} {\mathbf{ \gamma}}_j J_k$ and $\vec\gamma$ is an arbitrary three-vector. Hence, it can be twisted accordingly. We denote such an operator $\cO_{\vec J_a}$ that has been twisted by  $G(\vec\gamma\times\vec J_a)$, a $\gamma$-twisted operator. As we will now show, correlation functions between $\gamma$-twisted operators coincide with those of the $\gamma$-deformed theory.

\paragraph{The $\gamma$-deformed ${\cal N}=4$ SYM theory.}
The $\gamma$-deformed theory is defined starting with the Lagrangian of the ${\cal N}=4$ SYM theory and deforming the coefficients in front of the interaction vertices by the phase factors
\beq\label{eq:Qvertex}
 ( \text{\tt vertex} )\quad \rightarrow\quad e^{\frac{i}{2} \sum_{n < m} \vec Q_n \star\,\vec Q_m}\times (\text{\tt vertex} )\ ,
 \eeq
where $\vec Q_n\cdot\vec\Phi$ is the charge of the $n$-th field in the single-trace vertex under $G(\vec\Phi)$, (\ref{eq:gammatwist}), and the star product between two vectors is defined as 
 \beq
 \vec v\star\vec u=-\vec\gamma\cdot(\vec v\times \vec u)=-\epsilon^{ijk}\gamma_iv_ju_k\ .
 \eeq

\paragraph{Matching the correlators.}
\begin{figure}[t]
\centering
\def\svgwidth{12cm}
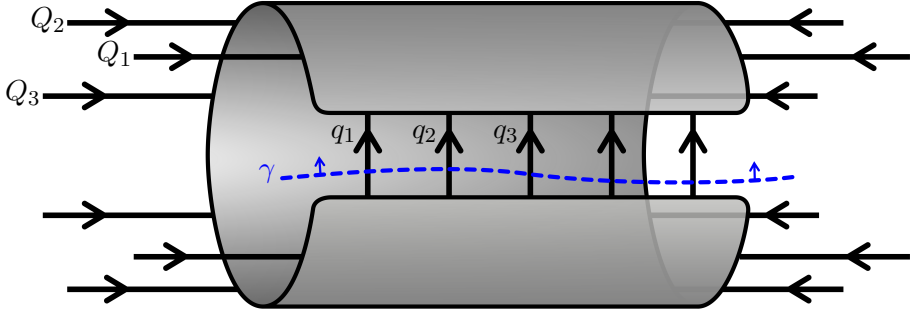
\caption{\small A planar diagram with cylinder topology, cut open into a disk along a set of internal propagators. The non-trivial effect of the $\ast$-deformation of the diagram (\ref{eq:Qvertex}) localises on the propagators crossing the cut.}\label{gammatocut}
\end{figure}
To show the equivalence between the correlators of the $\gamma$-deformed theory and the $\gamma$-twisted ones, we will use a result developed in the context of non-commutative field theory \cite{Filk:1996dm} known as the \emph{planar equivalence theorem}. This theorem states that planar Feynman graphs  with a {\it disk topology} in the star-deformed theory (\ref{eq:Qvertex}) are the same as those in the un-deformed theory, up to an overall phase factor
\beq\label{eq:disk}
 [\text{\small \tt deformed disk diagram}] = e^{\frac{i}{2} \sum\limits_{
 i < j   } \vec Q_i \star\,\vec Q_j}\times[\text{\small \tt undeformed disk diagram}]\ ,
\eeq
where $\vec Q_i$ is the incoming charge of the $i$-th external leg. The legs are 
cyclically ordered according to the colour-trace along the boundary of the disk. 

Now consider a planar diagram in the $\gamma$-deformed theory that contributes to the two-point function between two single-trace operators, $\<\cO_{\vec J}\ {\cO}'_{-\vec J}\>$. As discussed above, such a diagram has the topology of a cylinder. To apply the planar equivalence theorem to it we first cut it open into a disk along a set of internal propagators, see figure \ref{gammatocut}. We label the charges of these ordered internal propagators by $\vec q_1,\dots,\vec q_l$ and the charges on the external lines by $\vec Q_1,\dots,\vec Q_m$ and $\vec Q_{m+1},\dots,\vec Q_n$ for the propagators that connect to the two operators correspondingly. In total, the charges on the propagators around the resulting disk diagram belong to four ordered groups of charges
\beq\{\vec Q_1,\dots,\vec Q_m\}\cup\{{-}\vec q_1,\dots,{-}\vec q_l\}\cup\{\vec Q_1',\dots,\vec Q_n'\}\cup\{{+}\vec q_l,\dots,{+}\vec q_1\}\ .
\eeq
Due to charge conservation on the external legs, we have
\beq
\vec J=\sum\limits_{i=1}^m\vec Q_i=-\sum\limits_{i=1}^n\vec Q_i'\ .
\eeq
When plugging these ordered charges into the phase factor in (\ref{eq:disk}) we find a lot of cancellations. In total, we remain with the simple phase factor
\beq\label{eq:localise}
 \[\! \begin{array}{c}{\small \text{\tt cut-cylinder diagram}} \\ {\small \text{\tt deformed }}
 \end{array} \!
 \] =  e^{\frac{i}{2} \sum\limits_{i < j} \vec Q_i \star\,\vec Q_j+\frac{i}{2} \sum\limits_{i < j} \vec Q_i' \star\,\vec Q_j' {-}i\sum\limits_j \vec J\star\vec q_j} \times\[\! \begin{array}{c}{\small \text{\tt cut-cylinder diagram}} \\ {\small \text{\tt undeformed }}
 \end{array} \!
 \]\ .
\eeq
The first (second) phase factors depend only on the charges of the legs that are attached to the first (second) external operator. Each of these factors depends on the location on the trace where we have chosen the cut to start (end). It reproduces the dependence on the marked point for the twist cut and can be absorbed in the normalisation of the operator.

The third phase factor, ${-}i\vec J\star\vec q_j$, precisely reproduces the effect of the twist cut $G(\vec\gamma\times\vec J)$ in (\ref{eq:gammatwist}) on the $j$-th internal propagator crossing the twist cut. Hence, the two-point function in the theory deformed by (\ref{eq:Qvertex}) is the same as the one between $\gamma$-twisted operators in the undeformed theory.

The same proof generalises in a straightforward way to higher point correlation functions. When considering, for example, a three-point function, we have to cut a pants diagram open into a disk. This can be done along the same twist cuts as in figure \ref{3D3P}. Importantly, the twist $G(\vec\gamma\times\vec J_a)$ for different $\vec J_a$'s commute with each other. As a result, the monodromy condition (\ref{monodromy}) is satisfied trivially and does not depend on the ordering, provided that the total charge is conserved. We conclude that all planar correlation functions in the $\gamma$-deformed theory are the same as those between operators twisted by (\ref{eq:gammatwist}) in the un-deformed  $\mathcal{N}$=4 SYM theory. This equivalence can also be generalised order by order in the $1/N$ expansion as will be discussed briefly in section \ref{secdis}.\footnote{Strictly speaking the twisted theory could, in principle, get new type of divergences, which require regularisation in perturbation theory. For the $\gamma$-deformed theory, the presence of such divergences, which require double trace counterterms, was pointed out in \cite{Fokken:2013aea}. However, it was noticed in \cite{Grabner:2017pgm} that in the fishnet model those divergences disappear after the resummation of the perturbation theory and can be ignored at any finite non-zero coupling. }

\subsection{The limit that selects the fishnet diagrams.}\label{sec:fishnetlimit} 

The fishnet model was obtained in \cite{Gurdogan:2015csr} by starting with the $\gamma$-deformed theory and taking a double scaling limit. Hence, it is no surprise that the same scaling limit of a correlation function between $\gamma$-twisted operators results in those of the fishnet model. In fact, our picture where we twist the operators instead of the theory results in a more general set of fishnet-type correlators.  

For simplicity, we choose to focus on operators that consist of only two out of the three complex scalars of the ${\cal N}=4$ SYM theory, $Z$ and $X$, and an arbitrary number of derivatives
\beq\la{ZXop}
\cO_{\vec J}(x)=\tr\(Z(x)^{J_1}D^n\,X(x)^{J_2}\)+\text{permutations}.
\eeq

We choose the vector $\vec\gamma$ to be proportional to the unit vector $\hat\gamma\equiv(0,0,1)$ and consider the correlation function between a set of operators of the type (\ref{ZXop}), $\{\cO_{\vec J_a}(x_a)\}$, each twisted by ${\cal G}(\gamma_a)=G(\gamma_a\,\hat\gamma)$ with different $\gamma_a$'s but the same $\hat\gamma$. 
Next, we take the following double scaling limits 
\beq\la{dslimit}
\lambda=g_{YM}^2N_c\to0\ ,\qquad e^{-i\gamma_i}\to\infty\ ,\qquad\text{with}\qquad \xi^2_i\equiv \lambda\, e^{-i\gamma_i}=\text{fixed}\ ,
\eeq
 {where $\xi_i$ are $n$ independent parameters.} This limit has the effect of projecting out all Feynman diagrams that contribute to the correlation functions of the $\gamma$-twisted operators $\{\cO_{\vec J_a}^{\gamma_a}(x_a)\}$, except those of fishnet type. It is analogous to the limit that selects ladder diagrams for a Wilson loop with a cusp \cite{Correa:2012nk}.\footnote{The twist here plays the analogous role to the one played by the internal cusp angle in \cite{Correa:2012nk}. Also in that context, one can build correlation functions of different cusps, each with its own effective coupling  \cite{Cavaglia:2018lxi,Giombi:2018qox}.}

To see how the projection onto fishnet diagrams results from the limit (\ref{dslimit}), consider for example the two-point function
\beq\la{correlator}
\<\tr\(Z^J(x){\cal G}(\gamma)\)\,\tr\(\bar Z^J(y){\cal G}(-\gamma)\)\>\ .
\eeq
These operators have $J_2=J_3=0$ and since $\hat\gamma=(0,0,1)$, we have $\hat\gamma\times\vec J\propto(0,1,0)$, namely the twist transformation ${\cal G}(\gamma)$ only acts on the $X$ fields. For any twisted Feynman diagram, it measures the $U(1)_X$ charge of all propagators that cross the twist cut, but not the $U(1)_Z$ or $U(1)_Y$ ones. At tree-level there is a single diagram that contributes to (\ref{correlator}). It consists of $J$ free $Z\!\!\,-\!\!\,\bar Z$ propagators connecting the two operators. These propagators are not affected by the twist, which leaves the diagram invariant.
Next, consider loop diagrams. 
In order to maximise the contribution of a diagram, we have to ensure that the maximal amount of the $U(1)_X$ charge crosses the cut.
Otherwise, the diagram is projected out, as in (\ref{dslimit}) the 't Hooft coupling is sent to zero. For example, a gluonic exchange between two scalar lines is suppressed since gluon propagators are not affected by the twist ${\cal G}(\gamma)$. 
{For a $U(1)_X$ charge $q$ that crosses the cut, we get a factor of $e^{-iJq\gamma}$ from the twist. 
The unique diagram where this factor is weighted by exactly ${Jq}$ powers of $\lambda$ is that of the 
fishnet wheels made of the $X$ scalar. 
All other diagrams come with a power of $\lambda$ that is higher than $Jq$} and are projected out in the fishnet limit (\ref{dslimit}). 
For example, a fermion running around the operator only carries a half unit of charge ($q=1/2$) but is still weighted by at least $\lambda^J$. 
Similarly to (\ref{correlator}), for any other correlator one can easily check that in the limit (\ref{dslimit}) we remain with the fishnet diagrams only.\footnote{Note that in (\ref{dslimit}), we have taken the twist angle to be complex. It may be a little subtle how to complexify the symmetry.
While $G(\gamma)$ for such complex $\gamma$ is not an element of $SU(4)$, the action and the corresponding Feynman diagrams are invariant under $G(\gamma)$.  For example, $X$ and $\bar X$ have the opposite charge and therefore the kinetic term, $\tr(D_\mu XD^\mu \bar X)$, is invariant under the opposite rescaling
\beq
X\xrightarrow[G(\gamma)]{} \({\xi^2\over\lambda}\)^JX\ ,\qquad\bar X\xrightarrow[G(\gamma)]{}\({\lambda\over\xi^2}\)^J\bar X\ .
\eeq
Importantly, here we act on $\bar X$ with the same transformation $G(\gamma)$ and not with $G(\gamma^*)$. As a result, the transformed fields are no longer hermitian conjugate to each other. Twisted operators with complex $\gamma$ are still well-defined operators because in our definition in the previous section, where we only used the invariance of the action and the corresponding invariance of the propagators and interaction vertices under the action of $G(\gamma)$.}

In conclusion, in the limit (\ref{dslimit}) we remain with exactly the same Feynman diagrams that are generated by the fishnet Lagrangian \cite{Gurdogan:2015csr}  
\beq\la{bi-scalarL}
\mathcal L = N_c{\rm tr}\Big( \partial^\mu  \bar X \partial_\mu X+\partial^\mu \bar Z \partial_\mu Z 
+ (4\pi)^2 \xi^2  \bar X  \bar Z X Z\Big)\ .
\eeq
In these fishnet type Feynman diagrams, different scalars are circulating around different operators and are weighted by different effective couplings, $\xi_i^2$.\footnote{This is in analogy to the case of cusp correlators in the ladders limit studied in \cite{Cavaglia:2018lxi,Giombi:2018qox}.}

\subsection{Twist by rotation}
The second type of twist we consider is by a spatial rotation. We start with a general discussion of this type of twist in a four-dimensional CFT and then apply these considerations to the fishnet theory.
\subsubsection{General discussion}
\la{rottwist}
The most general four-dimensional rotation can be decomposed as a simultaneous rotation in two orthogonal planes by two rotation  angles, $\vec\theta=(\theta_1,\theta_2)$. This transformation rotates points as 
\beq\la{rotation}
x_\circlearrowleft=R_{\vec\theta}\circ x=(e^{i\theta_1} z_1,e^{-i\theta_1}\bar z_1,e^{i\theta_2} z_2,e^{-i\theta_2}\bar z_2)\ ,
\eeq
where $z_1=x_1+ix_2$ and $z_2=x_3+ix_4$ parametrise the two planes. The transformation (\ref{rotation}) has two fixed points, the origin and infinity.
More generally, we will consider the case where $R$ is any spacetime conformal symmetry transformation related to the one in (\ref{rotation}) by a conjugation with a conformal transformation $K \in SO(5,1)$,
\beq\label{KRK}
\tilde R_{\vec{\theta}}  = K \circ R_{\vec{\theta}} \circ K^{-1} , 
 \eeq
which in this case has fixed points $ x_0 = K(0)$ and $x_{\bar{0}} = K(\infty)$. 
 We will focus on the study of such twists in the case of the fishnet theory, but the  discussion of this section applies to any four dimensional CFT. 
\paragraph{Symmetry breaking pattern.}
For generic rotation angles, the type of conformal transformations in (\ref{KRK}) commute with a
\beq\la{twistsymm}
\underbrace{U(1)_1 }_{\text{rotation in plane (1,2)}}\times \underbrace{U(1)_2 }_{\text{rotation in plane (3,4)}}\times \ \mathbb{R}_{\text{dilatations}}\ \times\ (\mathbb{Z}_2 )_{\text{inversion}}
\eeq
subgroup of the conformal group $SO(5, 1)$. 
Operators that are twisted by $\tilde R_{\vec{\theta}}$ in (\ref{KRK}) are localised at $x_0$ and are characterised by their conformal dimension $\Delta$ and two spins $(S_1,S_2)$ in the two planes of rotation. We will denote such twisted operators as ${\cal T}_{\vec\theta;S_1,S_2,\Delta}(x_{0})$. Note that the translation symmetry is totally broken by the twist.  This implies that the notion of primary and descendant operator is no longer applicable. In the presence of the twist, all states are on an equal footing and the degeneracies of conformal multiplets are completely lifted.

\paragraph{Special cases.} There are a few special points in the space of rotation angles $\vec\theta=(\theta_1,\theta_2)$ where the residual symmetry is enhanced. One line of such points is given when the two angles are equal, $\theta_1=\theta_2$. In this case, the subgroup  of conformal transformations that commutes with the twist is (see appendix \ref{appSO4rep})  
\beq\la{twistsymm2}
SO(1,5)\quad \xrightarrow[\text{twist with }\theta_1=\theta_2]{} \quad
\(U(1)_L\times SU(2)_R\)/{\mathbb Z}_2\times  \mathbb{R}_{\text{dilatations}} \times (\mathbb{Z}_2 )_{\text{inversion}}\ .
\eeq
Another special case is when one of the two angles vanishes, where the twist transformation leaves invariant a two-dimensional plane. In this latter case, we recover a part of the structure of the descendants spectrum, associated to the translations in this plane.

\subsubsection{State invariance}\label{sec:invariance}
Local single-trace operators are classified by their irreducible representations of the rotation symmetry $SO(4)\simeq \(SU(2)_L\times SU(2)_R\)/{\mathbb Z}_2$. They are characterised by their spins $(j_L,j_R)$ and two angular momenta, $(m_L,m_R)=\((S_1+S_2)/2,(S_1-S_2)/2\)$, where $(S_1,S_2)$ are the angular momenta in the two planes $(x_1,x_2)$ and $(x_3,x_4)$. In appendix \ref{appSO4rep} we present a detailed construction of these representations.

Under the rotation (\ref{rotation}), such an  operator transforms by a phase $e^{i\vec\theta.\vec S}$, where $\vec S=(S_1,S_2)=(m_L+m_R,m_L-m_R)$. Hence, it can only be twisted by (\ref{rotation}) if $\vec\theta.\vec S$ is a multiple of $2\pi$. We notice, however, that one can relax this condition while keeping all spins integer, by compensating for this phase by adding an internal symmetry twist. We will now implement this in the fishnet model. 

Single-trace operators in the fishnet model are built out of the two complex scalars and derivatives. They take the schematic form
\beq\la{genop}
\cO(x)=\tr\(Z(x)^{J_1}D_{z_1}^{S_1}D_{z_2}^{S_2}\, X(x)^{J_2}\dots\)+\text{permutations}\ ,
\eeq
where the dots stand for any neutral combination of derivatives and scalars, $D_{z_1}D_{\bar z_1}$, $D_{z_2}D_{\bar z_2}$, $X\bar X$, and $Z\bar Z$. Such operators can carry arbitrary $U(1)_X\times U(1)_Z$ charges and integer angular momentum in the two planes, $(J_1,J_2,S_1,S_2)$. 
They are invariant under the combined action of
\beq\la{dressedtrans}
{\cal R}_{\vec\theta}\equiv R_{\vec{\theta}}\,.H(\vec\theta.\vec S)\ ,\quad\text{where}\quad H(\eta)\circ(Z,X) = (e^{-i\eta/J_1}Z,X)\ .
\eeq
Above, $H$ is our choice of compensating internal rotation making the state invariant. Importantly, here the $\theta_i$'s are arbitrary continuous parameters. As it was with the R-symmetry twist $G(\vec\gamma\times\vec J)$, also here the twist transformation is tailored to the charges of the operator, $J_i$ and $\vec S$.\footnote{When considering correlation functions of more than two operators in most situations one can  adjust the compensating rotations such that the monodromy condition (\ref{monodromy}) is satisfied without any additional constraint on spins.}

\paragraph{Alternative prescription for twisting fishnet diagrams.}
 It turns out that in the fishnet model we can introduce a simpler twisting prescription that is equivalent to (\ref{dressedtrans}) for operators of fixed spin. Without loss of generality, we consider an operator with $|J_1|\ge0$ and pick a marked point along the trace from which the twist cut will emerge. For any Feynman diagram that contributes to a correlation function of this operator with some other operators, we pick the unique cut that does not cross the $Z$ lines. Because this choice is well-defined at all orders, we do not need to consider the effect of topologically non-trivial deformations of the cut. Therefore, we do not need to require that the operator is invariant under the twist.

For operators of the type (\ref{genop}), this prescription is equivalent to (\ref{dressedtrans}) because the $Z\!-\!\bar Z$ propagators are not cut and therefore $H(\vec\theta.\vec S)$ in (\ref{dressedtrans}) does not play a role. Since the operators (\ref{genop}) form a complete basis, the two twisting prescriptions are equivalent. 
In what follows, we will always use this simpler prescription.

\paragraph{Operator length.} 

For single-trace operators in the fishnet theory, the length is defined by $L={\rm max }(|J_1|,|J_2|)$. It is the length of the corresponding spin chain state in the integrable formulation. 
Twisted operators can have length 
$L=0,1,2,\dots$, where the case $L=0$ corresponds to a twisted identity operator or, in other words, a {\it pure twist} operator. In the un-twisted case, the first non-trivial operator appears at $L=2$. In contrast, in the twisted case $L=1$ is already non-trivial. In this paper, we will focus on the simplest cases of colour-twist operators with length $L=0$ and $L=1$. In certain cases, this simplification will allow us to re-sum all diagrams.

 \section{Colour-twist operators with one scalar}\label{secOneScalar}

Twisted operators in the fishnet limit can come with an arbitrary number of scalars. The smaller that number, the simpler the corresponding Feynman diagrams. The simplest single-trace twisted operator has no scalars at all. That operator, however, turns out to be trivial, having zero conformal dimension and no loop corrections. Hence, here we will focus on the simplest non-trivial case of twisted operators with a single scalar and an arbitrary number of derivatives. Considering such short operators will allow us to obtain analytic results for their conformal dimension and some correlation functions at finite coupling.\footnote{Besides the results described in this paper, more general correlators will be reported in \cite{colortwist2}.} In the absence of a twist, such operators of the $U(N)$ theory decouple from the planar $SU(N)$ sector.  
They are however still very useful for understanding the general structure and  
are needed for the non-planar integrability of the model \cite{Bargheer:2017nne,Bargheer:2018jvq}.

\subsection{The $J=1$ CFT wave function}

Twisted operators of unit charge, spin $\vec S=(S_1,S_2)$ and conformal dimension $\Delta$ take the schematic form
\beq\la{J1op}
\cO_{\vec\theta,\Delta,\vec S}(x_0)\propto \tr\({\cal R}_{\vec\theta}\,\d_1^{S_1}\d_2^{S_2}\,(\d_1\bar\d_1)^{n_1}\,(\d_2\bar\d_2)^{n_2}\,Z(x_0) \) \ ,
\eeq
where $\d_i={\d\over\d z_i}$ are the derivatives in the two planes $z_1=x_1+ix_2$ and $z_2=x_3+ix_4$. 
Here, the twist transformation ${\cal R}_{\vec{\theta}}$ is the rotation (\ref{rotation}) dressed with an $SU(4)$ transformation as in (\ref{dressedtrans}), to ensure invariance of the operator under the twist. In this conformal frame, its two fixed points are $x_0=0$ and $x_{\bar 0}=\infty$. To change these fixed points, one has to apply a conformal transformation to (\ref{J1op}). 

In principle, one can also add a number of neutral pairs of $X$ and $\bar X$. Furthermore, using the equation of motion $\Box Z\propto \d_1\bar\d_1 Z+\d_2\bar\d_2 Z\,\propto \xi^2 X Z \bar X$, one can get rid of all powers of $\d_2\bar\d_2$ i.e. setting $n_2=0,\;n_1\equiv n$, by the price of introducing extra $X\bar X$.
Notice, however, that the operators containing $X$ or $\bar X$ will mix with each other as explained in \cite{Gromov:2017cja}
by means of three moves
$ZX\to XZ,\;\bar XZ\to Z\bar X$ or $\bar X Z X\to X Z\bar X$.
It is clear that the operation of applying those three moves is nilpotent and will necessary terminate after finitely many steps,
implying that the mixing matrix can be brought to an upper triangular form with zeros on the diagonal. From that simple argument, we conclude that all operators involving $X$ and $\bar X$ (or their derivatives)
belong to a logarithmic multiplet with zero anomalous dimension.
Thus, to get a non-trivial dimension, we will focus on the operators \eq{J1op} with $n_2=0$. 

At the loop levels, depending on the regularisation scheme, the  twisted operator in (\ref{J1op}) can still mix with the operators containing $X\bar X$ pairs.\footnote{For example one may perform a point-splitting regularisation by moving $Z$ slightly away from the fixed point $x_0$. }  In order to avoid this scheme dependent mixing problem at finite coupling it was suggested in~\cite{Gromov:2019bsj}
to consider the so-called ``CFT wave function".
The CFT wave function is a way to describe the local operator  by its correlation function with a point-split set of fundamental fields. In the present case it is given by the correlator 
\beq\la{wf}
\Psi
(x)\equiv\<\cO_{\vec\theta,\Delta,\vec S}(x_0)\,\tr(\bar Z(x)\,{\cal R}_{-\vec\theta})\> \ .
\eeq
Here, the operator $\tr(\bar Z(x)\,{\cal R}_{-\vec\theta})$ is a twisted trace made of a conjugate scalar at $x$ and the inverse twist transformation ${\cal R}_{-\vec\theta}$, which in particular has the same fixed points as in (\ref{J1op}). This operator is similar to the operator in (\ref{J1op}). The only difference is that now, instead of having derivatives, the field is separated from the twist fixed point, $x\ne x_{\bar 0}$.

Note that the
non-local operator $\tr(\bar Z(x)\,{\cal R}_{-\vec\theta})$ is not invariant under the dressed rotation ${\cal R}_{\vec \theta}$. As discussed in  section \ref{sec_ct}, the twisted correlator (\ref{wf}) is still well defined because it is sufficient that the operator $\cO_{\vec\theta,\Delta,\vec S}(0)$ is invariant. The non-local operator can be thought of as a generating function of local operators with different spins at $x_{\bar 0}$. In that sum, only the spin $-\vec S$ operator contributes to the correlator (\ref{wf}), while all other local operators are projected out.\footnote{For the $J=1$ case at hand, one may equivalently define the CFT wave function as the three-point function $$\Psi_{\vec\theta,\Delta,\vec S}(x)=\<\CO_{\vec\theta,\Delta,\vec S}(x_0)\,\tr(\bar Z(x))\,\tr(R_{-\vec\theta}(x_{\bar 0}))\>\ ,$$ where $\tr(R_{-\vec\theta})$ is the pure twist operator with no field insertions.}

\begin{figure}[t]
\centering
\def\svgwidth{9cm}
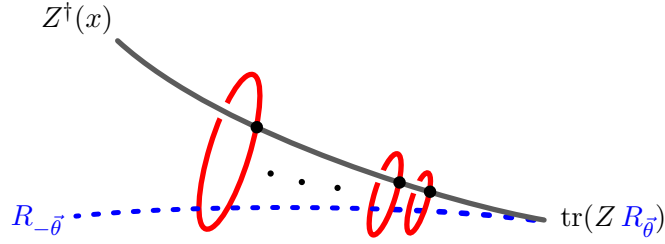
\caption{\small Feynman graphs contributing to the wave function. The graphs can be drawn on a cylinder. The blue dashed line represents the cut used to twist the diagrams according to the rules of section \ref{sec_ct}. The black line is made of $Z$-propagators, and the red lines represent twisted $X$-propagators. }\label{ZT1T0}
\end{figure}

\begin{figure}[t]
\centering
\def\svgwidth{15cm}
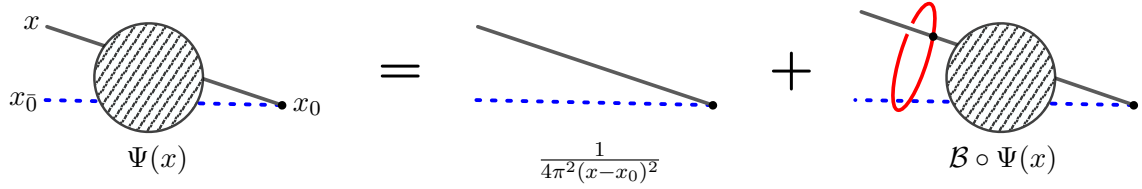
\caption{\small The correlator (\ref{wf}) satisfies a Dyson-type evolution equation. On the right hand side we have the tree-level propagator (first term) plus the correlator with one more wheel of the $X$-field (in red). The addition of a wheel is implemented by the graph building operator ${\cal B}$.  }\label{Dysonfig}
\end{figure}

The Feynman diagrams that contribute to the CFT wave function, given by the correlator (\ref{wf}), are of the iterative type drawn in figure \ref{ZT1T0}. At tree-level, we have a free scalar propagator between the origin and $x$, (the black line). At the $l$-th loop order, we have $l$ interaction vertices of the fishnet model inserted along the $Z\!-\!\bar Z$ line. Each vertex is contracted back with itself by an $X\!-\!\bar X$ propagator (the red lines). All of these $X\!-\!\bar X$ propagators cross the twist cut (the blue dashed line) and are therefore twisted. Namely, they connect the interaction point $y$ to its image under the twist, $y_{\circlearrowleft}$. 
More explicitly, for the choice $x_0=0$
and $x_{\bar 0}=\infty$ we have
\beq\label{eq:Dyson}
\Psi(x)={1\over4\pi^2(x-x_0)^2}+
16 \pi^2\xi^2\int d^4 y
{1\over4\pi^2(x-y)^2}
\frac{1}{4\pi^2(y- y_{\circlearrowleft})^2 }{1\over4\pi^2(y-x_0)^2}+\dots\;.
\eeq
Due to this iterative diagrammatic structure, the correlator (\ref{wf}) satisfies a Dyson-type evolution equation, see figure \ref{Dysonfig}, given by
\beq\la{Dyson} 
\Psi(x)={1\over4\pi^2(x-x_0)^2}+\cB\circ\Psi\ ,
\eeq
where $\cB$ is the ``graph building operator" \cite{Gurdogan:2015csr}  
\beq\la{evolution}
\mathcal{B} \circ \Psi(x) \equiv \frac{\xi^2}{\pi^2}\int d^4 y\frac{\Psi(y)}{(y- y_{\circlearrowleft})^2 ( x-y)^2}\ .
\eeq
Acting with this operator on the Feynman diagram in figure \ref{ZT1T0} would create a new diagram with one extra $X-\bar X$ wheel. By acting on both sides of (\ref{Dyson}) with $\cB^{-1}=-{1\over{4 {\xi^2}}}(x-x_{\circlearrowleft})^2\,\Box_x$ we arrive at the differential equation for $x\neq x_0$,
\beq\la{Binv}
-{1\over4}(x-x_{\circlearrowleft})^2\,\Box_x\,\Psi(x)=\xi^2\,\Psi(x)\ .
\eeq
We will now use its global symmetries to map this equation into a 1D Schr\"odinger equation. The operator ${\cal B}^{-1}$ commutes with dilatations and rotations in the two planes. Hence, its eigenfunctions can be characterised by two spins $S_1,\,S_2$ and the conformal dimension $\Delta$. These global charges determine the dependence of the wave function on $x$, up to a function of the ratio $r_1/r_2$ as
\beq\la{waveea}
\Psi(x)=\({x_1+ix_2\over r_1}\)^{S_1}\({x_3+ix_4\over r_2}\)^{S_2}{1\over(r_1^2+r_2^2)^{\Delta+1\over2}}\times \psi\(\log{r_1\over r_2}\) \ ,
\eeq
where $r_1^2= x_1^2 + x_2^2$, $r_2^2 = x_3^2 + x_4^2$ are the radial coordinates in the two planes of rotation. Here, the total scaling dimension of $\Psi$, equal to $\Delta+1$, is given by the sum of the scaling dimension $\Delta$ of the twisted operator 
${\cal O}$ and the protected scalar $\Delta_0=1$
in the correlator (\ref{wf}). By plugging this form of the wave function into the evolution equation (\ref{Binv}), and using that 
\beq\la{twisedprop}
(x-x_{\circlearrowleft})^2 = 4 \, \(r_1^2\sin^2{\theta_1\over2}+r_2^2\sin^2{\theta_2\over2}\)\ ,
\eeq
we arrive at the one-dimensional stationary  Schr\"odinger equation
\beq\la{Sch}
\[-\d_\sigma^2+V(\sigma)\] \psi(\sigma)=0\ ,\qquad \sigma=\log{r_1\over r_2}\ ,
\eeq
where the potential is given by
\beq\la{SchV}
V(\sigma)={1\over2\cosh\sigma}\[\(e^{-\sigma}S_1^2+e^{\sigma}S_2^2\)+{1-\Delta^2\over2\cosh\sigma}-{\xi^2 \over e^\sigma\sin^2{\theta_1\over2}+e^{-\sigma}\sin^2{\theta_2\over2}}\]\;.
\eeq
This equation looks like a stationary zero-energy Schr\"odinger problem for a potential well problem. Instead of the energy in the Schr\"odinger problem, which is set to zero, we should extract the conformal data $\Delta(\xi,S_1,S_2)$. To read the physical CFT spectrum, one has to tune $\Delta$ that enters the potential, so that a solution satisfying the relevant quantisation condition exists.
The relevant quantisation condition is in general different from the standard square integrability of $\psi(\sigma)$, as we describe below.

\paragraph{Quantisation condition.}

\begin{figure}[t]
\centering
\def\svgwidth{9.5cm}
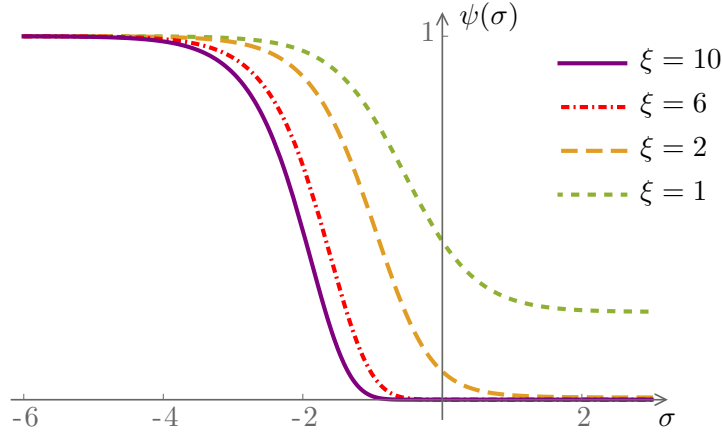
\caption{\small Numerical plot of the ground state wave function (normalised to unity at $\sigma = -\infty$), for zero spins, twist angles $(\theta_1,\theta_2)=(4/3,1/2)$ and coupling constants $\xi = 1 $ (green), $\xi=2$ (orange), $\xi=6$ (red), $\xi=10$ (purple). For zero spins, the wave function is not decaying but asymptotically constant at infinity. 
As we will discuss in more detail in section \ref{sec:fishchain}, it becomes more and more localised around $\sigma = -\infty$ as the coupling increases.}
\label{fig:wavefs}
\end{figure}

As is obvious from its definition, the CFT wave function $\Psi(x)$, given by the correlator (\ref{wf}), can only be singular at $x=x_0$. In particular, it is regular when the argument $x$ is placed on one of the two orthogonal planes, that is, when either $r_1=0$ or $r_2=0$ (but not both at the same time). In the coordinates we are using, these two cases correspond to $\sigma=\log\frac{r_1}{r_2}\to\pm\infty$. In these limits the ``potential" $V(\sigma)$ \eq{SchV} behaves as
\beq
\lim_{\sigma\to+\infty\;{\rm or}\;r_2\to 0} V(\sigma)=S_{2}^2+\cO\(e^{-2|\sigma|}\)\ ,\qquad 
\lim_{\sigma\to-\infty\;{\rm or}\;r_1\to 0} V(\sigma)=S_{1}^2+\cO\(e^{-2|\sigma|}\)\ .
\eeq
This implies the following possible asymptotics
\beq\la{qc0} 
\lim_{|\sigma|\to\infty}\psi(\sigma)\propto e^{\pm|S_i\sigma|}\(1+\cO(e^{-2|\sigma|})\)\ .
\eeq
The growing solutions would result in a singularity, thus we have to require that $\psi(\sigma)$ decays exponentially,
which for $\Psi(x)$ implies regular behaviour $|\Psi(x)|\sim r_i^{S_i}\to 0$ as $r_i\to 0$.
Similarly, for the case when one of the spins is zero, we get two solutions at infinity -- one asymptotically constant and  one linearly growing. Using the same principle, we have to exclude the linearly growing solution, as it would result in a singular $\Psi(x)$. 
In summary, we can 
 express the quantisation condition as
\beq\label{qc2}
\lim_{|\sigma |\rightarrow \infty} \partial_{\sigma}\psi(\sigma) = 0\ .
\eeq
Notice that, in the case where one of the spins is zero, the wave function is not square integrable with the naive flat measure, see figure \ref{fig:wavefs}. As we will explain shortly, the natural measure for the Schr\"odinger problem is in fact non-trivial.

\paragraph{Changing conformal frame.}
Above, we reduced the problem of computing the CFT wave function
to a one-dimensional stationary Schr\"odinger equation with a certain quantisation condition. However, for simplicity, we set $x_0=0$ and $x_{\bar 0}=\infty$. The general configuration with finite fixed points can be mapped to the standard one by a suitable conformal transformation $K$, such that $K(0)=x_0$, $K(\infty) = x_{\bar{0}}$, which relates the twist maps in the two frames as $R = K \circ R_{\vec{\theta}}\, \circ K^{-1}$. The simplest example of such a map 
is the special conformal transformation
\beq\label{eq:explK1}
\left(K^{-1} \circ x \right)^{\mu}= \frac{ x^{\mu} - x_{\bar{0}}^{\mu}}{(x - x_{\bar{0}} )^2} - \frac{ x_0^{\mu} - x_{\bar{0}}^{\mu}}{(x_0 - x_{\bar{0}} )^2}\ .
\eeq
In the following, we assume that this choice is made when discussing the correlator as a function of the fixed points of the twist, which we then can plug into \eq{eq:kinematics}.\footnote{The most general map is related to $K$ in (\ref{eq:explK1}) by rotation and dilatation, which are the conformal transformations that leave the origin and infinity invariant and would result with an extra scalar factor.} As shown in Appendix \ref{app:frame}, the wave function is given by 
\beq\label{eq:Psynew}
\Psi_{}(x)=\frac{{ \hat z_1}^{  S_1 }  \, {\hat z_2 }^{ S_2} }{| x - x_0 |^{\Delta+1} \, |x - x_{\bar 0} |^{1 - \Delta} \, | x_0 - x_{\bar{0}}|^{\Delta - 1}}\times \psi\(\sigma\)\ ,
\eeq
where $\sigma$ and $\hat z_i$ are
\beq\label{eq:defzsig1}
z_i = \left(K^{-1} \circ x \right) \cdot n_i\ , \qquad
\hat z_i = z_i/|z_i|\ ,\qquad \sigma = \log(| z_1/z_2| )\ ,
\eeq
with $n_1 = (1,i,0,0)$, $n_2 = (0,0,1,i)$. 

Note that, for zero spins, the wave function $\Psi(x)$ has the structure of the three-point correlator between an operator of dimension $\Delta$ at $x_0$, an operator of dimension $1$ at $x$ and an operator of dimension $0$ at $x_{\bar 0}$. However, an important difference as compared to the standard CFT case is that the coefficient $\psi(\sigma)$ carries an additional spatial dependence. 
\paragraph{The measure.} 
The measure for the functions of four  variables $\Psi(x)$, which plays an important role in what follows, is defined as\footnote{ On a solution to the evolution equation (\ref{Binv}), this measure can also be written also as $\langle\! \langle \Psi_1|\Psi_2^\text{on-shell} \rangle\! \rangle \propto \int d^4 x\, \overline{\Psi}_1(x) \, \Box_x\,\Psi_2^\text{on-shell}(x)$, which has the form of the ``CFT norm'' defined in \cite{Gromov:2019jfh}. }
\beq\label{innerCFT}
\langle\! \langle \Psi_1|\Psi_2 \rangle\! \rangle \equiv \frac{1}{\pi^2} \, \int d^4 x \frac{\left|{\d x_\circlearrowleft\over\d x}\right|^{\frac{1}{4} } }{(x - x_{\circlearrowleft} )^2} \, \overline{\Psi}_1(x) \, \Psi_2(x)\ .
\eeq
It is easy to check that 
with this measure the operator $\cB^{-1}$ (\ref{Binv}) is self-adjoint.\footnote{See appendix \ref{app:frame} for an explicit expression for the graph-building operator in a general frame.} 

For physical wave functions that correspond to operators in the theory, the wave function and the bar one are given by the correlators (\ref{wf}) and 
\beq
\overline{\Psi}(x)\equiv\<\overline{\cO}_{-\vec\theta,\Delta , \vec S}(x_{\bar 0})\,\tr( Z(x)\,R_{\vec\theta}(x_0))\>
\eeq
where $x_0$ and $x_{\bar 0}$ are the two fixed points of $R_{\vec\theta}$. By this definition, $\overline{\Psi}$ can be obtained from (\ref{eq:Psynew}) by interchanging the role of $x_0$ and $x_{\bar{0}}$, reversing the sign of $\vec \theta  \rightarrow - \vec \theta$, and taking the complex conjugate of the result.\footnote{Incidentally, $\vec \theta \rightarrow -\vec \theta$ and $\vec S \rightarrow - \vec S$ have no effect on the wave function in the present case, since the Schr\"odinger equation is even in the angles and spins.}
Correspondingly, we have 
\beq\la{waveeaconj}
\overline{\Psi}_{\vec \theta, \Delta, \vec S }(x)=\frac{{ \hat z_1}^{ - S_1 }  \, {\hat z_2 }^{ - S_2} }{| x - x_0 |^{-\Delta+1} \, |x - x_{\bar 0} |^{1 + \Delta} \, | x_0 - x_{\bar{0}}|^{\Delta - 1}}\times \overline{\psi}_{\vec \theta, \Delta, \vec S}\(\sigma\) ,
\eeq
where the bar over $\psi$ denotes 
 complex conjugation while treating $\xi$ as a real parameter. 
 By construction, $\overline{\Psi}$ satisfies the same  equation (\ref{evolution})  as the original wave function. By plugging (\ref{waveea}) and (\ref{waveeaconj}) into the ``CFT norm"~\eq{innerCFT}, we arrive at
\beq\la{CFTtoschmeasure}
\langle\! \langle \Psi_{\theta,\Delta,\vec S}|\tilde\Psi_{\theta,\Delta,\vec S'} \rangle\! \rangle=-\delta_{S_1,S'_1}\delta_{S_2,S'_2}\times{2\over\pi}{\log( \epsilon_{\text{UV}})\over|x_0-x_{\bar 0}|^{2\Delta}}\,
\<\psi_{\Delta,\vec S }|\tilde\psi_{\Delta,\vec S }\>\ ,
\eeq
where $\epsilon_\text{UV}$ is a small UV cutoff length scale and the corresponding Schr\"odinger measure is given by
\beq\label{innerSch}
\langle\psi|\varphi\rangle  ={\pi\over2}\int\limits_{-\infty}^\infty d\sigma\,{\overline{\psi}(\sigma)\,\varphi(\sigma)\over\(e^\sigma\sin^2{\theta_1\over2}+e^{-\sigma}\sin^2{\theta_2\over2}\)\cosh\sigma}\ .
\eeq
We see that the measure in (\ref{innerSch}) decays as $e^{-2|\sigma|}$ at large $|\sigma|$. This means that solutions that satisfy the quantisation condition (\ref{qc2}) are also normalisable with respect to the norm (\ref{innerSch}).

\paragraph{Relating the Schr\"odinger and the CFT normalisations.}
In CFT one usually normalizes the operator by setting its two point function to have the standard form $\frac{1}{|x_{1}-x_2|^{2\Delta}}$. This normalization can be related
to the normalization of the wave function according to the norm \eq{innerSch}.
In \cite{Gromov:2019bsj} the following relation between the CFT norm
\eq{innerCFT} and the two-point function of the operators was found 
\beq\la{totpf}
\langle\! \langle \Psi_{\Delta, \vec{S} }|\Psi_{\Delta, \vec{S} } \rangle\! \rangle=8\log(\epsilon_{\text{UV}} )\,  
( \partial_{\xi^2}  \Delta )\, \< \mathcal{O}_{1,\vec\theta,\Delta, \vec{S} }(x_0)\overline{\mathcal{O}}_{1,-\vec\theta,\Delta, \vec{S} }(x_{\bar 0})\>\ .
\eeq
To derive this relation, note that the twisted propagator in the CFT measure (\ref{innerCFT}) has the effect of introducing an extra graph-building operator into the diagrams that contribute to the two point function of $\mathcal{O}_{\Delta, \vec{S} }$. This can be interpreted as the integrated insertion of the interaction vertex in a two-point function and results in (\ref{totpf}).

By comparing (\ref{totpf}) with (\ref{CFTtoschmeasure}), we conclude that the normalisation of the Schr\"odinger measure is related to the normalisation of the twisted CFT operators as
\beq\la{normtoddelta}
\<\psi_{\Delta, \vec S }|{\psi_{\Delta, \vec S }}\>=-4\pi\,(\d_{\xi^2}\Delta)\times{\< \mathcal{O}_{1,\vec\theta,\Delta, \vec S }(x_0)\overline{\mathcal{O}}_{1,-\vec\theta,\Delta, \vec S }(x_{\bar 0})\>\over |x_0-x_{\bar 0}|^{-2\Delta}}=-4\pi\,(\d_{\xi^2}\Delta)\ ,
\eeq
where in the last equality we have fixed the standard CFT normalisation\footnote{One may be confused by the fact that the left-hand side of (\ref{CFTnormalization}) is the two-point function of operators with spins $\pm(S_1,S_2)$ while the right-hand side of that equation looks like a scalar. 
A general two-point function of twisted operators with spins $\vec S$ and $\vec S'$ is proportional to $\delta_{S_1,-S'_1}\delta_{S_2,-S'_2}(\epsilon_1.\epsilon_1')^{S_1}(\epsilon_2.\epsilon_2')^{S_2}$, where the $\epsilon_i$'s are the polarization vectors of the operators in the two planes. 
In our case however, that factor is equal to one by construction.
}
\beq\la{CFTnormalization}
\< \mathcal{O}_{1,\vec\theta,\Delta,\vec S }(0)\,\overline{\mathcal{O}}_{1,-\vec\theta,\Delta,\vec S }(x)\>={
1\over|x|^{2\Delta}}\ .
\eeq
This choice or normalisation can also be written as 
\beq\la{normalter}
\<\psi_{\Delta, \vec{S} }|\frac{e^{\hat \sigma} \sin^2\frac{\theta_1}{2}  + e^{-\hat \sigma} \sin^2\frac{\theta_2}{2} }{\cosh{\hat \sigma} }|{\psi_{\Delta, \vec S }}\>={\pi\over\Delta}\ .
\eeq
To relate (\ref{normalter}) with (\ref{normtoddelta}), we start from the Schr\"odinger equation and consider small variations of the potential with respect to $\xi^2$, which results in a small variation of $\Delta$. In that way we arrive at the relation 
\beq\la{dDeltadxi}
\d_{\xi^2}\Delta^2= -\frac{ \int d\sigma \psi_{\Delta, \vec{S} }(\sigma ) \overline{\psi_{\Delta, \vec S } (\sigma)}\; \partial_{\xi^2} V(\sigma) }{ \int d\sigma \psi_{\Delta, \vec{S} }(\sigma ) \overline{\psi_{\Delta, \vec S } (\sigma)} \; \partial_{\Delta^2} V(\sigma)} =  - \frac{\<\psi_{\Delta, \vec{S} }|{\psi_{\Delta, \vec S }}\>}{2\<\psi_{\Delta, \vec{S} }|\frac{e^{\hat \sigma} \sin^2\frac{\theta_1}{2}  + e^{-\hat \sigma} \sin^2\frac{\theta_2}{2} }{\cosh{\hat{ \sigma} }} |{\psi_{\Delta, \vec S }}\> }\ ,
\eeq
from which (\ref{normalter}) follows.

\begin{figure}[t]
    \centering
    \includegraphics[width=0.7 \textwidth]{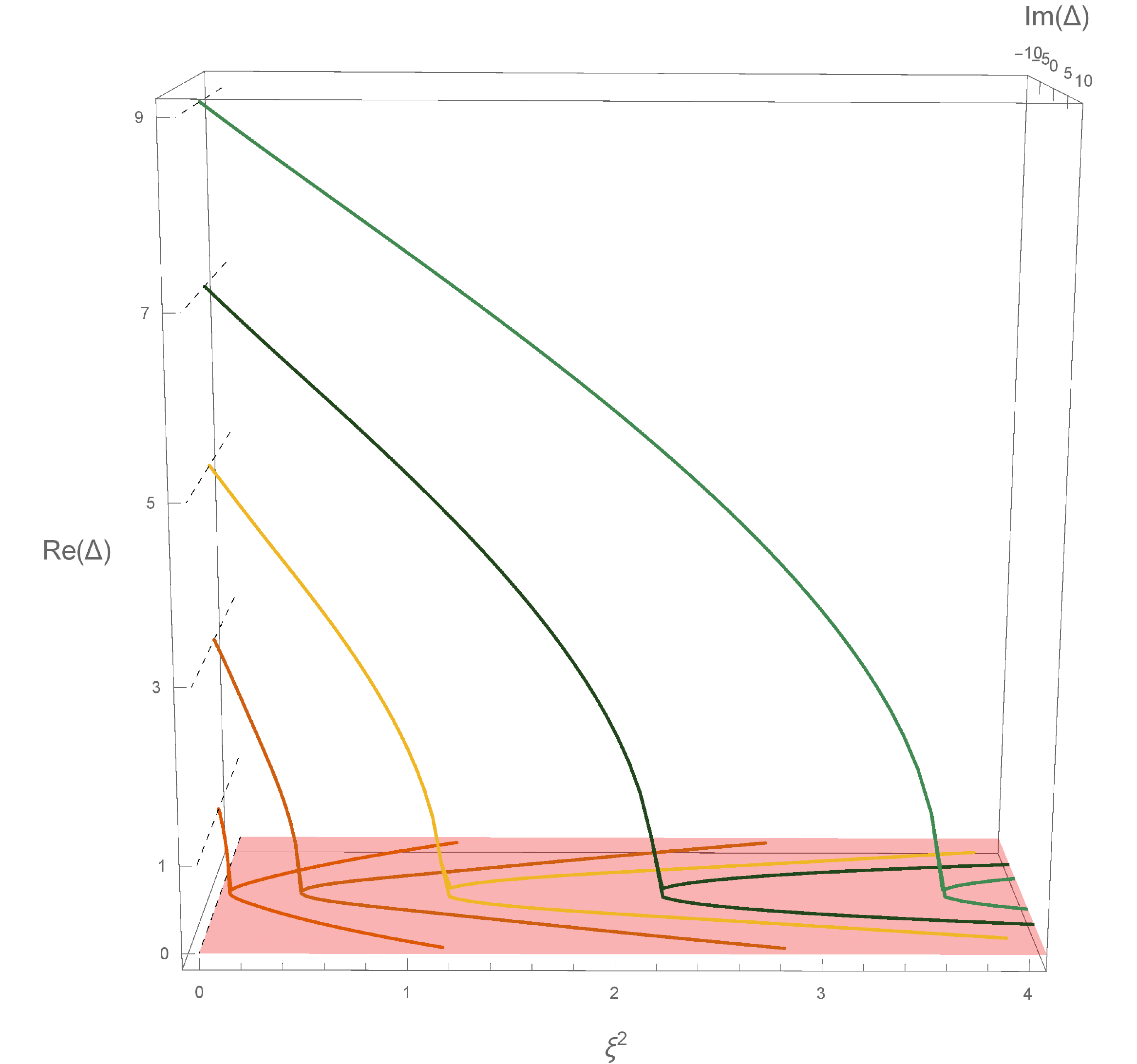}
    \caption{The first five states for $J_1=1,\; J_2=0$, $S_1=S_2=0$, with $\theta_1=5/6$ and $\theta_2=1/6$. All states reach a branch point where $\Delta  = 0$, after which the scaling dimension becomes purely imaginary and scales classically, $| \Delta| \sim \xi$ for $\xi \rightarrow \infty$. The strong coupling behaviour will be studied in section \ref{sec:fishchain}.  \label{fig:0mag}}
\end{figure}

\subsection{Structure of the spectrum: general situation}\label{sec:numerics0}
For two generic twist angles we cannot solve the Schr\"odinger equation analytically. Still, we have complete numerical control over it.
In figure \ref{fig:0mag} we have plotted the spectrum for $\vec S=0$, $\vec\theta=(5/6,1/6)$ and real $\xi$ obtained numerically by solving the stationary Schr\"odinger equation~\eq{Sch} with the boundary conditions described in the previous section.

As one can see from the plot, at zero coupling the spectrum is given by the bare operators of dimension
$1+2n+|S_1|+|S_2|$. We refer to $n$ as the excited state number. It can be related to the number of derivatives used to build the state at weak coupling,   \eq{J1op} with $n_2=0$ and $n_1=n$.

As we turn on the coupling $\xi$, the spectrum starts to deviate from the trivial one while the number of states is conserved. In particular,  
there is only one non-protected state for every choice of the spins and classical dimension. These states can be understood as broken conformal descendants.

At some critical real value of $\xi$, each of the trajectories $\Delta^2_{n}(\xi)$ passes smoothly through zero. After this point, the scaling dimension $\Delta_{n}(\xi)$ splits into two purely imaginary levels. At strong coupling, all levels scale as $\Delta_{n}(\xi^2) \sim \pm i \xi$.

\subsection{Special case of equal angles}

It turns out that for equal angles, $\theta_1=\theta_2\equiv\theta$, one can solve the Schr\"odinger problem analytically.\footnote{{This simple solvable case was also very useful  to test of the duality with the quantum fishchain model proposed in \cite{Gromov:2019aku}. In fact, after we had obtained the result for the spectrum presented below, it was reproduced by two of the present authors from the dual model in   \cite{Gromov:2019bsj}.}} The reason for that is the large amount of symmetry that is preserved by the twist in this case. As discussed in section \ref{rottwist}, for equal angles the twist preserves an extended subgroup of conformal symmetries (\ref{twistsymm2}). In particular, the rotation symmetry is only broken down to $SO(4)\to SU(2)\times U(1)$.  
For one scalar, that symmetry is enhanced even further, to the full $SO(4)$ group of rotations. To see that, we note that the twisted propagator (\ref{twisedprop}) now takes the form
\beq
(x - x_{\circlearrowleft} )^2=4\sin^2\(\theta/2\)\times|x|^2
\eeq
and depends only on the absolute magnitude of $x$, but not on its direction. 
The corresponding potential in the Schr\"odinger equation takes the form
\beq
V(\sigma)={S_1^2\over1+e^{2\sigma}}+{S_2^2\over1+e^{-2\sigma}}+{1-\Delta^2-\xi^2/\sin^2(\theta/2)\over4\cosh^2(\sigma)}\ .
\eeq
The solutions of the Schr\"odinger differential equation subject to the boundary conditions (\ref{qc2}) can be found explicitly. In order to satisfy the boundary conditions one has to restrict $\Delta$ to the following values 
\beq\label{eq:equalsp}
 \Delta_{n}  
=\sqrt{(1+2n+|S_1|+|S_2|)^2
-\xi^2/\sin^2(\theta/2)}
\ , 
\eeq
where $n$ is a non-negative integer that is equal to the excited state number introduced in the previous section. The corresponding wave functions are 
\beq\label{eq:psizero}
\psi_n(\sigma)={\,_2F_1\(-n, 1 + n + |S_1|+|S_2|, 1 + |S_2|,{1\over1+e^{2\sigma}}\)\over(1+e^{2\sigma})^{|S_2|\over2}(1+e^{-2\sigma})^{|S_1|\over2}}\equiv{P_n^{\vec S}(e^{2\sigma})\over(1+e^{2\sigma})^{n+{|S_2|\over2}}(1+e^{-2\sigma})^{|S_1|\over2}}\ ,
\eeq
where $P_n^{\vec S}(x)$ is a polynomial of degree $n$.\footnote{For example, for zero spins, $P^{0,0}_n(x) = (1 + x)^n \, P_n(\frac{x-1}{x + 1})$, where $P_n$ is the Legendre polynomial of degree $n$.} Notice that this wave function is independent of the coupling and therefore coincides with the tree-level one.  Explicitly, the wave function is fixed by the unbroken $SO(4)$ symmetry and is given by
\beq\label{eq:Psider}
\Psi_{\Delta_{n}(\xi),\vec S}(x)\ \propto \ 
x^{-\gamma}(\partial_1 \bar{\partial}_1 )^{n} \, \bar{ \partial}_1^{S_1} \bar{\partial}_2^{S_2} {1\over x^{2}}\ ,
\eeq
where $\gamma=\Delta_n(\xi)-\Delta_n(0)$, $\bar{\partial}_a^{-|k|} \equiv \partial_a^{|k| }$ and the $SO(4)$ spin is $S= 2 n + |S_1| + |S_2|$.

\subsection{Strong coupling}\label{sec:fishchain}
In this section, we study the strong coupling limit $\xi^2 \rightarrow +\infty$ of the Schr\"odinger equation. 
We will show that the states $\Delta_n$ with some fixed spins $S_1,S_2$
can be described quasi-classically (i.e. by means of WKB analysis) even for the lowest lying states. The strong coupling behaviour depends on the values of the angles parameters. Without loss of generality, in this section we assume that $\sin^2{\theta_1\over2}\ge \sin^2{\theta_2\over2}$. 
 
The problem under consideration is a  Schr\"odinger equation
\beq\la{Schoffshell}
\[-\d_\sigma^2+V(\sigma)\] \psi(\sigma)=E(\Delta) \, \psi(\sigma)\ ,
\eeq
where the scaling dimension $\Delta$ and the coupling $\xi$ enter as a parameters in the potential (\ref{SchV}). The physical scaling dimension is obtained by tuning $\Delta$, so that the Schr\"odinger energy vanishes
\beq\la{SchE}
E(\Delta_\text{phys})=0\ .
\eeq
From the numerical results described in section \ref{sec:numerics0}, we know that at strong coupling and for fixed values of the spins, $\Delta$
become purely imaginary and
$\Delta \sim \mathcal{D} \xi$, where $\mathcal{D}^2 < 0$. 
 With this scaling, the strong coupling potential takes the form,
\beq\label{potentialstrong}
V(\sigma) = {\xi^2\over2\cosh\sigma}\[\(e^{-\sigma} \mathcal{S}_1^2+e^{\sigma} \mathcal{S}_2^2\)-{\mathcal{D}^2\over2\cosh\sigma}-{ 1 \over e^\sigma\sin^2{\theta_1\over2}+e^{-\sigma}\sin^2{\theta_2\over2}}\]\ , 
\eeq
where we also re-scaled the spins as $S_i=\xi{\cal S}_i$. 
In the following, we will drop these parameters and only consider the case of zero classical spins,  $\mathcal{S}_i = 0$. 
\begin{figure}[t]
\centering
\includegraphics[scale=0.53]{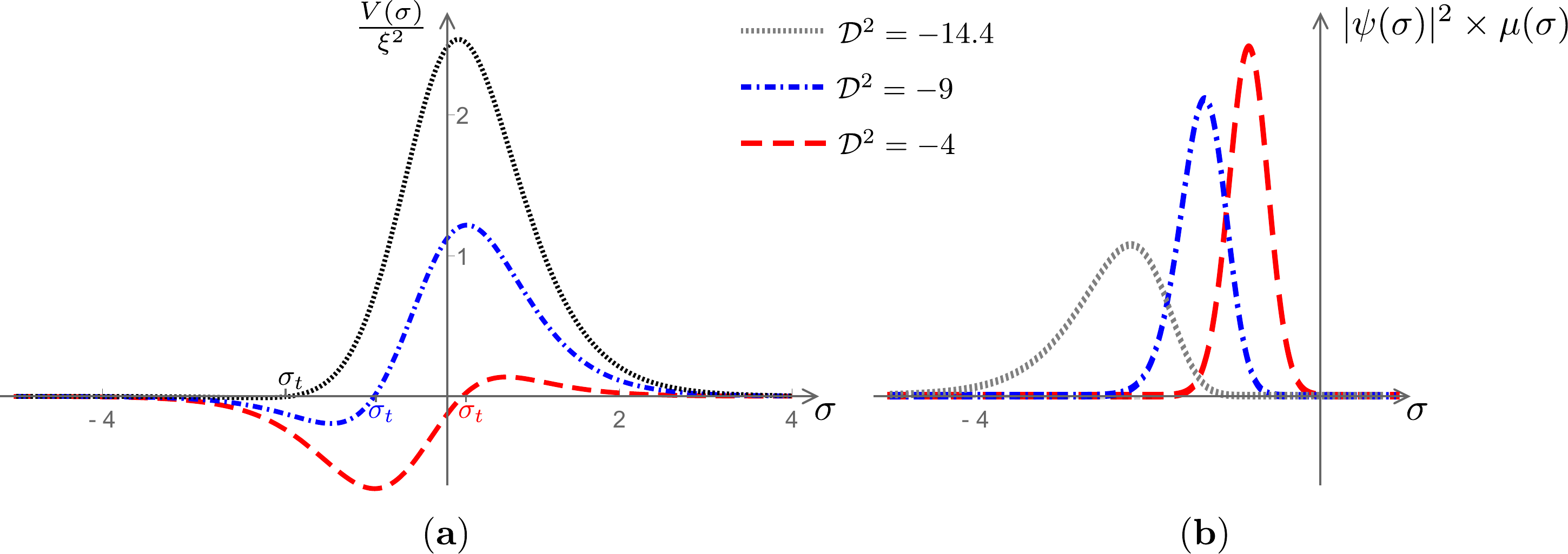}
\caption{
\small {\bf a}) The typical shape of the strong coupling potential (\ref{potentialstrong}), drawn here for $(\theta_1,\theta_2)=(4/3,1/2)$, zero spins and ${\cal D}^2=-4$ (red), ${\cal D}^2=-9$ (blue), ${\cal D}^2=-14.4$ (black). At large coupling, the potential becomes very high and the wave function is localised, with the appropriate measure (\ref{innerSch}), around the minimum. 
We then tune ${\cal D}^2$ so that the minimum is at zero Schr\"odinger energy (\ref{SchE}). This is achieved at ${\cal D}^2=-1/\sin^2(\theta_2/2)\simeq-16.3$ (\ref{eq:scD}), where the minimum moves to $\sigma=-\infty$. {\bf b}) Numerical plot of $|\psi(\sigma)|^2\times\mu(\sigma)$, where $\mu(\sigma)={1\over\(e^\sigma\sin^2{\theta_1\over2}+e^{-\sigma}\sin^2{\theta_2\over2}\)\cosh\sigma}$ is the Schr\"{o}dinger measure. We see that the ground state ``probability" is localised around the minimum of the potential, which moves towards $\sigma_{\ast}=-\infty$ as ${\cal D}^2$ approach its physical value.}
\label{strongcouplingfig}
\end{figure}
The potential (\ref{potentialstrong}) is plotted in figure \ref{strongcouplingfig} (a), for some values of $\mathcal{D}^2$ such that it admits bound states. In this case, the potential displays a minimum around which it is negative. 

As $\xi^2 \rightarrow \infty$, the potential  becomes very deep and the wave function (weighted by the appropriate measure (\ref{innerSch})) is supported around the minimum, see figure \ref{strongcouplingfig} (b). 
Hence, to leading order, the Schr\"odinger energy $E$ is given by the value of the potential at the minimum. 
To impose $E=0$, we look for the minimum of the potential and tune ${\cal D}^2$ such that at this point the potential is also zero, $V'(\sigma_*)=0$ and $V(\sigma_*)=0$.   
This leads to the result
\beq\label{eq:scD}
{\cal D}^2=-{1\over{\sin^2{\theta_2\over2}}}. 
\eeq 
This result describes the leading strong coupling behaviour $\Delta_n \sim \mathcal{D} \xi$ of the ground state, as well as low-lying excited states with fixed excitation numbers $n/\xi \sim 0$ and spins $S_i/\xi \sim 0$. 

 At the physical value (\ref{eq:scD}), both $\sigma = \pm \infty$ are minima of the potential. However, the potential well at $\sigma_{\ast}=-\infty$ is steeper, meaning that $V(\sigma) \sim e^{- 4 |\sigma|}$ rather than $V(\sigma ) \sim e^{-2 |\sigma|}$ there. This is the minimum around which the wave functions  are asymptotically concentrated, see figure \ref{strongcouplingfig} (b).   
  
 We can also analyse the excited states.
 The scaling dimensions of operators corresponding to different excited states start to differ at order $\hbar=1/\xi$. These excited states  correspond to semi-classical solutions that come from $\sigma=-\infty$ and bounce against the potential at some finite turning point 
 $\sigma_{\text{t}}$ which scales as $O(\log(1/\xi^2))$.  The turning point separates the \emph{classically allowed region} of the potential ($V< 0$), from the \emph{classically forbidden region} ($V>0$), see figure \ref{strongcouplingfig} for an illustration. These points are fixed by the Bohr-Sommerfeld quantisation condition
 
\beq\label{eq:BohrSomm}
\int\limits_{-\infty }^{\sigma_{\text{t}} }\!d\sigma\, \sqrt{{V}(\sigma) } 
= (n + 1/2 ) \, {i\pi\over\xi}\ ,
\eeq  
where $n \in \mathbb{N}$ is the excitation level above the strong coupling vacuum. By expanding this equation we find
\beq\label{eq:angl1}
\Delta^2_n=  - \frac{\xi^2 }{\sin^2{\frac{\theta_2}{2} } } \,\left(1 -  \frac{(2 + 4 n) \, \sqrt{ \sin^2\frac{\theta_1}{2} - \sin^2\frac{\theta_2}{2} }}{\xi}\, + O\({1\over\xi^2}\) \right)\ ,\qquad \xi^2 \rightarrow + \infty\ .  
\eeq
The same equation can be used to calculate further terms. The cases where $\sin^2{\theta_2\over2}>\sin^2{\theta_1\over2}$ and $\xi^2<0$ are treated in an analogous way.

\subsection{Dual  description}\la{sec:dual}
In the rest of this section, we interpret these results from the point of view of the holographic description of the fishnet model introduced in \cite{Gromov:2019jfh}. The dual model consists in  a chain of particles with nearest neighbour interactions propagating in $AdS_5$, where $\xi$ is identified with $1/\hbar$. In the strong coupling limit, we are interested in its classical dynamics, which takes place near the boundary of $AdS_5$. For the present case, there is a single particle, which is classically confined to the light-cone in $\mathbb{R}^{1,5}$, described by coordinates
\beq\label{eq:lc}
X^2 = 0 , \,\,\,\, X \equiv \left( X^{-1}, X^0, \dots, X^4 \right)\in \mathbb{R}^{1,5}\ .
\eeq
In these coordinates, the twist transformation (\ref{rotation}) takes the form
\beq
X_{\circlearrowleft} = R_{\vec\theta} \circ X\ ,\qquad R_{\vec\theta} = \left(
\begin{array}{cccccc}
1 & 0 & 0 & 0 & 0 & 0 \\
0 & 1 & 0 & 0 & 0 & 0 \\
0 & 0 & \cos{\theta_1} & - \sin{\theta_1} & 0 & 0 \\
0 & 0 & \sin{\theta_1} &  \cos{\theta_1} & 0 & 0 \\
0 & 0 & 0 & 0 &\cos{\theta_2} & - \sin{\theta_2} \\
0 & 0 & 0 & 0 & \sin{\theta_2} &  \cos{\theta_2}  
\end{array} 
\right)\ ,
\eeq
{and as above we will assume $\sin^2\frac{\theta_1}{2} > \sin^2\frac{\theta_2}{2}$.} 
Following~\cite{Gromov:2019jfh}
we write the action of the model
\beq\label{eq:Lfish}
S = \xi \int L dt\ ,\qquad 
L = - \left[ \frac{\dot X^2 }{2 \alpha} -  {\alpha}\,{(X \cdot X_{\circlearrowleft} )^{-1}}   \right]\ ,
\eeq
where $\dot X = \partial_t X$ and $\alpha$ is an auxiliary field, related to the worldline metric. 
This action is invariant under worldline time reparametrisation symmetry as well as time-dependent rescaling of $X$. As explained in \cite{Gromov:2019bsj}, it is convenient to fix these gauge redundancies by imposing
$\alpha = 1$ and $L = m^2$, which leads to the constraints,
\beq\label{eq:cons2}
\dot X^2 = -2 (X \cdot X_{\circlearrowleft} )^{-1} = m^2\ .
\eeq
It can be verified that, for the present case, the equations of motion arising from the Lagrangian  are a trivial consequence of the constraints and of the $SO(1,5)$ charge conservation. 
The classical spins and scaling dimension in this description are given by
\beq\label{eq:symclass}
 \mathcal{D} = -i \mathcal{Q}^{-1,0}\ ,\qquad \mathcal{S}_1 = \mathcal{Q}^{1,2}\ ,\qquad \mathcal{S}_2 = \mathcal{Q}^{3,4}\ ,
\eeq
where $ \mathcal{Q}^{M, N} = 2 (\dot X^M X^N - \dot X^N X^M )$. In the following discussion, for simplicity, we set the classical spins to zero. 
We parametrise the solution using four functions of time, $\rho(t)$, $s(t)$,  $\varphi_1(t)$ and $\varphi_2(t)$ as
\beq
\begin{array}{ll}
 X^{-1}  =  \rho\, ( 2 \cosh \sigma )^{\frac{1}{2} } \, \cosh s\ ,\quad &X^0 =\rho\,( 2 \cosh \sigma)^{\frac{1}{2} }  \sinh s\ , \\
 X^{1} = \rho\,e^{\sigma/2 } \, \cos \varphi_1\ ,  &X^2  =\rho\,e^{\sigma/2 } \, \sin \varphi_1\ , \\
 X^3 =\rho\,e^{-\sigma/2 } \, \cos \varphi_2\ ,  &X^4 =  \rho\,e^{-\sigma/2 } \, \sin \varphi_2 \  .
\end{array}
\eeq
By combining the constraints  (\ref{eq:cons2}) and (\ref{eq:symclass}), one obtains the following equation for $\sigma(t)$
\beq\label{eq:classfish}
p_{\sigma}^2(t) + \mathcal{V}(\sigma(t) ) = 0\ ,\qquad  p_\sigma(t) = \frac{ \dot \sigma(t) }{ 2  m^2 \cosh(\sigma(t) ) \, \left( e^{\sigma(t) } \sin^2\frac{\theta_1}{2}  + e^{-\sigma(t) } \sin^2\frac{\theta_2}{2}  \right) }\ ,
\eeq
where $p_{\sigma}$ is the conjugate momentum variable, and $\mathcal{V}(\sigma) = \lim_{\xi\rightarrow \infty} V(\sigma)/\xi^2$ is the classical limit of the potential (\ref{potentialstrong}). 
Equation (\ref{eq:classfish})  shows that the classical motion is restricted to the region where $\mathcal{V}(\sigma) \leq 0$. Such a region exists only for $\mathcal{D}^2 \geq  -1/\sin^2\frac{\theta_2}{2} $, which is the classically allowed range for the scaling dimension. The bottom of this range coincides with (\ref{eq:scD}), and gives the classical dimension of the ground state at the leading strong coupling limit.

 The classical solution corresponding to the ground state is particularly simple, and stays at $\sigma = \sigma_{\ast}$ at all times, where $\sigma_{\ast}=- \infty$ as   in the previous section.  Introducing the convenient four-dimensional coordinates $\vec x \equiv \frac{1}{X^{-1} + X^0}\,\left(X^1,X^2,X^3,X^4\right)$, the classical solution corresponding to the ground state is
\beq
\vec x(t)  =e^{- m^2 \sin\frac{\theta_2}{2} \,t } \, \vec{x}(0)\ ,\qquad \text{ where }(\vec x )_1 = (\vec x)_2 = 0 . 
\eeq
To describe excited states, one can use semi-classical arguments. The solutions corresponding to excited states are periodic orbits oscillating between $\left\{\sigma_{\ast}, 0\right\}$ and the turning point $\left\{\sigma_{\text{t}}, 0\right\}$ in the coordinates $\left\{\sigma, p_{\sigma} \right\}$. The semi-classical  Bohr-Sommerfeld quantisation rule $\oint p_{\sigma} d\sigma \in 2 \pi \hbar (\mathbb{N} + 1/2) $  leads to the same condition (\ref{eq:BohrSomm}) as the WKB study of the Schr\"odinger equation.

\section{Twisted operators with two orthogonal scalars (one-magnon case)}\label{sec:onemagnon}
\begin{figure}[t]
\centering
\includegraphics[scale=0.23]{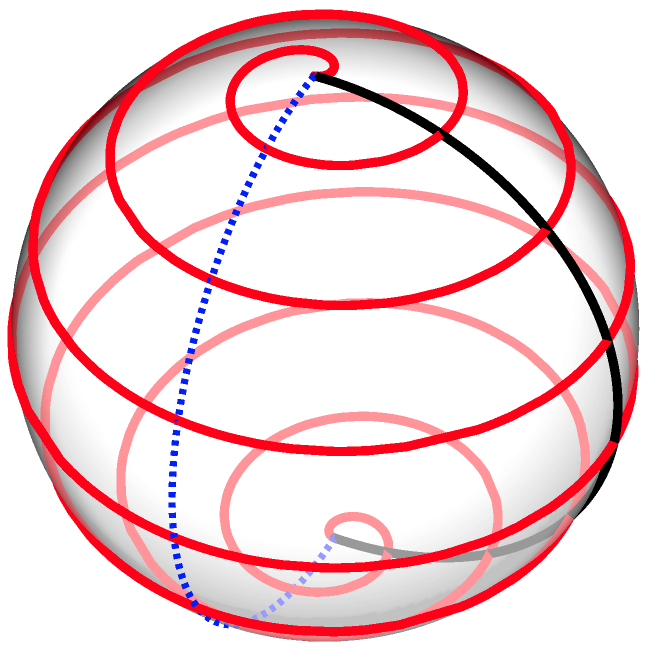}
\caption{\small Feynman diagram contributing to the correlation function between one operator of the type (\ref{J1J2op}) and its conjugate. These operators consist of a $Z$ and an $X$ field and any configuration of derivatives. At $l$ loops the red $Z$ line spirals $l$ times around the black  $X$ line. Along the way, it also crosses the dashed blue twist cut $l$ times.}
\label{spiral}
\end{figure}

In this section, we consider the next simplest example of rotation twisted operators in the fishnet model. This consists of operators with charges $J_1 = J_2 = 1$. Such operators are built with one $X$ scalar and one $Z$ scalar and any distribution of derivatives. They take the schematic form
\beq\la{J1J2op}
\cO_{\vec\theta,\Delta,\vec S}(0)=\tr\(R_{\vec \theta}\ \d_1^{S_1}\d_2^{S_2}\,(\d_1\bar\d_1)^{n_1}\,(\d_2\bar\d_2)^{n_2}\,X(0) Z(0)\)+{\rm permutations\;of\;derivatives}\ .
\eeq
As in the previous section, we define the CFT wave function of this operator as a two-point function
\beq\la{wavefunc2}
\Psi(x)=\<
\cO_{\vec\theta,\Delta,\vec S}(0)
\tr(R_{-\vec\theta}X(x)Z(x))
\>\;.
\eeq
The Feynman diagrams contributing to this correlator look like spirals, see figure \ref{spiral}. Like in the previous case, we can use the iterative structure of the diagram to write an integral equation for the CFT wave function. 
As the same symmetry considerations apply, a decomposition of the wave function as in \eq{waveea} is still valid. However, an important difference is that the graph-building operator $\cB$ cannot be inverted as a differential operator, making this case considerably more complicated. As a result there is no simple differential equation which determines the remaining function $\psi(\sigma)$. As was noticed in \cite{Gromov:2019jfh}, this is a common feature of all fishnet operators with $|J_1|=|J_2|$.

Below, we study the spectrum perturbatively at one loop in section \ref{sec:1loop1mag}. We  then compute the finite coupling spectrum at equal angles in section \ref{sec:allloop1mag}. After that, in section \ref{sec:integrability} we   demonstrate a precise match between these results and the predictions of the  integrability formalism, which will allow us to  extend our results to non-equal angles at finite coupling.

\subsection{Structure of the spectrum}
Here we describe the general properties of the spectrum of this family of operators, which we then explore perturbatively at one loop in the next section.

There are plenty of operators of the type \eq{J1J2op}, which can mix with each other in perturbation theory. 
Similarly to the case considered in the previous section with a single scalar, we have to solve a mixing problem between all operators with the same spins $S_1$, $S_2$ and the same tree-level scaling dimension $\Delta(0)=2+S_1+S_2+2(n_1+n_2)$. As there is a possibility to choose where the derivatives act, this is a very large family of operators.
However, due to a peculiarity of the fishnet theory, only a small subset of these operators
can receive a non-trivial anomalous dimension.

First, for the purposes of counting non-trivial operators, let us temporarily remove the twist. Consider an operator
\beq\la{example}
\tr\(\bar\d_1XZ\)-\tr\(X\bar\d_1Z\)\ .
\eeq
When such an operator appears in a loop diagram (e.g. figure \ref{onemagtree}(b)), it gets contracted with the interaction vertex which is symmetric in $X$ and $Z$, so that the result is immediately zero. 

To make this consideration more formal, one can effectively replace the part of the interaction vertex which contracts with the operator by $\tr (\bar X(x)\bar Z(x))$, generating the same two propagators connecting the operator, at tree-level:
\beq\la{treecor}
\<{\cal O}(0)\tr (\bar X(x)\bar Z(x))\>_{\rm tree}\;.
\eeq
Now looking at \eq{treecor}, we can immediately conclude that ${\cal O}(0)$ should be a descendant of
$\cO_0=\tr ( X(x) Z(x))$
in the free theory in order for \eq{treecor} to be non-zero. In other words, we can divide 
the whole space of operators \eq{J1J2op} into two families -- 
${\cal O}_0$ and its descendants, and other primaries and their descendants. According to the argument above, any loop diagram will automatically project onto the operators in the first group. The same considerations apply in the twisted case too, if the cut is chosen so that it does not cross the two propagators connecting to the operator. This is always possible to do without changing the result, as is demonstrated in figure \ref{onemagtree}.b and in the next section. The only difference is that, in presence of the twist, the projection only applies to one of the two operators. Hence, the anomalous dimension matrix has off-diagonal Jordan elements between the two families of operators. These Jordan cells, however, are irrelevant for the computation of the eigenvalues. Thus, in this section we restrict ourselves to the operators in the first family, which are total derivatives of  $\cO_0$.\footnote{This general argument explains some of the results obtained previously in the literature \cite{Gromov:2017cja}.}

This non-trivial sub-class of operators thus consists of all operators of the form
\beq\la{O0}
\cO_{\vec\theta,\Delta,\vec S}(0)=\sum_{m=0}^{n}\alpha^{(m)}_{\vec\theta,\Delta,\vec S}\ \bar\d_1^{S_1}\bar\d_2^{S_2}\,(\d_1\bar\d_1)^{n-m}\,(\d_2\bar\d_2)^{m}\,\cO_0\qquad\text{where}\qquad\cO_0\equiv \tr({\cal R}_{\vec\theta}\,ZX)\ .
\eeq
However, the basis above is not easy to work with, as it is not orthogonal w.r.t. the tree-level contractions.  The orthogonal combinations are quite non-trivial; for example, for the case of zero spins, one can show that 
the following basis is orthogonal under the tree-level contractions\footnote{In order to verify the orthogonality, one should
compute the tree-level contraction with a conjugate operator sitting at infinity. The conjugate operator is obtained by applying an inversion conformal transformation to ${\cal O}(x)$ and then taking the limit $x\to 0$.}
\beq\la{O01}
\cO_{S_1=0,S_2=0,n,m}\equiv\Box^{n-m}
\(
\sum_{k=0}^m\binom{m}{k}^2(-\d_1\bar\d_1)^{k}(\d_2\bar\d_2)^{m-k}
\)
\,\cO_0\ ,\qquad m=0,\dots,n\ ,
\eeq
so for fixed bare dimension $\Delta(0)=2+2n$ there are $n+1$ operators 
$\cO_{S_1,S_2,n,\circ}$ that can mix with each other.
For  general spins, in order to construct the tree-level orthogonal operators, it is convenient to introduce lowering generators $J_L^-$ and $J_R^-$ for the two $SU(2)$ subgroups of the $SO(4)$ rotation symmetry
as defined in appendix \ref{appSO4rep}. 
Specifically, one can define
\beq\la{opmn}
\cO_{S_1,S_2,n,m}=\Box^{n-m}\[(J_L^-)^{m}(J_R^-)^{m+S_2}\bar\d_1^{2m+S_1+S_2}\]\cO_0\;\;,\;\;m=0,\dots,n\ ,
\eeq
where in our notations the generators $J_L^-$ and $J_R^-$ only rotate the highest weight ``state" $\bar\d_1^{2m+S_1+S_2}{\cal O}_0$ of $SO(4)$ spin $j=2m+S_1+S_2$.
Since this formula uses the symmetry generators, orthogonality is guaranteed by construction.
In particular for $S_1=S_2=0$ one
reproduces \eq{O01} up to a numerical factor. 

In the next section we demonstrate how to resolve the mixing problem at one loop. In section \ref{sec:integrability}
we solve the problem numerically, using integrability,
and reproduce the correct $(n+1)$-degeneracy in the spectrum as deduced in this section.

\subsection{The one-loop spectrum}\label{sec:1loop1mag}

\begin{figure}[t]
\centering
\def\svgwidth{14cm}
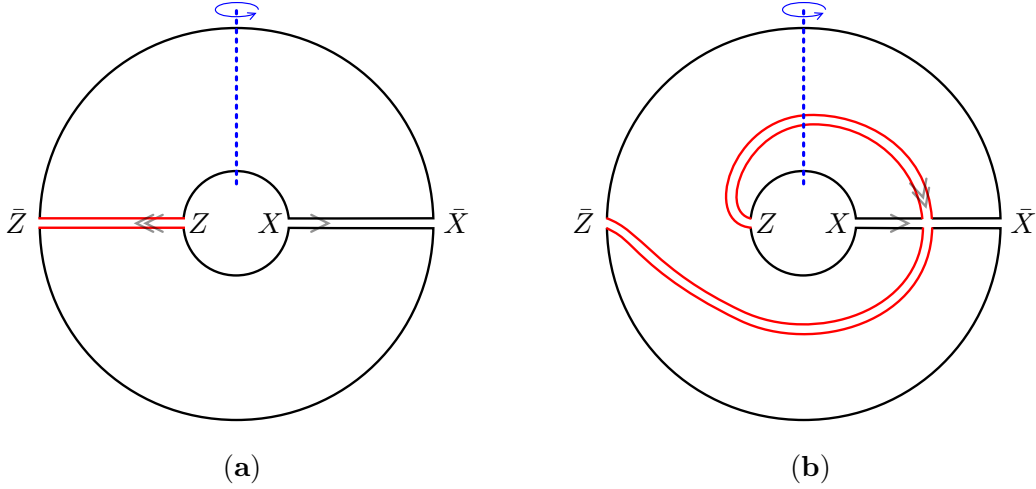
\caption{\small {\bf a}) The unique tree-level diagram that contributes to the correlation function between a twisted operator of the type (\ref{J1J2op}) and a conjugate operator from the same family. 
The diagram is plotted in double line notation, where the external and internal circles represent the colour-trace in (\ref{J1J2op}). Here, we have made a conventional choice for the twist cut. {\bf b}) The unique one-loop diagram that contributes to the same correlator.}
\label{onemagtree}
\end{figure}

At one-loop order, there is only one diagram that contributes to the correlation function between two operators of the type (\ref{J1J2op}), see figure \ref{onemagtree}.b. As usual, we can read off the one-loop anomalous dimension of the operators from the logarithmically divergent piece of that diagram. 

First, let us consider the case without the twist. As all the operators we consider are total derivatives of ${\cal O}_0=\tr(X(0)Z(0))$, they will 
all have the same anomalous dimension.

The sum of the tree-level and one-loop diagram for the correlator of ${\cal O}_0$ with $\overline{\cal O}_0$ is 
\beq
\frac{1}{16\pi^4(x_0-x_{\bar 0})^4}+
16\pi^2\xi^2\int d^4  x\frac{1}{16\pi^4(x_0-x)^4}\frac{1}{16\pi^4(x-x_{\bar 0})^4} \ .
\eeq
The integral, indeed, is $\log$-divergent at the locations of the operators. Introducing an $\epsilon$ 
cutoff around $x_0$ and $x_{\bar 0}$ we get
\beq
\frac{1}{16\pi^4(x_0-x_{\bar 0})^4}\(1+
\frac{\xi^2}{\pi^2}\;2\times 2\pi^2\log\(\frac{|x_0-x_{\bar 0}|}{\epsilon}\)\) \ ,
\eeq
from where we read off the one-loop anomalous dimension $\gamma = -2\xi^2$, which is in agreement with the ABA result
of \cite{Caetano:2016ydc}\footnote{ We analyse the zero twist limit in more detail in the next section. We will see that at two loops the perturbation theory breaks down in this case.}
\beq
\Delta^{ABA}=J+\sqrt{1-4\xi^2}\simeq 2-2\xi^2+{\cal O}(\xi^4)\;.
\eeq

Now we place the operators 
at zero and at infinity and
introduce the twist by the rotation around the origin.
This will result in the twist cut going through one of the two $Z$-field propagators as in figure \ref{onemagtree}.b.
We notice, however, that at one loop, the effect of the twist is very innocent, as one can simply move the starting point of the cut to the left from $Z$ so that it does not affect the propagator anymore. As the $Z$-field is sitting at the origin, it is invariant under the twisting rotation, so for the operator $\tr(X(0)Z(0))$ there will be no difference and our calculation above is still valid. Thus we conclude that
\beq\la{O0g}
\gamma_{S_1=0,S_2=0,n=0}=-2\xi^2+{\cal O}(\xi^4)\;.
\eeq
We see below in section \ref{sec:allloop1mag} that the two-loop term does have a non-trivial $\theta$-dependence.

Now consider an example with non-trivial angle dependence. Let us take the operator $S_1=1,\;S_2=0,\;n=0$ 
\beq
{\cal O}_{1,0,0}=\bar\d_1\cO_0=\tr\(R_{\vec\theta}\,\bar\d_1XZ\)+\tr\(R_{\vec\theta}\,X\bar\d_1Z\)\ .
\eeq
This time, when moving $R_{\vec\theta}$ to the left from $Z$ we introduce an additional phase factor due to the derivatives
\beq
{\cal O}_{1,0,0}=\bar\d_1\cO_0=\tr\(\bar\d_1X R_{\vec\theta}Z\)+e^{i\theta_1}\,\tr\(X R_{\vec\theta}\bar \d_1Z\)\ .
\eeq
After we move the twist marker as above, it can be removed, as the twist cut will no longer cross any of the propagators for each of the two terms above. This transformation defines a linear operator ${\bf R}_{\vec\theta}$ 
\beqa\la{hatRexample0}
{\bf R}_{\vec\theta}\circ\bar\d_1\tr\(XZ\)&=&\tr\(\bar\d_1XZ\)+e^{i\theta_1}\tr\(X\bar\d_1Z\)\ .\nn
\eeqa
Finally, following the previous discussion, we know that only the descendants of ${\cal O}_0$ of the free theory
survive in the one-loop diagram, so we write
\beqa\la{hatRexample0}
{\bf R}_{\vec\theta}\circ\bar\d_1\tr\(XZ\)
&=&{1+e^{i\theta_1}\over2}\(\tr\(\bar\d_1XZ\)+\tr\(X\bar\d_1Z\)\)+{1-e^{i\theta_1}\over2}\(\tr\(\bar\d_1XZ\)-\tr\(X\bar\d_1Z\)\)\ .\nn
\eeqa
The second operator, with anti-symmetric combination of derivatives, is a new primary of the free theory and is not a descendant of ${\cal O}_0$. Its presence leads to an off-diagonal Jordan element of the anomalous dimensions matrix $\hat\Gamma_{\vec\theta}$ and therefore can be projected out. Hence, we conclude that we get the same result as in the un-twisted case multiplied by $(1+e^{i\theta_1})/2$,
giving for the scaling dimension of ${\cal O}_{1,0,0}$ the result $\Delta=3-\xi^2\(1+e^{i\theta_1}\)+O(\xi^4)$.

One can treat the general case in the same way. First, one has to deduce the operator $\bf R_{\vec\theta}$ arising from moving the twist mark point by using
\beq\la{twist}
\tr[R_{\vec\theta}\dots\d_1^{n_1}\bar\d_1^{\bar n_1}\d_2^{n_2}\bar\d_2^{\bar n_2}Z(0)]=e^{i(\bar n_1-n_1)\theta_1+i(\bar n_2-n_2)\theta_2}\tr[\dots R_{\vec\theta}\d_1^{n_1}\bar\d_1^{\bar n_1}\d_2^{n_2}\bar\d_2^{\bar n_2}Z(0)]\ .
\eeq
After that, one can remove the twist marker and compute the divergent part of the diagram.
For that one should project onto the free theory descendants of $\cO_0$ and read off the mixing matrix. A convenient way of projecting back to this class of operators is to contract (\ref{opmn}) with $\tr\(\bar X(x)\bar Z(x)\)$ at tree-level, as in (\ref{treecor}).

For example, in the sector $S_1=S_2=0$ and $n=1$ we have two operators \eq{O01}
\beq\la{M2}
\cO_{0,0,1,1}\propto\(\d_1\bar\d_1-\d_2\bar\d_2\)\cO_0\ ,\qquad\cO_{0,0,1,0}\propto\(\d_1\bar\d_1+\d_2\bar\d_2\)\cO_0\ .
\eeq
Following the above procedure we get a non-trivial mixing matrix
\beq\la{mix}
-\xi^2\(\!\begin{array}{cc}{1\over3}\(4+\cos\theta_1+\cos\theta_2\)&{1\over\sqrt 3}(\cos\theta_1-\cos\theta_2)\\{1\over\sqrt 3}(\cos\theta_1-\cos\theta_2) &\cos\theta_1+\cos\theta_2
\end{array}\!\)\ .
\eeq
The corresponding anomalous dimensions, found as eigenvalues of the mixing matrix \eq{mix}, are
\beq
\label{eq:gamma1loop}
\gamma_{0,0,1 }^{(\pm)}=-{2\over3}\xi^2\(1+\cos\theta_1+\cos\theta_2\pm \sqrt{1+\cos^2\theta _1+\cos^2\theta _2-\cos\theta_1-\cos\theta _1\cos\theta _2-\cos\theta_2}\)\ .
\eeq
Note that for $\theta_2=\theta_1=\theta$,
the operators \eq{M2} do not mix and the anomalous dimensions simplify to
\beq
\label{eq:gamma1loopeq}
\gamma_{0,0,1,1}=-\xi^2\frac{2}{3}(2+\cos\theta)\;\;,\;\;
\gamma_{0,0,1,0}
=-2\xi^2\cos\theta\ .
\eeq
We give further examples and summarise the one-loop results in  appendix \ref{oneloopdata}.

As in the case of one scalar field, the calculation simplifies considerably in the case where the twist angles are equal and the current results can be pushed to arbitrary loop level. This is considered in the next section.

\subsection{All-loop spectrum at equal angles}\la{sec:allloop1mag}

Similarly to (\ref{wf}), the starting point for the all-loop calculation is the CFT wave function \eq{wavefunc2}.
The diagrams that contribute to the correlator \eq{wavefunc2} are generated by the graph-building operator
\beq
\mathcal{B} \circ \Psi(x) \equiv \frac{\xi^2}{\pi^2}\int d^4 y\frac{\Psi(y)}{(x- y_{\circlearrowleft})^2 ( x-y)^2}\ ,
\eeq
which adds one more spiral to the diagram in figure \ref{spiral}. Following the same procedure as was exemplified in detail in \cite{Gromov:2018hut}, the sum of all diagrams can be written as a simple geometric sum of this operator. In particular, physical twisted operators correspond to stationary wave functions\footnote{
Notice that the tree-level term on the r.h.s. of the Dyson-type evolution equation (analogous to (\ref{eq:Dyson})) is suppressed due to the wave function renormalisationion at finite coupling. The same also applies for the case of a single scalar considered in the previous section.}
\beq\label{eq:BStwo}
 \mathcal{B} \circ \Psi_{\vec \theta, \Delta, \vec{S} }(x) = \Psi_{\vec \theta, \Delta, \vec S }(x)\ .
\eeq

\begin{figure}
    \centering
    \def\svgwidth{12cm}
    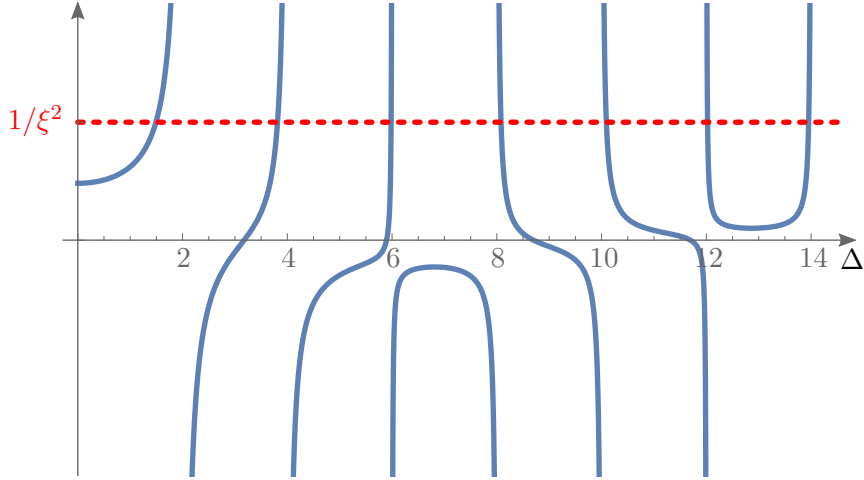
    
    \caption{The eigenvalue of the graph building operator $E(\Delta,0,0)$ in (\ref{E1int}) at $\theta=1$. It has poles at the positive real even integers $\Delta_0=2+2n$. The spectrum is giving by equating $E(\Delta,0,0)$ with $1/\xi^2$ and is represented by the intersection with the red dashed line. The solution between the $n$-th and the $(n-1)$-th poles correspond to the operator $\Box^n\,\tr\(R_{\vec\theta}ZX\)$, or, equivalently, the wave function (\ref{eq:wavefonemag}).}
    \label{fig:E1}
\end{figure}

\begin{figure}[t!]
    \centering
    \includegraphics[scale = 0.6]{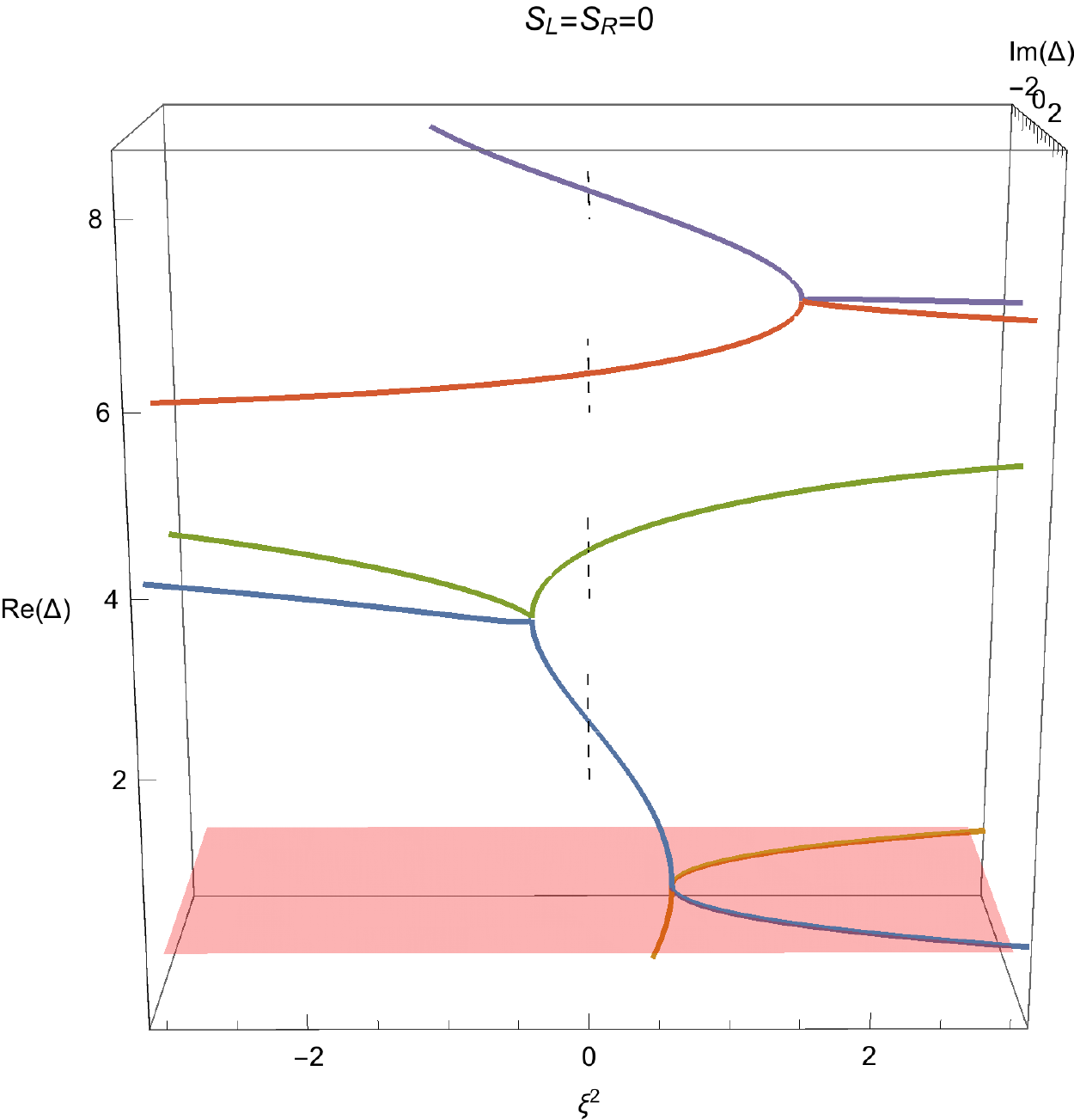}
    \caption{The conformal dimensions of the one-magnon operators at twist angle $\theta=2.0$ and real $\xi^2$. The dimensions start real for small $\xi^2$ and then split into pairs of complex conjugate $\Delta$'s. We see how the scaling dimensions of different operators are connected to each other through analytic continuation in the complex $\xi^2$ plane. }
    \label{fig:spec1a}
\end{figure}

\begin{figure}[t!]
    \centering
    \def\svgwidth{10cm}
    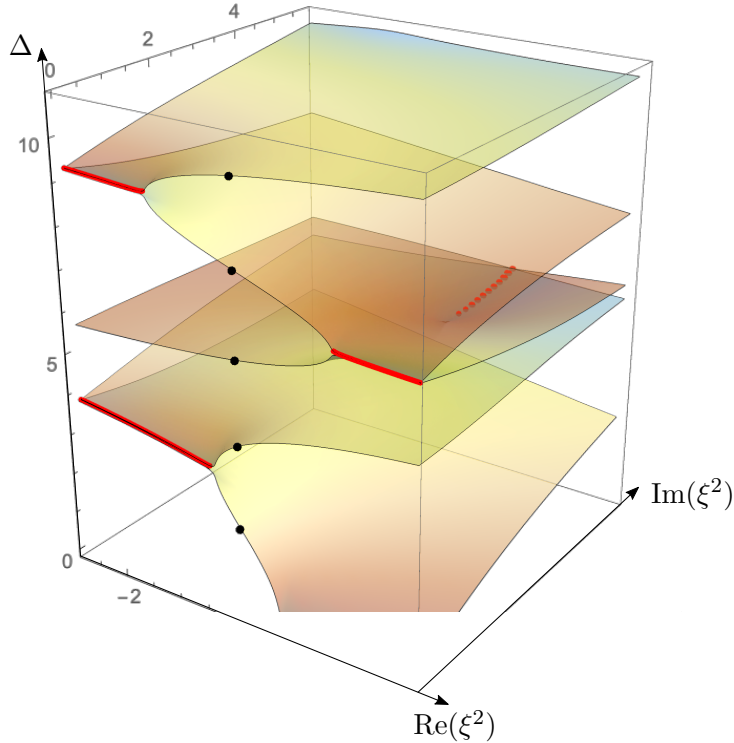
    
    \caption{Spectrum of the one-magnon family of operators as a function of a complexified coupling $\xi^2$. All the operators, corresponding to different combinations of fields at weak coupling (black dots) in fact belong to the same Riemann surface and can be obtained as an analytic continuation in $\xi^2$ from one another. The plot is done for $\theta=2$.}
    \label{fig:spec1}
\end{figure}

We do not know how to directly diagonalise $\cB$ as it is a rather complicated integral operator, which cannot be easily inverted as in the previous case. However, in the special case $\theta_1=\theta_2\equiv\theta$ the eigenfunctions are completely fixed by the enhanced symmetry (\ref{twistsymm2})
and all we need is to find the eigenvalue of ${\cal B}$ and impose it to be one according to \eq{eq:BStwo}. We parametrise the wave functions by their corresponding $SO(4)\simeq SU(2)_L \times SU(2)_R$ charges, $\{\Delta,S,m_L,m_R\}$.\footnote{
In the vector representation, the Casimir operators of both $SU(2)$ subgroups are equal to $j(j+1)$ with $j=S/2$. 
The $(S+1)^2$ states in the multiplet are
labelled by 
$-j\leq m_L,m_R\leq j$.
See appendix \ref{appSO4rep} for details.} 
As the symmetry preserved in the case $\theta_1=\theta_2$ is $U(1)_L\times SU(2)_R$, the eigenvalue should not depend on $m_R$
and we can consider the $SU(2)_R$ highest weight state with $m_R=S/2\equiv j$. It takes the form
\beq\label{eq:wavefonemag}
\Psi_{\Delta,S,m_L} \propto 
z_1^{m_L+S/2} \bar{z}_2^{S/2-m_L} \left(z_1 \bar{z}_1+z_2
   \bar{z}_2\right){}^{-\frac{\Delta }{2}-S/2-1}\ .
\eeq
Since the wave function is explicitly given by (\ref{eq:wavefonemag}), the problem is reduced to the calculation of the eigenvalue
\beq\label{E1int}
\mathcal B \circ \Psi_{\Delta,S,m_L}(x) =  \xi^2\,E(\Delta,S,m_L) \,\Psi_{\Delta,S,m_L}(x)\ .
\eeq
The evaluation of this integral for several choices of the spins is given in Appendix~\ref{app:E1}.
In the simplest case $S=0$, the result is\footnote{The function $\Phi(z,1,x)=\sum_{n=0}^\infty \frac{z^n}{n+x}$, or \verb|HurwitzLerchPhi[z, 1, x]| in {\it Mathematica}.}
\beq\la{E1res1}
E(\Delta,0,0) = 
i\frac{\Phi \left(e^{-i \theta },1,-\frac{\Delta
   }{2}\right)-\Phi \left(e^{-i \theta },1,\frac{\Delta }{2}\right)-\Phi \left(e^{i
   \theta },1,-\frac{\Delta }{2}\right)+\Phi \left(e^{i \theta },1,\frac{\Delta
   }{2}\right)}{\Delta \sin \theta }\ .
\eeq
This function is plotted in figure \ref{fig:E1} for real positive $\Delta$. It has simple poles at $\Delta=2+2n$ for non-negative integer $n$ and is smooth between these poles.  
Next, from (\ref{eq:BStwo}), we can extract the spectrum using the condition
\beq\la{eqanspec}
E(\Delta, 0, 0)=\xi^{-2}\ .
\eeq
The resulting spectrum is plotted in figures \ref{fig:spec1a} and \ref{fig:spec1} for real and complex coupling. 

At zero coupling, the solutions to (\ref{eqanspec}) are localised at the poles. Hence, in the free theory we have $\Delta_n=2+2n$, which correspond to the operators $\cO_n=\Box^n\cO_0$. Expanding (\ref{eqanspec}) at weak coupling gives 
\beq
\begin{split}
\Delta_{n=0}&= 2-2 \xi ^2\qquad\;+\xi ^4 \left(4 \log[2 \sin \tfrac{\theta}{2}]-2\right)\qquad\qquad\qquad\;\;\;\,+{\cal O}\left(\xi ^6\right)\\
\Delta_{n=1}&= 4-2 \xi ^2 \cos \theta +\xi ^4 \cos \theta  \left(\cos \theta \left(4 \log[2 \sin \tfrac{\theta}{2}]-1\right)+2\right)+O\left(\xi^6\right)
\end{split}\ ,
\eeq
in agreement with \eq{O0g} and (\ref{eq:gamma1loopeq}) at one-loop order. We see that the limit $\theta\to 0$ of the two loop coefficient is singular.\footnote{
The expansion in $\xi$ and $\theta\to 0$ limit do not commute.
Fixing $\xi$ and then analytically continue the solution for $\Delta$ from some finite $\theta$ to  zero we find that $\Delta \to 2 {\mathbb Z}^{\ast}$. Except for the state $n=0$, where depending on the initial value of $\xi$ one can either get $2$ or $\pm i\infty$ as a limit.}

For each given spin $S$
the integral in (\ref{E1int}) can also be computed. Unfortunately, we were not able to obtain a closed expression for the eigenvalue for all spins and values of $m_L$. For $m_L=S/2$ we found
\beq
{E(\Delta,S,\tfrac S2)}=
\frac{1}{\left(1+e^{-i \theta }\right)^{S}} \sum _{k=0}^{S} \binom{S}{k} M (\Delta -2 k+S)+R(\Delta,S)\ ,
\eeq
where $R(\Delta,S)$ is a rational function, which removes all simple poles in the first term in $\Delta$
inside the interval $(-1-S,1+S)$. For example, $R(\Delta,0)=0$ and 
\beq
R(\Delta,1)=\frac{2 }{(\Delta -1) \left(1+e^{-i \theta }\right)}-\frac{2}{(\Delta +1) \left(1+e^{-i \theta }\right)}\ ,
\eeq
and so on.
For $m_L=S/2-1$ we get
\beq
{E^{(1,1)}(\Delta,S,\tfrac S2-1)} \simeq
\frac{e^{-i \theta}}{ \left(1+e^{-i \theta }\right){}^{ S} }\sum _{k=0}^{S}2\left(\frac{S}{2} \binom{ S}{k}- k \left(\cos \theta _1+1\right)
   \binom{ S-1}{k}\right) M(\Delta -2
   k+S) \ ,
\eeq
where $\simeq$ means that again, we have to subtract simple poles in the interval $(-1-S,1+S)$.
Finally, for $ E^{(1,1)}(\Delta,S,S/2-2)$ we found 
\beqa
\frac{1}{2} \frac{e^{-2 i \theta }}{\left(1+e^{-i \theta }\right){}^{ S}}  \sum _{k=0}^{
   S} M(\Delta -2 k+ S) \left(-\left(\cos \theta +1\right) \left(6
   k^2-6 (k+1) S+4 S^2+2\right) \binom{ S-2}{k-1}\right.\nn\\
   \left.+2 (k-1) k \left(\cos\theta+\cos \left(2 \theta\right)\right) \binom{ S-2}{k}+S ( S-1) \binom{
   S}{k}\right)\ . \nn 
\eeqa

In the next section we explain how to compute the spectrum from integrability.

\section{The spectrum via integrability}\label{sec:integrability}
In this section we connect our construction with the integrability approach. Our main claim is that the twist we introduced can be studied by means of the twisted ABA of \cite{Beisert:2005if} in the asymptotic regime, or exactly using the twisted QSC
 construction of \cite{Gromov:2015dfa,Kazakov:2015efa}. To demonstrate this is the case, we consider several examples in the fishnet model.

We start with a leading order perturbative test of the equivalence by considering the ``vacuum" operator at length $J$
\beq\la{OJ}
\cO_J=\tr\(Z^J(x_0)R_{\vec\theta}\)\;.
\eeq
The Feynman diagrams that contribute to the two-point function of these operators are all wheel graphs, see for example figure \ref{fig:figure1}. From the integrability perspective, these correspond to wrapping corrections. Using this fact, we will perform a test of the integrability at the first L\"uscher order $O(\xi^{2J})$. 

Next, we review the finite coupling twisted fishnet Baxter equations that were presented in \cite{Gromov:2019jfh} for generic operators. We present the corresponding quantisation condition, which was used in \cite{Gromov:2019jfh} for solving numerically the spectrum of length-three operators. Finally, we will match and generalise the field theory results obtained in the previous sections at finite coupling. 

\subsection{One-wheel diagram vs. first L\"uscher correction}
\subsubsection{One-wheel diagram}
\begin{figure}[t]
\centering
\def\svgwidth{11cm}
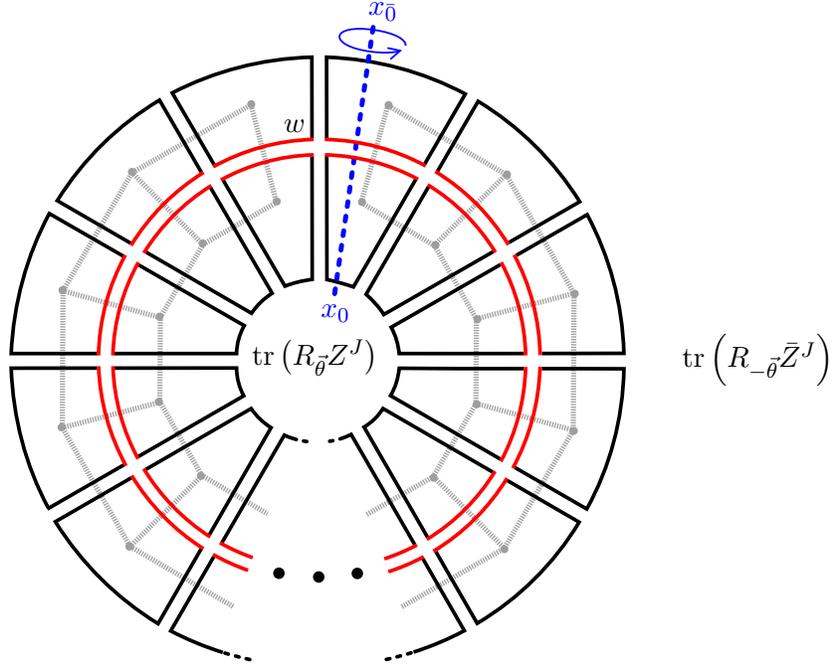
\caption{The first loop correction to the two-point function $\<\cO_J\bar\cO_J\>$ in (\ref{OJ}) comes from the diagram in the figure. It is composed of one wheel of the $X$ scalar (in red) crossing all the $Z$-propagators (in black). After factoring out one of the loop integrations over an internal vertex ($w$ in the figure), we remain with a known ladder integral (dashed grey) times two propagators (between $w$ and $x_0$, $x_{\bar0}$). In appendix \ref{app:oneloop}, we use this representation of the integral to extract the $J$-loop anomalous dimension of these operators (\ref{eq:GammaJ1}). 
\label{fig:figure1}
}
\end{figure}

The first loop correction to the dimension of $\cO_J$ comes from the single-wheel graph in figure \ref{fig:figure1}. It is given by the coefficient in front of the logarithmic divergence of that graph. 
This coefficient is computed in appendix \ref{app:oneloop} by expressing this $J$-loop integral as a single 4D integral over a known $(J-1)$-loop ladder function \cite{Usyukina:1993ch}. 
The resulting anomalous dimension is
\beq\label{eq:GammaJ1}
\gamma_{J}^{\text{1-wheel}}=\xi^{2 J} \(\!\begin{array}{c}2J-2\\ J-1\end{array}\!\){{\rm Li}_{2J-1}(e^{i\theta_1})+{\rm Li}_{2J-1}(e^{-i\theta_1})-{\rm Li}_{2J-1}(e^{i\theta_2})-{\rm Li}_{2J-1}(e^{-i\theta_2})\over\cos\theta_2-\cos\theta_1}\ . 
\eeq

\subsubsection{First L\"uscher correction}
 We will now reproduce the result (\ref{eq:GammaJ1}) from integrability. 
 We will do that by starting with the scalar operator $\tr(Z^JR)$ in $\mathcal{N}=$4 SYM theory, twisted by a generic element $R \in PSU(2,2|4)$. We will then take the appropriate double scaling limit.
 \paragraph{Diagonal twist in $\mathcal{N}$=4 SYM.}
 When studying the spectrum, we consider twist symmetries that commute with dilatations. Such a transformation $R \in PSU(2,2|4) $ can always be brought to a diagonal form with a symmetry transformation, after which it depends on six independent parameters,  
\beq
 R \equiv \text{diag}\left(  \by_1, \by_2, \by_3, \by_4 | \bx_1 , \bx_2 , \bx_3 , \bx_4  \right) \in PSU(2,2 | 4)\ ,
 \eeq
with $\prod_{i=1}^4 \bx_i = \prod_{i=1}^4 \by_i = 1$. As a group element, this twist map can be written as
\beq
R =\underbrace{\left(\frac{ \by_1 \by_2}{\by_3 \by_4} \right)^{\hat D/2 } \, \left(\frac{\by_2 \by_4}{\by_1 \by_3}  \right)^{ \hat S_1/2} \, \left(\frac{\by_2 \by_3}{\by_1 \by_4}  \right)^{ \hat S_2/2} }_{\text{conformal transf.}\,\in \, \text{SU(2,2) }}\, \underbrace{\left(\frac{ \bx_1 \bx_2}{\bx_3 \bx_4} \right)^{\hat J_1/2 } \, \left(\frac{\bx_1 \bx_3}{\bx_2 \bx_4}  \right)^{ \hat J_2/2} \, \left(\frac{\bx_2 \bx_3}{\bx_1 \bx_4}  \right)^{ \hat J_3/2} }_{\text{R-symmetry transf.}\, \in \,\text{SU(4) } }\ ,\label{eq:RPSU224}
\eeq
where $\hat D$ is the dilatation operator, $\hat S_i$ generate rotations in two orthogonal planes, and $\hat J_i$ are Cartan generators of the R-symmetry group $SU(4)$. 
As we discussed in section \ref{sec_ct}, a state twisted by $R$ must be invariant under this transformation. Denoting as $(\Delta, S_1, S_2 , J_1,J_2, J_3)$ the charges of the state, this constraint reads
\beq\label{eq:invariantxy}
 \left( \frac{ \by_1 \by_2}{\by_3 \by_4} \right)^{\Delta/2 } \, \left(\frac{\by_2 \by_4}{\by_1 \by_3}  \right)^{  S_1/2} \, \left(\frac{\by_2 \by_3}{\by_1 \by_4}  \right)^{  S_2/2} \, \left(\frac{ \bx_1 \bx_2}{\bx_3 \bx_4} \right)^{  J_1/2 } \, \left(\frac{\bx_1 \bx_3}{\bx_2 \bx_4}  \right)^{  J_2/2} \, \left(\frac{\bx_2 \bx_3}{\bx_1 \bx_4}  \right)^{  J_3/2} = 1\ .
\eeq
Twisted Quantum Spectral Curve equations describing the full spectrum of scaling dimensions in the presence of generic twists were proposed in \cite{Gromov:2015dfa,Kazakov:2015efa}.
The condition \eq{eq:invariantxy}
seems to be omitted there.
We believe that it may play an important role in the QSC approach. It could be that the QSC can only have   solutions when this condition is satisfied. This point should be further investigated.

\paragraph{L\"uscher corrections.}
 To match the result (\ref{eq:GammaJ1}), we use the method of L\"uscher corrections (see \cite{Luscher:1985dn} and \cite{Janik:2010kd} for a review).\footnote{Similar computations with the L\"uscher method were previously made for the case of twists corresponding to the $\gamma$-deformation in  \cite{Ahn:2010yv,Gromov:2010dy,Ahn:2011xq,Kazakov:2015efa}, including at higher loops.} This approach is very convenient to study the operator $\tr(Z^J R)$, which is a protected operator in the limit where the twists are sent to zero. The L\"uscher method gives in one go the twisted anomalous dimension at $J$ loops. 
  In this setup, one considers the dual worldsheet theory. The anomalous dimension at this order arises from the elastic interaction between the state and a virtual  particle travelling a closed loop around the cylinder in the mirror channel obtained by double Wick rotation. This process gives an energy shift, equivalent to the anomalous dimension, described by the formula \cite{Luscher:1985dn,Janik:2010kd}:
\beq\label{eq:Luscher}
\gamma_{J}^{\text{1-wheel}}  = \delta E = -\sum_{a=0}^{\infty}\int \frac{du}{\pi} e^{-J \,\widetilde E_a(u) } \, T_{a,1}^L \, T_{a,1}^R\ ,
 \eeq
 where the sum runs over bound states in the mirror channel,  $\tilde E_a(u)$ is the dispersion relation for mirror particles, and $T_{a,1}^{L/R}$ are asymptotic large-volume transfer matrix eigenvalues \cite{Gromov:2009tv}. For the twisted vacuum state, they are independent of the spectral parameter and can be expressed in terms of $PSU(2,2|4)$ characters of the twist matrix \cite{Gromov:2010vb}. This leads to the solution
  \beqa
&&T_{a,1}^L \, T_{a,1}^R  =  \sum_{i=1,2} \sum_{j=3,4} (\by_i/\by_j)^{a-1} \,\kappa_{ij}( \left\{ \bx \right\} , \left\{\by\right\})\ ,\\
 &&\kappa_{ij}( \left\{ \bx \right\} , \left\{\by \right\}) =\frac{(-1)^{i+j} \by_j\,(\by_i - \bx_1 ) (\by_i-\bx_2 ) (\by_j^{-1} - \bx_3^{-1}) (\by_j^{-1}-\bx_4^{-1} ) }{\by_i \, (\by_1-\by_2) (\by_3^{-1} - \by_4^{-1})}\ ,
 \eeqa
 where the eigenvalues of the twist matrix are parametrised as in (\ref{eq:RPSU224}).
 The state we are considering has charges $J_1 = J$, $J_2 = J_3= S_1 = S_2 = 0$. To satisfy the constraint 
 of state invariance under the twist for general $\Delta$, we should restrict the twists to $\bx_1 \bx_2 = \bx_3 \bx_4 = \by_1 \by_2 = \by_3 \by_4 = 1$.  
 We will consider the special choice
 \beq\label{eq:mulam}
 \left(\by_i | \bx_i \right) = \left(e^{i \frac{\theta_1 + \theta_2}{2} } , e^{-i \frac{\theta_1 + \theta_2}{2} } , e^{i \frac{\theta_1 - \theta_2}{2} } , e^{-i \frac{\theta_1 - \theta_2}{2} } |  e^{i \gamma J/2} ,  e^{-i \gamma J/2} ,  e^{i \gamma J/2},  e^{-i \gamma J/2} \right)\ , 
 \eeq
which corresponds to the twist matrix $$ R = R_{\vec\theta}\cdot \mathcal{G}(\gamma)\ ,$$
namely the product of a spacetime rotation $R_{\vec{\theta}} $  defined in  (\ref{rotation}) and an internal rotation $\mathcal{G}(\gamma)$,  defined  in  section \ref{sec:fishnetlimit}.
 Plugging the weak coupling expansion $e^{-\widetilde E_a(u) J } \sim \left(4 g^2/(a^2 + 4u^2 ) \right)^J$ into (\ref{eq:Luscher}), computing the integrals and summing the series, we find
\beq 
\delta E =-2 g^{2 J} \, \(\!\begin{array}{c}2J-2\\ J-1\end{array}\!\) \sum_{i=1,2} \sum_{j=3,4}
\kappa_{ij}( \left\{ \bx \right\} , \left\{\by\right\}) \,{\rm Li}_{2J-1}\(\frac{\by_i }{ \by_j }\)\ ,
\eeq
where  $g^2 \equiv  N_c g^2_{\text{YM} }/(16 \pi^2) = \lambda/(16 \pi^2)$ is proportional to the 't Hooft coupling. With the choice of twists (\ref{eq:mulam}), the result reduces precisely to (\ref{eq:GammaJ1}), but with a redefined coupling constant
\beq
\xi^{2 J} \rightarrow 16 g^{2 J} \,   \sin\left( \frac{\gamma J - \theta_1 - \theta_2}{4} \right) \sin\left( \frac{\gamma J - \theta_1 + \theta_2}{4} \right) \sin\left( \frac{\gamma J + \theta_1 - \theta_2}{4} \right) \sin\left( \frac{\gamma J + \theta_1 + \theta_2}{4} \right)\ .
\eeq
 In the double scaling limit  which selects the fishnet diagram of Figure \ref{fig:figure1}, $g \rightarrow 0$, $g^2 e^{i \gamma} \rightarrow \xi^2$, we perfectly recover the result of the field theory computation. 
 
 It should be possible to reproduce the result at finite value of $\gamma$ by a direct diagrammatic calculation in $\mathcal{N}$=4 SYM using the methods introduced in this paper. 

\subsection{Baxter equations and Q-functions}
The most powerful method with which to study the spectrum of a quantum integrable model are the so-called Baxter TQ equations or Quantum Spectral Curve. These remarkable equations reduce the diagonalization problem of a quantum integrable Hamiltonian, which is a complex many-body problem, to equations in a single variable. The solutions of Baxter equations are known as Q-functions. It is expected that the Q-functions give access to the wave function of the system in a very special set of coordinates (the Separated Variables), where it becomes completely factorised \cite{Sklyanin:1995bm}. In the next section we will see explicitly the link between Q-functions and the wave function.  
 
The form of the Baxter equation for an arbitrary state in the fishnet theory was determined in  \cite{Gromov:2010kf} using the dual fishchain model, and in \cite{UpcomingFishnet} from the diagrammatic formulation of the quantum field theory. 
They can also be obtained starting from the Quantum Spectral Curve for $\mathcal{N}$=4 SYM theory \cite  {Gromov:2013pga,Gromov:2014caa}  with generic twists \cite{Kazakov:2015efa}, and taking the opportune double scaling limit as done in \cite{Gromov:2017cja}. 
Here, we review the main features to make the discussion internally consistent, and then discuss the example of length-one operators. 

Based on the $SU(2,2)$ symmetry of the model, the Baxter equation for the fishnet model is a fourth order difference equation for the Q-functions depending on the spectral parameter $u$,
\beq\label{eq:TQ}
\sum_{n = 0}^4 a_{n}^{[2 - n]}( u  ) q^{[+4 - 2 n]}(u)  = 0\ ,
\eeq
where $f^{[n]}(u) \equiv f(u + i n/2 )$. 
 Above, $q(u)$ is the Q-function, while the coefficients $a_n(u)$ of the equation are related to the eigenvalues of transfer matrices with antisymmetric representations in the auxiliary space (see e.g. the review \cite{Gromov:2010kf}). As such, these coefficients  are (related to) polynomials in $u$. Their form was fixed in  \cite{Gromov:2019bsj,UpcomingFishnet}, and is given by
 \begin{align}
a_0(u) &= a_4(u) = u^{J} (u-i)^{M}  , &  a_2(u) &= u^{M - J} \, P_{2 J}^{\mathbf{6}}(u) , \\ a_1(u) &= - (u-i/2)^{M} \, P_J^{\mathbf{4}}(u) , & a_3(u) &= - (u-i/2)^{M}  \, P_J^{\bar{\mathbf{4}}}(u) ,
 \end{align}
where $P_n^{\mathbf{4}}$, $P_n^{\bar{\mathbf{4}}}$ and $P_n^{\mathbf{6}}$ are polynomials of degree $n$. Here $J = \text{max}\left(| J_1|, |J_2|\right)$ and $M = \text{min}\left(|J_1| , |J_2| \right)$.\footnote{As was shown in \cite{Gromov:2019bsj}, there exist equivalent forms of the Baxter equation, with the same spectrum, containing an additional anti-magnon number $\bar{M}$. We are writing here the representative equation with $\bar{M}=0$.}
  Being a fourth order equation, (\ref{eq:TQ}) has in general four independent solutions for the Q-functions. They can be distinguished by their large-$u$ asymptotics, which are related to the twist and to the $SU(2,2)$ charges as
\beq\label{eq:Qasy}
q_i(u)\simeq \by_i^{-i u}  \, u^{ \hat M_i } , \,\,\,\,\, u \rightarrow +\infty ,
\eeq
where 
\beq\label{eq:Mh}
\hat M_i = \left(\frac{\Delta-S_1-S_2-D_0}{2}, \frac{\Delta+S_1+S_2-D_0}{2} , \frac{ -\Delta-S_1+S_2-D_0 }{2}  , \frac{-\Delta+S_1-S_2-D_0 }{2}\right)\ ,
\eeq
with $D_0 = J_1 + J_2$, and where $\by_i$ are the eigenvalues of the twist transformation represented as a $SU(2,2)$ matrix, see (\ref{eq:RPSU224}).
 In the following, we will restrict to the twist by a rotation (\ref{rotation}). In this case the eigenvalues $\by_i$ are given in (\ref{eq:mulam}). 
 
 The physical solutions to the Baxter equation, and therefore the spectrum, are determined by imposing two additional constraints. The first is a \emph{quantisation condition}. We propose here for the first time\footnote{Different types of quantisation conditions existed previously. The first condition was derived in \cite{Gromov:2017cja} for a particular case $J_1=3,\;J_2=0$. The general $J_1$ case was developed in \cite{UpcomingFishnet}, based on the field theoretical derivation. We propose here the most general quantisation condition which can be understood as a consequence of the QSC for the full $\mathcal{N}$=4 SYM, and also uses some ideas of \cite{UpcomingFishnet}. The method presented here was used and verified in \cite{Gromov:2019jfh} for several non-trivial cases with $|J_1|\neq |J_2|$. } a new simple quantisation condition which is expected to be valid for all states in the theory.  Enforcing this condition constrains the Q-functions and the coefficients of the polynomials $a_n(u)$ to a discrete set of solutions, which correspond to physical states. 
 Furthermore, we will use an extra equation, derived in \cite{UpcomingFishnet}, to introduce the coupling constant of the fishnet theory in the problem for $|J_1|=|J_2|$ case (the only special case not covered by \cite{Gromov:2019bsj}). 

\subsubsection{Quantisation conditions}\label{sec:quantize}

The starting point is the observation that, for any fixed choice of the coefficients $a_n(u)$, we can define two alternative sets of solutions of the same Baxter equation. The first solution, denoted as $q^{\downarrow}_i(u)$, is obtained by requiring analyticity in the upper-half plane. The asymptotics (\ref{eq:Qasy}) are taken to be valid in this region. Notice that, as a consequence of the Baxter equations, these functions will then in general have poles in the lower half plane at positions $ u \in -i \mathbb{N}$. \\ 
 The second solution $q_i^{\uparrow}(u)$, instead, is defined by requiring analyticity in the lower half plane. The asymptotics (\ref{eq:Qasy}) will be valid in this region. 

Notice that both $q_i^{\uparrow}(u)$ and $q_i^{\downarrow}(u)$ are specified  uniquely as functions of the parameters entering the Baxter equations -- in particular, they can be computed numerically with the method of \cite{Gromov:2015wca}, which we review below in section \ref{sec:numerics}. As observed in \cite{Gromov:2014caa} in the case of the full  $\mathcal{N}$=4 SYM theory, since  $q_i^{\uparrow}(u)$ and $q_i^{\downarrow}(u)$ are solutions of the same finite difference equation, they must be related by a linear transformation $i$-periodic in $u$: 
\beq
\label{eq:Qbar}
\qc_i(u)= {\Omega}_{i}^{j}(u)q_j^{\downarrow}(u)\ ,
\eeq
where 
\beq
 { \Omega}_{i}^{j}(u+i)= { \Omega}_{i}^{j}(u)\ .
\eeq
We found that a sufficient \emph{quantisation condition} is the following  constraint on this matrix:
\beq
\label{eq:NumConstr}
{\Omega}_{1}^{2}(u)= {\Omega}_{2}^{1}(u)= {\Omega}_{3}^{4}(u)= {\Omega}_{4}^{3}(u)=0\ .
\eeq
 This condition was also obtained independently in \cite{UpcomingFishnet}, and can be justified in various ways. In particular, one can argue that the same condition is valid for the Quantum Spectral Curve of $\mathcal{N}=4$ SYM, see Appendix \ref{app:justification}, and therefore it should be inherited by the fishnet model. 
 For the states with one scalar, we also found, together with F. Levkovich-Maslyuk, a direct proof\footnote{This proof is similar to one found for the cusp operators in the ladders limit in \cite{Cavaglia:2018lxi}.}  that (\ref{eq:NumConstr}) is equivalent to the quantisation conditions for the Schr\"odinger equation (\ref{qc2}), see section \ref{secmap}.
 The same quantisation condition was also used to study different operators with length three in \cite{Gromov:2019jfh}. It is expected to be valid for arbitrary operators. 

\subsubsection{Introducing the coupling constant}
\label{sec:Qplus}
For states with $J_1 > J_2$ it can be shown (see, \cite{Gromov:2019jfh,UpcomingFishnet}) that the coupling constant appears as a coefficient of a pole in the Baxter equation:
\beq\label{eq:xipole}
\xi^{2 J_1} = \lim_{u \rightarrow 0} u^{J_1 - J_2} a_2(u).
\eeq
The case with $|J_1| = |J_2| $ is more complicated. 
  The method to take into account the coupling constant for generic states was found in \cite{UpcomingFishnet} and we borrow this unpublished result here.\footnote{We thank the authors of \cite{UpcomingFishnet} for sharing this result with us.} 
 We construct a function $Q_+(u)$, defined as 
\beq
\begin{split}
\label{eq:DefQplus}
Q_+(u) &= A \left(  q_1^{\downarrow}\left(u-\frac{i}{2}\right)\qc_2 \left(u+\frac{i}{2}  \right) -  q_2^{\downarrow}\left(u-\frac{i}{2}\right)\qc_1 \left(u+\frac{i}{2}  \right) \right)\\
&+ q_3^{\downarrow}\left(u-\frac{i}{2}\right)\qc_4 \left(u+\frac{i}{2}  \right) - q_4^{\downarrow}\left(u-\frac{i}{2}\right)\qc_3 \left(u+\frac{i}{2}  \right) \ .
\end{split}
\eeq
 The coefficient $A$ should be chosen in such a way that this combination of Q-functions does not have a pole at $u \sim \pm i/2$. It can be shown that this can be done only when the quantisation condition (\ref{eq:NumConstr}) is satisfied, see Appendix \ref{app:justification}. 
Once this condition is enforced, we can relate the function $Q_+(u)$ to the coupling constant. For the states studied in this paper this relation is
\beq
\label{eq:Qplus0magnons}
\xi^2=\lim_{\epsilon \to 0} \epsilon \frac{Q_+\left( \frac{3i}{2} -i \epsilon \right)}{Q_+\left( \frac{i}{2}\right)}\ .
\eeq

\subsubsection{Baxter equations at length one}
 For the families of length-one operators described in this paper, the form of the Baxter equation is almost entirely fixed by the structure described above. 
\paragraph{Baxter equation for $J_1=1$, $ J_2= 0$.}
Requiring compatibility of the asymptotics  (\ref{eq:Qasy}) with the polynomial-type ansatz for the coefficients $a_n(u)$ fixes
\beqa\label{eq:Baxter0magnon}
a_{0}(u) &=& a_4(u) = u\ , \quad a_1(u) = a_+(u)\ , \,\, a_3(u) = a_-(u)\ ,\nn\\
a_2(u) &=& \frac{ \kappa }{u} + 2 u \, \left( 1 + \cos\theta_1 + \cos\theta_2 \right) + 2\, S_1 \sin\theta_1 +2 \, S_2 \sin\theta_2\ ,\\
a_\pm(u)&=&-2\(\cos\tfrac{\theta_1}{2},\sin\tfrac{\theta_1}{2}\).\(\begin{array}{cc}2u&S_2\\S_1&\mp i\Delta\end{array}\).\(\cos\tfrac{\theta_2}{2},\sin\tfrac{\theta_2}{2}\)^\intercal\ ,\nn
\eeqa
for the family of operators with $J_1=1$, $J_2=0$. Notice that only one coefficient, $\kappa$, is left unfixed. Using (\ref{eq:xipole}), it can be directly related to the coupling constant, $\kappa = \xi^2 $. 
Therefore, the Baxter equation contains both $\Delta$ and $\xi^2$ as parameters. 
Their mutual dependence  for physical states is fixed by the  quantisation condition (\ref{eq:NumConstr}). 

\paragraph{Baxter equation for $J_1=1$, $J_2=1$}
With similar considerations one can see that, for the states built with two scalars and charges $J_1=1$, $J_2=1$, the Baxter equation takes the form
\beqa
\label{eq:Baxter1magnon}
a_{0}(u) &=& a_4(u) = u (u-i),    \quad a_1(u) = b_+(u) (u-i/2 ) , \,\, a_3(u) = b_-(u) (u-i/2),\nn\\
a_2(u) &=& \rho + 2\left(u^2-i u\right) ( 1 + \cos\theta_1 + \cos\theta_2 ) + (2\, u -i)\, \left( S_1 \sin\theta_1 +  S_2 \sin\theta_2 \right)\ ,\\
b_{\pm}(u) &=&-{2}\(\cos\tfrac{\theta_1}{2},\sin\tfrac{\theta_1}{2}\).\(\begin{array}{cc}2u-i&S_2\\S_1&\mp i\Delta\end{array}\).\(\cos\tfrac{\theta_2}{2},\sin\tfrac{\theta_2}{2}\)^\intercal  \ . \nn
\eeqa
Notice that also in this case there is precisely one unfixed coefficient, denoted as $\rho$. 
In this case the relation between this coefficient and the coupling constant is much less straightforward. It is, however, determined implicitly by equation (\ref{eq:Qplus0magnons}). 
The physical states are again selected by the quantisation condition (\ref{eq:NumConstr}). 

\subsection{Exact solutions for $\theta_1 = \theta_2$}
Before discussing numerical results  at generic values of the twists, let us consider the case $\theta_1 = \theta_2 \equiv \theta$ which is solvable analytically. 
In this case, there are no unfixed parameters in the Baxter equation. The  quantisation condition (\ref{eq:NumConstr}) is automatically satisfied and the spectrum can be found explicitly. 
For $\theta_1 = \theta_2$, we have an enhanced symmetry $U(1)_L \times SU(2)_R$. In terms of the related spins, $S_R = \frac{S_1 - S_2}{2}$, $S_L = \frac{S_1+ S_2}{2}$, the asymptotics of the Q-functions are
\beq\label{eq:equalasy}
q_i \sim \left( e^{\theta u}  u^{\frac{\Delta-2S_L -D_0}{2}}, e^{- \theta u} \, u^{\frac{\Delta+2S_L-D_0}{2} } , u^{\frac{ -\Delta-2S_R -D_0 }{2} } , u^{\frac{-\Delta+2S_R -D_0 + 2 }{2}}\right)\ .
\eeq
Notice that, as compared to (\ref{eq:Qasy}), the asymptotics of the last  Q-function is modified by a power of $u$. This is a typical effect of removing one twist \cite{Kazakov:2015efa}. 
\paragraph{States with $J_1=1$, $J_2=0$.}
The compatibility of the asymptotics (\ref{eq:equalasy}) with the Baxter equation (\ref{eq:Baxter0magnon}) at $\theta_1 = \theta_2$ fixes 
\beq\label{eq:kappaeq}
\kappa = -\sin^2\frac{\theta}{2} \, ( \Delta^2 - (1 + 2 S_R )^2 )\ .
\eeq
This new constraint is a special feature of the equal angles limit. 
Together with the condition $\kappa = \xi^2$, it gives precisely the exact spectrum (\ref{eq:equalsp}) determined earlier from Feynman diagrams.

\paragraph{States with $J_1=1$, $J_2=1$.}
In this case the asymptotics (\ref{eq:equalasy}) for equal angles imposes the condition 
\beq
\rho = \frac{-4 S_R^2 (t-1)^2-4 S_R (t-1)^2+\Delta ^2 (t-1)^2-4 (t^2+1)}{4 t}, \,\,\,\, t=e^{i \theta}\ .
\eeq
To compute the spectrum, we must map this parameter to the coupling via (\ref{eq:Qplus0magnons}), and to do this we need to solve explicitly the Baxter equations. 
The solution can be found after transforming the finite-difference equation to a differential equation through a Mellin transform. 
 Imposing analyticity in the upper half plane and the asymptotics (\ref{eq:equalasy}), we find the solutions\footnote{We have written the solutions with a different normalisation for $q_3$, as compared to (\ref{eq:Qasy}).}
 \beqa\label{eq:onemagqs}
q_1^{\downarrow}(u) &=&   \mathcal{F}(S_R, S_L, \Delta, t, u ) ,  \,\,\,\,\,
q_2^{\downarrow}(u)  = \mathcal{F}(S_R, -S_L, \Delta, 1/t, u ), 
\\
q_4^{\downarrow}(u)&=&\mathcal{G}(S_R, S_L, \Delta, t, u ), \,\,\,\,\,
q_3^{\downarrow}(u) \propto \mathcal{G}(S_R, -S_L, \Delta, 1/t, u ) -  \mathcal{G}(S_R, S_L, \Delta, t, u )\ , \nn
 \eeqa
 with 
 \beqa\label{eq:defFG}
&&\frac{\mathcal{F}(S_R, S_L, \Delta, t, u)}{\Gamma(-i u) t^{-i u -1/2} \,  (it - i)^{-\frac{\Delta}{2} + S_L + 1}  } \\
&&\equiv\sum_{l = 0}^{S_R - S_L} (-1)^l \frac{\Gamma( S_R - S_L + 1 ) \Gamma(S_R + S_L + 1 - l)  }{\Gamma(l+1) \Gamma(S_R + S_L + 1) \Gamma(S_R - S_L -l + 1 )} \, (1/t+1)^{l} \, \nn \\
 && \hspace{3cm}\times \; {}_2\tilde F_1\left(-i u , S_L - \frac{\Delta}{2} + 1 - l , -i u + S_L - \frac{\Delta}{2} + 1 - l , t  \right)\ , \nn
 \eeqa
 and
 \beqa
&&\frac{\mathcal{G}(S_R, S_L, \Delta, t, u) }{\Gamma(-i u ) \, (it - i)^{\frac{\Delta}{2} - S_R } } = \\
&=&\sum_{l = 0}^{S_R  + S_L} \frac{\Gamma( S_R + S_L + 1 ) \Gamma(2 S_R - l + 1) \Gamma^2(S_R - \frac{\Delta}{2} + 1)  }{\Gamma(l+1) \Gamma(2 S_R + 1) \Gamma(S_R + S_L -l + 1 ) \Gamma^2(S_R-\frac{\Delta}{2} - l +1 ) } \, (t+1)^{l} \, \nn \\
 && \hspace{3cm}\times \; {}_2\tilde F_1\left(-i u , -S_R  + \frac{\Delta}{2} +  l , -i u - S_R  + \frac{\Delta}{2} + l , t  \right)\ , \nn
 \eeqa
 where $ {}_2 \tilde F_1(a,b,c,z) \equiv {}_2 F_1(a,b,c,z)/\Gamma(c)$.
 One can verify explicitly that the  quantisation condition  (\ref{eq:NumConstr}) is automatically satisfied. 
All we need to do is compute $Q_+(u)$ and use equation (\ref{eq:Qplus0magnons}) to obtain the coupling dependence of the spectrum. 

For instance, in the case $S_R = S_L = 0$, we find that the constant $A$ ensuring the absence of poles in $Q_+(u)$ at $u \sim i/2$ is
\beq
A = \frac{ \Gamma^2( 1- \frac{\Delta}{2} ) }{\Gamma^2(\frac{\Delta}{2} ) }\, \left(4 \sin^2 \frac{\theta}{2} \right)^{\Delta - 1}\ , 
\eeq
and we find, with the definition (\ref{eq:Qplus0magnons}) and using ${}_2 F_1(1,n, n+1, z) = n \Phi(z, 1, n ) $, 
\beqa
&&\lim_{u \rightarrow 0 } u\, Q_+(u + 3/2 i )\equiv- 2\Delta \sin\theta \, \sin\tfrac{\theta}{2} \, A^{\frac{1}{2} } ,\\
&&Q_+(i/2)=-2 \sin \frac{\theta}{2} \, A^{\frac{1}{2}} \, \left( \Phi(e^{i \theta}, 1, -\tfrac{\Delta}{2} ) - \Phi( e^{-i \theta}, 1, -\tfrac{\Delta}{2} )  - \Phi( e^{i \theta}, 1, \tfrac{\Delta}{2} )  + \Phi( e^{-i \theta}, 1, \tfrac{\Delta}{2} )  \right)\ , \nn
\eeqa
from where we see that the result (\ref{E1res1}) earlier computed from Feynman diagrams is exactly reproduced. We can similarly compute the result for any  values of the spins. We did not manage to find a general closed form expression for generic spins, but we list several results in Appendix \ref{app:onemag}. 
\begin{figure}[t!]
    \centering
    \includegraphics[width=0.7 \textwidth]{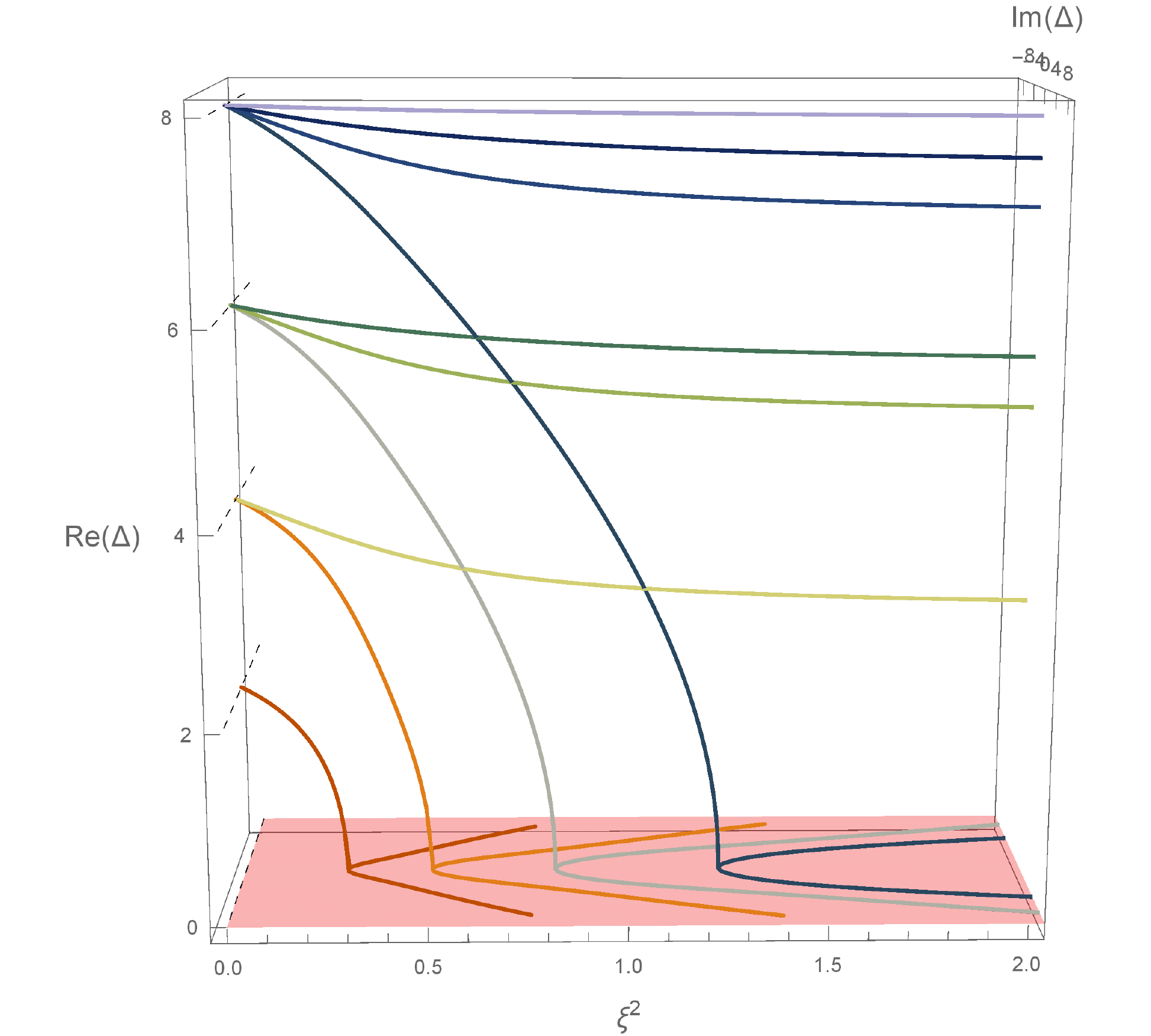}
    \caption{The states with $\Delta(0)=2,4,6,8$ for $J=1,\;M=1$ with $\theta_1=5/6$ and $\theta_2=1/6$, $S_1=S_2=0$. The number of states and weak coupling  match precisely with the predictions obtained in section \ref{sec:onemagnon}. Among the states with a given $\Delta(0)$, one level reaches a square-root branch point where $\Delta=0$, and  the scaling dimension splits into two imaginary levels after this point, with a classical scaling at strong coupling, $\Delta( \xi^2) \sim \pm i \xi$, $\xi \rightarrow \infty$. The   states that do not reach a branch point appear to stay real for all values of the coupling, and tend to constant values at strong coupling. \label{fig:1magA}}
\end{figure}

\subsection{Numerics for generic angles}\label{sec:numerics}
 
\paragraph{Method.}
In this section, we briefly discuss how to find  the spectrum and Q-functions numerically, at a finite value of the coupling constant.  The numerical method is essentially the same as the one  introduced in  \cite{Gromov:2015wca} in the context of $\mathcal{N}$=4 SYM. In this approach, one regards the coefficients entering the polynomials $a_n(u)$ in the Baxter equation as variational parameters. These coefficients are tuned using a generalisation of Newton's method in order to find a solution satisfying the quantisation condition (\ref{eq:NumConstr}). This quantisation condition fixes all the coefficients in the Baxter equation except for one, which is a continuous independent parameter of the solution, and can be related to the coupling constant using the equations (\ref{eq:xipole}) or (\ref{eq:Qplus0magnons}).

To implement the method, we need a routine to   compute the Q-functions $q_i^{\downarrow}$ and $q_i^{\uparrow}$ for any given choice of parameters. Consider for instance the solution analytic in the upper half plane. It satisfies a large-$u$ expansion:
\begin{equation}\label{eq:qlarge}
    q_i^{\downarrow}(u)\sim \by_i^{-i u}u^{\hat{M}_i}\left( 1 +\sum_{n=1}^{\infty} \frac{B_{i,n}}{u^n} \right)\ ,\;\; \text{Im}(u) \rightarrow \infty , 
\end{equation}
where the coefficients $B_{i,n}$ are completely determined by the coefficients in the Baxter equation. 
 With a suitable choice of truncation for the sum, (\ref{eq:qlarge}) gives an arbitrarily good approximation to the solution for large enough values of $\text{Im}(u)$ in the upper half plane.  
We can then use the Baxter equation to translate down the Q-function from this  region to generic points in a strip around the real axis.  
 
From the Q-functions, one can then construct the matrix $  \Omega_i^j(u)$  as a ratio of determinants. It is given by\footnote{This relation can be obtained using the $i$-periodicity of $\Omega$ and the Baxter equation.}
\begin{equation}\label{eq:Omdet}
  \Omega_{a}^b(u) = u^{J_1} (u+i)^{J_1 + J_2} (u + 2 i )^{J_2} \,\frac{ \epsilon^{b b_1 b_2 b_3 } \,\underset{ n=0,1,2,3}{\text{det}}\left\{ q^{\uparrow}_a(u - i n) q^{\downarrow}_{b_1}(u - i n) q^{\downarrow}_{b_2}(u - i n) q^{\downarrow}_{b_3}(u - i n) \right\} }{ 3! \,  \prod_{i<j} ( \by_i - \by_j )}. 
\end{equation}
To turn the quantisation condition (\ref{eq:NumConstr}) into a numerical condition, it is convenient to expand   (\ref{eq:Omdet}) around $u=0$. For the operators studied in this paper, every matrix element of $\Omega$ has at most a single pole at the origin. 
We impose the vanishing of the residue of the pole for $\Omega_1^2$, $\Omega_2^1$, $\Omega_4^3$, $\Omega_3^4$ (it is sufficient to  consider only one of these matrix elements for the convergence of the algorithm). This procedure fixes the trajectory 
 $\Delta(\kappa)$ for $J_2=0$, or $\Delta(\rho)$ for $J_2=1$, where $\kappa$ and $\rho$ are the parameters in the Baxter equations (\ref{eq:Baxter0magnon}),(\ref{eq:Baxter1magnon}). 
We can then compute the associated value of the coupling constant using the on-shell condition \eqref{eq:Qplus0magnons}.\footnote{Alternatively, one can fix the desired value of the coupling constant and use the quantisation condition, together with (\ref{eq:Qplus0magnons}), to fix all parameters in the Baxter equation as functions of the coupling.} 

\begin{figure}[t!]
    \centering
    \includegraphics[width=0.7 \textwidth]{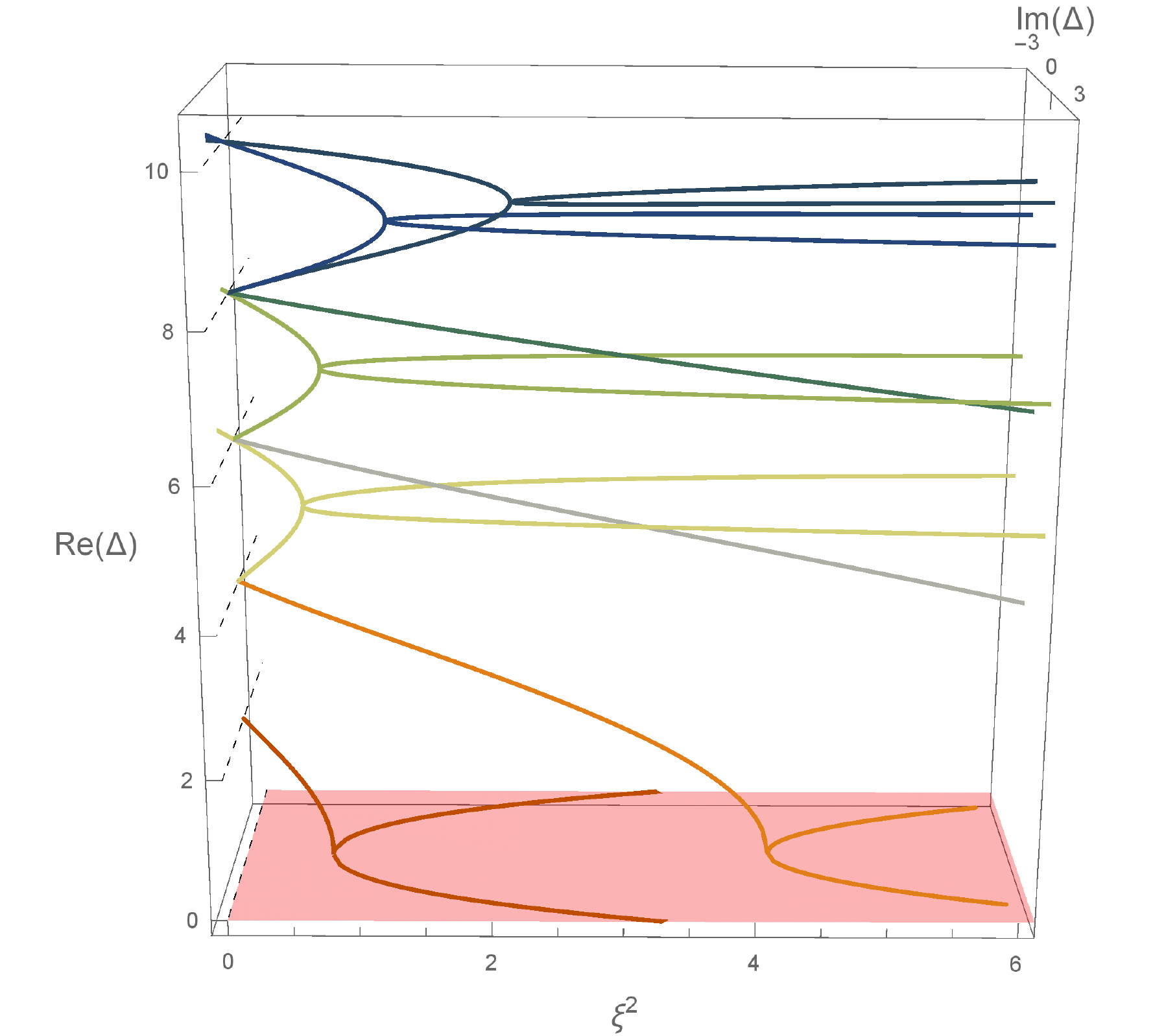}
    \caption{Some states with $\Delta(0)=4,6,8,10$ for $J=1,\;M=1$ with $\theta_1=11/4$ and $\theta_2=9/4$, $S_1=S_2=0$. The number of states and weak coupling expansions again match the predictions in section \ref{sec:onemagnon}, however, for this choice of angles the spectrum is qualitatively different. As for the case of figure \ref{fig:1magA}, exactly one level, for any given $\Delta(0) = 2 n$, reaches a branch point with $\Delta_c = 0$, and then splits into a pair of imaginary levels which show classical behaviour at strong coupling. All other states  merge in pairs at branch points with $\Delta_c \neq 0$. After these branch points, they split into two complex conjugate levels with a nonzero real part, which appear to tend to constant values at strong coupling. \label{fig:1magB}}
\end{figure}
\paragraph{Results.} 
We computed the scaling dimension as a function of the coupling for several states in the length-one sectors, see figures \ref{fig:0mag}, \ref{fig:1magA}, \ref{fig:1magB}. 
In the $J_2 = 0$ case, these results agree with the spectrum of the Schr\"odinger problem which was already discussed.

Some findings for the more complicated one-magnon case are shown in figures  \ref{fig:1magA} and \ref{fig:1magB}. Notice that, at weak coupling, our numerical results confirm very clearly the pattern of degeneracies due to the mixing between different combinations of  derivatives, described in section \ref{sec:1loop1mag}. 
For this case where $J_1=J_2=1$, at strong coupling, we see that most of the states do not display a classical scaling of the type $|\Delta| \sim \xi$, but instead tend to constant values. This non-classical  behaviour is a generic feature of the states with $|J_1| = |J_2|$, and is reminiscent of the behaviour of states in the SYK model. Presently, these states elude a dual description in terms of the fishchain model of \cite{Gromov:2019aku}.  

However, we find that, for any group of states
with a given $\Delta(0)=2n$,  there is exactly one state which behaves classically at strong coupling. For these special \emph{classical states}, the scaling dimension, as a function of the coupling constant, has qualitatively the same behaviour as for the solutions of the zero-magnon case: the dimension decreases monotonically with the coupling, until it reaches a branch point   where $\Delta=0$, and then  splits into a pair of purely imaginary levels. At strong coupling, they scale as $\Delta \sim \pm i \xi$. It would be interesting to determine whether these special states can be captured by the fishchain model. 

All the other levels, instead, reach a constant plateau at strong coupling, which may be either real or complex, depending on the  values of the two angles. We find that, depending on these parameters, the structure of the spectrum is qualitatively different, compare figures \ref{fig:1magA} and \ref{fig:1magB}. 
In particular, while in figure \ref{fig:1magA} all the non-classical states have a real scaling dimension, in the case studied in figure \ref{fig:1magB} we see that all of them fuse pairwise at square-root branch points that occur for $\Delta \neq 0$. After these branch points, the levels split into complex conjugate pairs, which approach complex constant values at strong coupling. 
It might be interesting to study the boundaries of these different phases, determined by the values of the angles. We leave these investigations for future studies.

\subsection{Mapping the Q-functions to the CFT wave functions}\la{secmap}

So far, we had two different descriptions of the same physics. One is in terms of the CFT wave function (\ref{wf}). Its dynamics are governed by the Schr\"odinger equation (\ref{Sch}) with boundary conditions (\ref{qc2}). The other is in terms of the SoV wave function, namely the Q-functions. They are fixed 
by the Baxter equation (\ref{eq:TQ}), (\ref{eq:Baxter0magnon}) and quantisation condition (\ref{eq:NumConstr}). In this section we close the circle by explicitly mapping one to the other. We will be short, and only consider the simplest non-trivial case of $J_1=1$ and $J_2=S_1=S_2=0$. In \cite{colortwist2} we will extend this map and study it in greater detail. We thank Fedor Levkovich-Maslyuk for collaboration on this part. 

Our starting point is the Baxter equation (\ref{eq:TQ}), (\ref{eq:Baxter0magnon}). After setting $S_1=S_2=0$ this equation becomes 
\beq
\begin{split}
0=&\(2u(1+\cos\theta_2+\cos\theta_2)+{\xi^2\over u}\)q(u)\\
&-2\((2u+i)\cos\tfrac{\theta_1}{2}\cos\tfrac{\theta_2}{2}-i\Delta\sin\tfrac{\theta_1}{2}\sin\tfrac{\theta_2}{2}\)q(u+i)+(u+i)q(u+2i)\\
&-2\((2u-i)\cos\tfrac{\theta_1}{2}\cos\tfrac{\theta_2}{2}+i\Delta\sin\tfrac{\theta_1}{2}\sin\tfrac{\theta_2}{2}\)q(u-i)+(u-i)q(u-2i)\ .
\end{split}
\eeq
To map this equation to the Schr\"odinger equation (\ref{Sch}), we first use a Mellin transform similar to the one introduced in \cite{Cavaglia:2018lxi}
\beq
\Theta(\sigma)=\int\limits_{c-i\infty}^{c+i\infty}{du\over2\pi i u}q(u)\,w(\sigma)^{iu}\ ,\qquad w(\sigma)={e^{-\frac{i }{2}\theta_2} \sin \frac{\theta_1}{2}-e^{2\sigma}\, e^{+\frac{i}{2}\theta_1}\sin\frac{\theta_2}{2}\over e^{+i\frac{\theta_1}{2}} \sin\frac{\theta_1}{2}-e^{2\sigma}\,e^{-i\frac{\theta_2}{2}} \sin\frac{\theta_2}{2}}\ ,
\eeq
where $c>0$. We then plug the result into the integral
\beq\la{map}
\psi(\sigma)=\int\limits^{V(\sigma)}_{V_-} {dV'\over V'}{V\over V'}\,\({V-V'\over V'}\)^{{\Delta-3\over2}}\Theta(\sigma(V'))\ ,
\eeq 
where
\beq
V(\sigma)=\tanh\sigma-{\sin^2{\theta_1\over2}+\sin^2{\theta_2\over2}\over \sin^2{\theta_1\over2}-\sin^2{\theta_2\over2}}\ ,\qquad V_-={2\sin^2{\theta_1\over2}\over\sin^2{\theta_1\over2}+\sin^2{\theta_2\over2}}\ ,
\eeq
and ${\rm Re}\,\Delta>1$. It can be checked the $\psi(\sigma)$ in (\ref{map}) satisfies the Schr\"odinger equation (\ref{Sch}), and obeys the boundary conditions (\ref{qc2}) exactly when the quantisation conditions for the Q-functions (\ref{eq:NumConstr}) are satisfied.

Importantly, this relation and its generalisations allows one to explicitly express the results of correlation functions that are computed using field theory techniques in terms of the SoV variables \cite{colortwist2}.

\section{Conclusions and Discussion}\la{secdis}

In this paper, we have introduced a novel type of operators, which can be constructed in any theory with a 't Hooft large-$N$ limit by twisting the colour-trace: \emph{colour-twist operators}.  
We have studied some simple examples of these operators  perturbatively, at finite coupling, and in the strong coupling limit in the fishnet model. 
This was done in two complementary ways, first by a direct field theory calculation
and second, using the integrability based Quantum Spectral Curve (QSC)  technique. A perfect match was obtained for the spectrum of scaling dimensions computed in the two approaches. Finally, we explicitly mapped the wave function in the separation of variable basis, known as the Q-function, to the CFT wave function.

This paper sets the ground for the computation of planar correlation functions in terms of the same objects that are used for computing the quantum spectrum of anomalous dimensions, namely in terms of the QSC Q-functions. Further progress in this direction will be reported in \cite{colortwist2}. We end this paper by enumerating some of the many future directions.
\\

It would be interesting to study colour-twist operators in other integrable quantum field theories and in ${\cal N}=4$ SYM theory in particular. We expect that our field theory definition will give results that match with the twisted Bethe ansatz of \cite{Beisert:2005if}, and, more generally, with the QSC predictions \cite{Kazakov:2015efa}.

In this paper, we have only considered colour-twist operators in the leading planar limit. It would be interesting to extend our definition beyond the planar limit and study $1/N$ corrections to the correlation functions between colour-twist operators. Specifically, the topology of planar Feynman diagrams is that of a sphere with punctures. 
At higher orders in the $1/N$ expansion, the Feynman diagrams have topologies with higher genus. To extend our definition to these cases, one should allow the twist cuts to wind 
around the cycles of the diagrams. For twist transformations of the type (\ref{eq:gammatwist}), such a cutting prescription can be obtained starting from the non-planar Feynman diagrams of the $\gamma$-deformed theory and cutting them into disks in a similar way to the cylinder-cut of section \ref{gamma-deformetion}.

As explained in section \ref{sec_ct}, an operator can only be twisted by a symmetry transformation that leaves it invariant. The twist-symmetry transformations we have considered in this paper are continuously connected to the identity. Namely, they are of the form $R(\theta)=e^{i\theta \hat J}$. Correspondingly, the operators we studied have zero charge under the symmetry generator $\hat J$. Another way of making an operator invariant under $R(\theta)$ is to have its charge quantised in units of $1/\theta$, but not necessarily equal to zero. In particular, when the twist transformation is a spacetime rotation or boost, the corresponding colour-twist operator would have non-integer spin. Operators with non-integer spin exist in any CFT \cite{Kravchuk:2018htv}. These so-called light-ray operators give a continuous interpolation between different operators that are on the same Regge trajectory. They appear as intermediate states in correlation functions, but otherwise they annihilate the vacuum; therefore, their correlators in the un-twisted theory are vanishing. 
It would be interesting to see if colour-twist operators with non-integer spin are somehow related to the light-ray operators in the planar limit. If so, the twisting procedure may  provide a direct way of studying them and  their correlation functions at large $N$. 

Finally, our main motivation for studying the colour-twist comes from the separation of variables (SoV) approach. The twist removes 
the degeneracy in the spectrum, allowing for a one-to-one correspondence between the integrability description (say in terms of the Q-functions) and the actual states in the theory. In the case of integrable spin chains, the twisting procedure is essential for the SoV to be well defined. Equally, in the fishnet theory, or ${\cal N}=4$ SYM theory, we expect the twist to play an essential role for a possible SoV construction for the correlation functions. An example of SoV structures was recently found in \cite{Cavaglia:2018lxi,McGovern:2019sdd,Giombi:2018qox, Giombi:2018hsx} for correlation
functions of cusp operators, and following this paper for single trace operators \cite{colortwist2}.

\section*{Acknowledgements}
We thank Fedor Levkovich-Maslyuk for collaboration on section \ref{secmap}. We thank V.~Kazakov and A.~Zhiboedov for invaluable discussions. A.C. was supported by the STFC grant (ST/P000258/1) Fundamental Physics from the Planck Scale to the LHC. D.G. was supported by the EPSRC Research Studentship (EP/N509498/1). N.G. was supported by the STFC grant (ST/P000258/1). A.S. was supported by the I-CORE Program of the Planning and Budgeting Committee, The Israel Science Foundation (1937/12).

\appendix

\section{$SO(4)$ representations}\label{appSO4rep}

In the main body of the paper, we compute the spectrum of operators in the fishnet model that are twisted by rotation. For that aim, it is useful to explicitly realise the symmetries that are preserved by (\ref{KRK}) and to decompose the $SO(4)$ irreducible representations into representations of the residual symmetry (\ref{twistsymm}) or (\ref{twistsymm2}). 

We work in the canonical frame where the twist transformation is given by $R_{\vec{\theta}}$ in (\ref{rotation}). In this conformal frame, the two fixed points of the twist are $x_0=0$ and $x_{\bar 0}=\infty$. 
Any point $x\in{\mathbb R}^4$ can be represented as a $2\times2$ matrix
\beq
x=\(\!\!\begin{array}{cc}\ \ z_1&z_2\\-\bar z_2&\bar z_1\end{array}\!\),
\eeq
where, as before, $z_1 = x_1 + i x_2$ and $z_2 = x_3 + i x_4$. In this convention, rotations around the origin act as $SO(4)\simeq \(SU(2)_L\times SU(2)_R\)/{\mathbb Z}_2$ right and left multiplications
\beq\la{glr}
x\rightarrow g_L\cdot x\cdot g_R\ ,\qquad g_{L/R}\in SU(2)_{L/R}\ .
\eeq
These $SU(2)_{L/R}$ transformations are generated by the  differential operators
\beq\label{eq:generators}
\begin{array}{ll}
J_L^+=z_2\d_{\bar z_1}-z_1\d_{\bar z_2}  \ ,&J_R^+=z_1\d_{z_2}-\bar z_2\d_{\bar z_1}\ ,\\
J_L^-= \bar z_1\d_{z_2}-\bar z_2\d_{z_1}\ , &J_R^-=z_2\d_{z_1}-\bar z_1\d_{\bar z_2}\ ,\\
J_L^3=(z_1\d_{z_1}-\bar z_1\d_{\bar z_1}+z_2\d_{z_2}-\bar z_2\d_{\bar z_2})/2\ ,\quad &J_R^3=(z_1\d_{z_1}-\bar z_1\d_{\bar z_1}-z_2\d_{z_2}+\bar z_2\d_{\bar z_2})/2\ .
\end{array}
\eeq

Irreducible representations of $SO(4)$ are characterised by two irreducible representations of these two commuting $SU(2)$-factors, and are labelled by the corresponding spins $(j_L,j_R)$. The allowed representations for fields built with scalars and their derivatives (such as in the fishnet model that we consider in this paper) are the ones with $j_L+j_R\in{\mathbb Z}$. Moreover, operators with a single scalar can only have $j_L=j_R\equiv S/2$, 
which correspond to traceless symmetric tensors with $S$ indices. 
Operators in such a $(2j_L+1)(2j_R+1)$-dimensional representation are labelled by the eigenvalues of $(J^{3}_L,J^3_R)$ in (\ref{eq:generators}), and $(m_L,m_R)$, that take values in the range $-j_{L/R}\leq  m_{L/R} \leq j_{L/R}$. The rest of the generators act on these operators in the standard way
\beq\la{basis}
\begin{split}
J^{\pm}_L\circ\cO_{\Delta,j_L,j_R,m_L,m_R}(x)&=\sqrt{j_L(j_L+1)-m_L(m_L\pm1)}\,\cO_{\Delta,j_L,j_R,m_L\pm1,m_R}(x)\ ,\\
J^{\pm}_R\circ\cO_{\Delta,j_L,j_R,m_L,m_R}(x)
&=\sqrt{j_R(j_R+1)-m_R(m_R\pm1)}\,\cO_{\Delta,j_L,j_R,m_L,m_R\pm1}(x)\ .
\end{split}
\eeq
In this representation, the rotation symmetry (\ref{rotation}) that we will use to twist the operators takes the form
\beq\la{rotategen}
R_{\vec{\theta}}= e^{i \theta_1 (J^3_L + J^3_R ) + i \theta_2 (J^3_L - J^3_R )} .
\eeq
In particular, for $\theta_1=\theta_2=\theta$, (\ref{rotategen}) reduces to $\exp(2i\theta J_L^3)$, which makes it clear why the remaining symmetry is enhanced from (\ref{twistsymm}) to (\ref{twistsymm2}). 

\section{Wave function in a generic frame}\label{app:frame}
In this section, we compute the wave function in a generic frame, using two distinct but equivalent points of view. 
\paragraph{From covariance of correlation functions}
The wave function is defined as a particular correlator (\ref{wf}):
\beq\label{eq:newWf}
\Psi_{\tilde R}(x) \equiv \langle \mathcal{O}_{\tilde R}(x_0)  \, \tr\left( \bar{Z}(x) \, \tilde R^{-1} \right)\rangle.
\eeq
In this general definition, $R$ is a conformal transformation with fixed points $x_0$, $x_{\bar{0}}$, which we assume to take the form $\tilde R = K \circ R_{\vec \theta} \circ K^{-1}$, for a conformal map $K$ satisfying $K(0) = x_0$, $K(\infty) = x_{ \bar{0}}$. 
As a result of a generic conformal change of coordinates, correlation functions transform as explained in section \ref{sec:kinematics}.
 This allows us to relate the wave function (\ref{eq:newWf}) to the one in the frame where the  fixed points are $0 $ and $\infty$. In fact, the transformation rule (\ref{eq:kinematics}) gives 
\beq\label{eq:Psitransf}
\Psi_{\tilde R}(x) = \Psi_{R_{\vec \theta}}( K^{-1} \circ x ) \,  \left| \frac{\partial K^{-1} }{\partial y} \right|_{y = x_0 }^{\frac{\Delta}{4} }  \,  \left| \frac{\partial K^{-1} }{\partial y} \right|_{y = x }^{\frac{1}{4} } .
\eeq
To evaluate the Jacobian determinants on the right hand side, we can use the explicit form of the change of frame given in  (\ref{eq:explK1}).\footnote{Here we do not lose generality. The most general change of frame satisfying $K(0)=x_0$, $K(\infty)=x_{\bar{0}}$ is related to (\ref{eq:explK1}) by a rotation $r \in SO(4)$ and dilatation $d$, $K \rightarrow d \circ r \circ K$. The effect of this modification on the map $\tilde R$ only affects the final result (\ref{eq:Psynew}) by redefining the null vectors $n_i$ 
as $n_i \rightarrow r \circ n_i $. Alternatively, we can say that the geometry of the map is fully specified by the data $\left\{ x_0 , x_{\bar{0}} , n_i \right\}$, and this fixes the form of the wave function (\ref{eq:Psynew}). }

For the special conformal transformation (\ref{eq:explK1}), we have, for any $x$, $y$, 
\beq \left| \frac{\partial K^{-1} }{\partial y} \right|^{\frac{1}{4}} = \left| \frac{\partial (K')^{-1} }{\partial y} \right|^{\frac{1}{4}} = \frac{1}{(y - x_{\bar{0}} )^2 }, \quad \quad(K^{-1} \circ x )^2 = \frac{(x_0 - x)^2}{(x_0 - x_{\bar{0}})^2 \, (x- x_{\bar{0}})^2}. 
\eeq
Using these identities, (\ref{eq:Psitransf}) leads  to the result (\ref{eq:Psynew}). 

\paragraph{From the Dyson-type equation in a generic frame} 
Alternatively, the wave function can be characterised as a solution of the Dyson-type equation
\beq\label{eq:newBS} 
\tilde{\mathcal{B}} \circ \Psi_{\tilde R}(x) = \Psi_{\tilde R}(x) ,
\eeq
where $\tilde{\mathcal{B}}$  is the graph-building operator. According to the perturbative rules explained in this paper, for a generic twist map $\tilde R$ the graph-building operator is given by
\beq\label{eq:tilB}
\tilde{\mathcal{B}} \circ  \tilde F(y) \equiv \frac{\xi^2}{\pi^2} \, \int \frac{ d^4 x \, \left|{\d  \tilde R (x) \over\d x}\right|^{\frac{1}{4} } }{(y - x)^2 (x- {\tilde{R}}(x) )^2 } \tilde  F(x) .
\eeq
To relate this expression to the Dyson-type equation in the original frame, we use  (\ref{eq:covariance}) with $x_{\circlearrowleft} \equiv K^{-1}(x)$, which leads to
\beq\label{eq:thisid}
\frac{d^4 x \, \left|{\d  \tilde R (x) \over\d x}\right|^{\frac{1}{4} }  }{(x- {\tilde{R}}(x) )^2} = \frac{d^4 x'   }{(x'- R_{\vec \theta}(x') )^2} \,  \left|{\d  K^{-1} \over\d x}\right|^{-\frac{1}{2} } ,
\eeq
with $x' = K^{-1} \circ x$. Changing integration variables $x \rightarrow x'$ in  (\ref{eq:tilB}), with the help of (\ref{eq:thisid}) we find
\beq
\left(\tilde{ \mathcal{B} } \circ \tilde F \right)(y) =  \,  \left|{\d  K^{-1} \over\d y}\right|^{\frac{1}{4} } \,\left(\mathcal{B} \circ  F \right)( K^{-1} \circ y ) ,
\eeq
where $\mathcal{B}$ is the graph-building operator  defined in the frame with fixed points at $0$ and $\infty$, and the relation between $F$ and $\tilde F$ is
\beq
 \tilde F(x) =  \left|{\d  K^{-1} \over\d x}\right|^{\frac{1}{4} } \, F( K^{-1} \circ x)  .
\eeq
Thanks to this property, the solution of the Dyson-type equation (\ref{eq:newBS}) can be constructed as
 $\Psi_{\tilde R}(x) \propto \Psi_{R_{\vec \theta} }(K^{-1} \circ x) \, \left|{\d  K^{-1} \over\d x}\right|^{\frac{1}{4} }  , $
where  $\Psi_{R_{\vec \theta}}$ is the wave function stationary under the graph-building operator $\mathcal{B}$. Therefore, we have rederived the transformation rule (\ref{eq:Psitransf}), apart from a proportionality factor (independent of $x$). This factor is simply $1/|x_{0 \bar{0}}|^{2 \Delta}$, and can be recovered by demanding that the wave function reduces to $\Psi_{R_{\vec \theta} }$  for $x_{0 \bar{0}} \rightarrow \infty$.

\section{One-magnon eigenvalue}
\la{app:E1}
In this appendix, we describe the details of the calculation of the integral 
\eq{E1int} for the eigenvalue of the graph-building operator at $J_1=J_2=1$ and $\theta _2=\theta _1\equiv\theta$. States in this case are parametrised by their scaling dimension $\Delta$ as well as their $U(1)_L\times SU(2)_R$ quantum numbers, $(m_L,j_R,m_R)$. The method we use is applicable for any values of these spins. For simplicity, we only consider the case where $m_L=j_R=m_R=0$. In this case, the wave function (\ref{eq:wavefonemag}) simplifies to $\Psi_{\Delta,0,0,0}(x)=|x|^{-2\Delta-2}$. The action of the graph building operator on this function is given by the integral (\ref{E1int}) 
\beqa
E_{1,1}={\mathcal B \circ \Psi_{\Delta,0,0,0}(x)\over\xi^2\,\Psi_{\Delta,0,0,0}(x)} &=&\frac{x^{2\Delta+2}}{\pi^2}\int \frac{d^4 w}{(x- w_{\circlearrowleft})^2 ( x-w)^2 w^{2\Delta+2}}\\
&=&
\int\frac{(r_1dr_1d\phi_1)\,(r_2dr_2 d\phi_2)\, (r_1^2+r_2^2)^{-\Delta/2 -1}}{\pi ^2 \left(r_1^2+r_2^2+1-2 r_1
    \cos \phi _1\right) \left(r_1^2+r_2^2+1-2 r_1  \cos \left(\theta-\phi
   _1\right)\right)}\ ,\nn
\eeqa
where, without loss of generality, we have fixed $x=(1,0,0,0)$ and parametrised the integration point as $w=(r_1\cos\phi_1,r_1\sin\phi_1,r_2\cos\phi_2,r_2\cos\phi_2)$.
The integral over $\phi_2$ factorises trivially. The integral over $\phi_1$ can be done explicitly by changing variables to $z=e^{i\phi_1}$, blowing up the contour of integration, and picking up the poles. This results in
\beq
 E_{1,1}=\int\limits_0^\infty\!d\rho\, \int\limits_0^{\pi/2}\!d\sigma\, 
\frac{2 \left(\rho^2+1\right)\rho^{1-\Delta } \sin (2 \sigma )}{\sqrt{\rho^4-2\rho^2 \cos (2
   \sigma )+1} \left(\rho^4-\rho^2 \left(2 \cos ^2\frac{\theta}{2} \cos (2
   \sigma )+\cos \theta-1\right)+1\right)}\ ,
\eeq
where we have used the parametrisation $r_1+i\,r_2=\rho\, e^{i\sigma}$. 
The integration in $\sigma$ can be performed explicitly and gives
\beq
 E_{1,1}={i\over\sin\theta}\int\limits_0^\infty\!{d\rho\over\rho^{\Delta+1}}\log {\rho^2-e^{i\theta}\over\rho^2-e^{-i\theta}}\ .
\eeq
Finally, by integrating over $\rho$, we arrive at \eq{E1res1}.
\label{sec:intro}

\section{One-loop one-magnon data}\la{oneloopdata}

In the table below we present the one-loop scaling dimensions of operators with $J_1=J_2=1$ and $\theta_1=\theta_2\equiv\theta$. Operators in that group can only have an anomalous dimension if they are the twist of a conformal descendant of $\cO_0=\tr(XZ)$. Such operators are characterised by the number of boxes in $\Box^n\cO_0$ and the $U(1)_L\times SU(2)_R$ representation, $(m_L,j_R)$.
\beq\la{onelooptable}
\begin{array}{|c|c|c|l|}\hline
j_R&m_L&
n & \Delta \\ \hline
 0 & 0 & 0 & 2-2 \xi ^2 \\
 0 & 0 & 1 & 4-2 \xi ^2 \cos (\theta ) \\
 \frac{1}{2} & \frac{1}{2} & 0 & 3-2 e^{\frac{i \theta }{2}} \xi ^2 \cos
   \left(\frac{\theta }{2}\right) \\
 \frac{1}{2} & \frac{1}{2} & 1 & 5-\frac{2}{3} e^{\frac{i \theta }{2}} \xi ^2
   \left(\cos \left(\frac{\theta }{2}\right)+2 \cos \left(\frac{3 \theta
   }{2}\right)\right) \\
 1 & 0 & 0 & 4-\frac{2}{3} \xi ^2 (\cos (\theta )+2) \\
 1 & 0 & 1 & 6-\frac{2}{3} \xi ^2 (2 \cos (\theta )+\cos (2 \theta )) \\
 1 & 1 & 0 & 4-\frac{2}{3} e^{i \theta } \xi ^2 (2 \cos (\theta )+1) \\
 1 & 1 & 1 & 6+\frac{1}{3} e^{i \theta } \xi ^2 (-2 \cos (\theta )-3 \cos (2 \theta
   )-1) \\
 \frac{3}{2} & \frac{1}{2} & 0 & 5+\frac{1}{3} e^{\frac{i \theta }{2}} \xi ^2 \left(-5
   \cos \left(\frac{\theta }{2}\right)-\cos \left(\frac{3 \theta }{2}\right)\right) \\
 \frac{3}{2} & \frac{1}{2} & 1 & 7+\frac{1}{15} e^{\frac{i \theta }{2}} \xi ^2 \left(-7
   \cos \left(\frac{\theta }{2}\right)-17 \cos \left(\frac{3 \theta }{2}\right)-6 \cos
   \left(\frac{5 \theta }{2}\right)\right) \\
 \frac{3}{2} & \frac{3}{2} & 0 & 5-2 e^{\frac{3 i \theta }{2}} \xi ^2 \cos
   \left(\frac{\theta }{2}\right) \cos (\theta ) \\
 \frac{3}{2} & \frac{3}{2} & 1 & 7+e^{\frac{3 i \theta }{2}} \xi ^2 \left(-\frac{3}{5}
   \left(\cos \left(\frac{\theta }{2}\right)+\cos \left(\frac{3 \theta
   }{2}\right)\right)-\frac{4}{5} \cos \left(\frac{5 \theta }{2}\right)\right) \\
 2 & 0 & 0 & 6-\frac{2}{15} \xi ^2 (6 \cos (\theta )+\cos (2 \theta )+8) \\
 2 & 0 & 1 & \frac{1}{5} \xi ^2 (-5 \cos (\theta )-4 \cos (2 \theta )-\cos (3 \theta
   ))+8 \\
 2 & 1 & 0 & 6+\frac{1}{5} e^{i \theta } \xi ^2 (-6 \cos (\theta )-\cos (2 \theta )-3)
   \\
 2 & 1 & 1 & 8-\frac{2}{15} e^{i \theta } \xi ^2 (4 \cos (\theta )+7 \cos (2 \theta )+2
   \cos (3 \theta )+2) \\
 2 & 2 & 0 & 6-\frac{2}{5} e^{2 i \theta } \xi ^2 (2 \cos (\theta )+2 \cos (2 \theta
   )+1) \\
 2 & 2 & 1 & 8-\frac{2}{15} e^{2 i \theta } \xi ^2 (4 \cos (\theta )+4 \cos (2 \theta
   )+5 \cos (3 \theta )+2) \\
 \frac{5}{2} & \frac{3}{2} & 0 & 7-\frac{2}{15} e^{\frac{3 i \theta }{2}} \xi ^2
   \left(7 \left(\cos \left(\frac{\theta }{2}\right)+\cos \left(\frac{3 \theta
   }{2}\right)\right)+\cos \left(\frac{5 \theta }{2}\right)\right) \\
 \frac{5}{2} & \frac{3}{2} & 1 & 9-\frac{2}{105} e^{\frac{3 i \theta }{2}} \xi ^2
   \left(27 \cos \left(\frac{\theta }{2}\right)+27 \cos \left(\frac{3 \theta
   }{2}\right)+41 \cos \left(\frac{5 \theta }{2}\right)+10 \cos \left(\frac{7 \theta
   }{2}\right)\right) \\
 \frac{5}{2} & \frac{5}{2} & 0 & 7-\frac{2}{3} e^{\frac{5 i \theta }{2}} \xi ^2 \cos
   \left(\frac{\theta }{2}\right) (2 \cos (2 \theta )+1) \\
 \frac{5}{2} & \frac{5}{2} & 1 & 9+\frac{2}{21} e^{\frac{5 i \theta }{2}} \xi ^2
   \left(-5 \left(\cos \left(\frac{\theta }{2}\right)+\cos \left(\frac{3 \theta
   }{2}\right)+\cos \left(\frac{5 \theta }{2}\right)\right)-6 \cos \left(\frac{7 \theta
   }{2}\right)\right) \\
 3 & 2 & 0 & 8-\frac{2}{21} e^{2 i \theta } \xi ^2 (8 \cos (\theta )+8 \cos (2 \theta
   )+\cos (3 \theta )+4) \\
 3 & 2 & 1 & 10+\frac{1}{21} e^{2 i \theta } \xi ^2 (-10 \cos (\theta )-10 \cos (2
   \theta )-14 \cos (3 \theta )-3 \cos (4 \theta )-5) \\
 3 & 3 & 0 & 8-\frac{2}{7} e^{3 i \theta } \xi ^2 (2 \cos (\theta )+2 \cos (2 \theta
   )+2 \cos (3 \theta )+1) \\
 3 & 3 & 1 & 10+\frac{1}{14} e^{3 i \theta } \xi ^2 (-6 \cos (\theta )-6 \cos (2 \theta
   )-6 \cos (3 \theta )-7 \cos (4 \theta )-3) \\  \hline
\end{array}
\eeq

Another example is given by $\{m_L,m_R,S+2n\}=\{0,0,4\}$. In this case, there are three operators $\vec{\mathcal{O}} \equiv \left\{ (\partial_1 \bar{\partial}_1 )^2 , \, \partial_1 \partial_2 (\bar{ \partial}_1 \bar{ \partial}_2 ) , \,  (\partial_2 \bar{\partial}_2 )^2 \right\} \, \mathcal{O}_0$. The corresponding mixing matrix is
\beq
\hat\Gamma_{\vec{\theta}}\,\vec{\mathcal{O}} = \xi^2  \left(\begin{array}{ccc}  -\frac{32 s_1^2}{5}  + 8 s_1  -2 & \frac{16 s_1 }{15} (5 - 6 s_1  ) & -\frac{16 s_1^2  }{15} \\
\frac{4 s_2 }{15} (5 - 6 s_1  )   & -\frac{64 s_1  s_2 }{15 } + \frac{8}{3} (s_1  + s_2  ) -2& \frac{4 s_1 }{15} (5 - 6 s_2  )  \\
-\frac{16 s_2^2 }{15}   & \frac{16 s_2 }{15} (5 - 6 s_2  )  &  -\frac{32 s_2^2}{5}  + 8 s_2  -2  
  \end{array} \right)  \vec{ \mathcal{ O } }\ , 
  \eeq
 where $ s_i = \sin^2\frac{ \theta_i }{2} $. It is straightforward to construct the mixing matrix for any $m_L$, $m_R$ and $M$.

\section{Single-wrapping anomalous dimension from fishnet diagrams}\label{app:oneloop}
The leading diagram contributing to the two-point function $\langle \mathcal{O}_J(x_0) \, \bar{\cO}_J(x_{\bar{0}}) \rangle $, where $\mathcal{O}_J(x_0) = \tr\left( Z^J(x_0) R_{\vec{\theta}} \right)$, is exemplified in figure \ref{fig:figure1}. The goal of this section is to extract the logarithmic divergence of the diagram, which corresponds to the anomalous dimension of the operator. 

We introduce a point-splitting regularisation by separating the $Z$ scalar from the fixed point of the twist transformation $x_0$
\beq
\mathcal{O}_J^{\text{reg}}  = \tr\left( Z^J(x) R_{\vec{\theta}} \right)\ ,\qquad\bar{\cO}_J ^{\text{reg}} =  \tr\left( \bar{Z}^J(y) R_{-\vec{\theta}}\right)\ .
\eeq
We will be working in the frame with $x_0 = 0$, $x_{\bar{0}} = \infty$ and $x^2=1/y^2=\epsilon^2$.

The anomalous dimension can be read off from
\beq
{\langle \mathcal{O}_J^{\text{reg}}\bar{\cO}_J^{\text{reg}}\rangle_{J\text{-loop}}\over\langle \mathcal{O}_J\bar{\cO}_J\rangle_\text{tree}}\sim  \log(\epsilon) \, \gamma^{\text{1-wheel}}_J\ .
\eeq
As illustrated in figure \ref{fig:figure1}, we can isolate the integration over one of the internal vertices. We remain with a ladder subdiagram times two propagators
 \beq
\langle \mathcal{O}_J^{\text{reg}}\bar{\cO}_J^{\text{reg}}\rangle_{J\text{-loop}}=  \frac{\xi^{2}}{ \pi^2} \,\int d^4 w \, \frac{\mathcal{G}_{J-1}^{\text{ladder}}(x, y, w, w_{\circlearrowleft} ) }{(x - w)^2 \, (y - w)^2}\ .\label{eq:Gint}
 \eeq
 The ladder integral (grey in figure \ref{fig:figure1}) was evaluated in 
\cite{Usyukina:1993ch} and is given by
\beq
\mathcal{G}_{n}^{\text{ladder}}(x_1,x_2,x_3,x_4)= \xi^{2 n} \, {1\over x_{12}^{2n}x_{34}^2}{(1-z)(1-\bar z)\over z-\bar z}L_n(z, \bar{z} )\ ,
\eeq
where  $z$ and $\bar{z}$ are the two conformal cross ratios
\beq\label{eq:crossratios}
z\bar z={x_{14}^2x_{23}^2\over x_{13}^2x_{24}^2}\ ,\qquad(1-z)(1-\bar z)={x_{12}^2x_{43}^2\over x_{13}^2x_{24}^2}\ ,
\eeq
and the ladder function $L_n$ is given by
\beq
L_n=\sum_{j=n}^{2n}{j![-\log(z\bar z)]^{2n-j}\over n!(j-n)!(2n-j)!}\[{\rm Li}_j(z)-{\rm Li}_j(\bar z)\]\ .
\eeq
The form (\ref{eq:Gint}) is convenient, since we can directly set $x=0$  and $y \rightarrow \infty$ in the integrand without encountering any singularity. The UV divergence will be produced only after performing the integration over $w$. 
 Defining the integral with a cutoff around $0$ and $\infty$, we have
\beq\la{relationto4point} 
 {\langle \mathcal{O}_J^{\text{reg}}\bar{\cO}_J^{\text{reg}}\rangle_{J\text{-loop}}\over\langle \mathcal{O}_J\bar{\cO}_J\rangle_\text{tree}}
 =\frac{\xi^{2 J} }{\pi^2 }\int d^4 w \frac{1}{(w - w_{\circlearrowleft})^2 } \, {(1-z)(1-\bar z)\over z-\bar z}L_n(z, \bar{z} )\ , 
\eeq
where the explicit form of the cross ratios in terms of the integration variable is
\beq
z\bar z = 1\ ,\qquad(1-z)(1-\bar z)={(w-w_{\circlearrowleft})^2\over w^2} = 4{ r_1^2 \sin^2{\theta_1\over2}+r_2^2\sin^2{\theta_2\over2}\over ( r_1^2 + r_2^2)}\ ,
\eeq
 with $r_1^2 = w_1^2 + w_2^2$, $r_2^2 = w_3^2 + w_4^2$ being the square radial coordinates in the two rotation planes. 
 
 The simplest case to consider is when the twist involves equal angles $\theta_1 = \theta_2 = \theta$. In this case, the cross ratios are  independent on the integration variable and given by
 \beq
z = e^{i \theta}\ ,\qquad \bar{z} = e^{- i \theta}\ .
\eeq
Therefore, the radial integration factors out and we can immediately extract the anomalous dimension
\beq
\gamma_J^{\text{1-wheel}}(\theta, \theta) =  - {i \xi^{2 J }\over\sin\theta} \, \(\begin{array}{c}2J-2\\J-1\end{array}\)\[{\rm Li}_{2J-2}(e^{-i\theta})-{\rm Li}_{2J-2}(e^{i\theta})\]\ .
\eeq
For generic angles $\theta_1 \neq \theta_2$, the cross ratios instead have a non-trivial dependence on  $r_1$, $r_2$. Introducing $\varphi$ defined as
\beq\la{ztovarphi}
z=e^{-i\varphi}\ ,\quad\bar z=e^{i\varphi}\ ,\quad(1-z)(1-\bar z)=4\sin^2{\varphi\over2} ,
\eeq
 and changing integration variables to $(r_2^2, \varphi)$, we find that the integral factorises as
\beqa\la{relationto4point2}
&&{\langle \mathcal{O}_J^{\text{reg}}\bar{\cO}_J^{\text{reg}}\rangle_{J\text{-loop}}\over\langle \mathcal{O}_J\bar{\cO}_J\rangle_\text{tree}}
\sim\frac{\xi^{2 J} }{4 i}  \(\!\begin{array}{c}2J-2\\ J-1\end{array}\!\)\!\int\limits_{ \epsilon^2 < r_1^2 + r_2^2 < 1/\epsilon^2 }\!\!\!\!\!\!\!\!\!\!{d(r_1^2)d(r_2^2) \,  \[{\rm Li}_{2J-2}(e^{i \varphi} )-{\rm Li}_{2J-2}( e^{- i \varphi})\] \, \tan\frac{\varphi}{2} \over(r_1^2+r_2^2)(r_1^2\sin^2{\theta_1\over2}+r_2^2\sin^2{\theta_2\over2})}  \nn\\
& = & \underbrace{  \left(-\frac{1}{2} \int\limits_{\epsilon< r_2< 1/\epsilon} {d (r_2^2)\over r_2^2} \right) }_{ -2\log\epsilon } \times \underbrace{\left(  - i \xi^{2 J} \(\!\begin{array}{c}2J-2\\  J-1\end{array}\!\)\int\limits_{\theta_2}^{\theta_1} d\varphi\,{{\rm Li}_{2J-2}(e^{i\varphi})-{\rm Li}_{2J-2}(e^{-i\varphi})\over\cos\theta_1-\cos\theta_2} \right) }_{ \gamma_J^{\text{1-wheel}} (\theta_1, \theta_2) }\ , 
\eeqa
yielding the anomalous dimension in (\ref{eq:GammaJ1}).

\section{Justification of the quantisation condition of Baxter equations from $\mathcal{N}=4$
 SYM} \label{app:justification}
 As shown in \cite{Gromov:2017cja,UpcomingFishnet}, the Baxter equations for the fishnet model can be derived by applying the double scaling limit to the Quantum Spectral Curve of $\mathcal{N}=4$ SYM.  
 These results show that the Q-functions of the fishnet model are directly connected to a  quadruplet of Q-functions of $\mathcal{N}=4$ SYM,  denoted as $\bQ_i$. For example, for the states with charges $J_1 = J$, $J_2=0$, we have the relation \cite{UpcomingFishnet}:
 \beq
\bQ_i(u)  \rightarrow q^{\downarrow}_i(u)/u^{J/2}  \quad \quad    \text{(double scaling limit)}\ .
 \eeq
 A dual set of Q-functions, denoted as $\bQ^i(u)$, also plays an important role. In the double scaling limit, they are related to $3 \times 3$ determinants constructed out of the $q$-functions. 
 Given this relation, it is not surprising that there is a way to derive the quantisation condition (\ref{eq:NumConstr}) from the properties of the QSC, which we now explain. 
 
 The two sets of Q-functions $\bQ_i$ and $\bQ^i$ each satisfy a fourth order Baxter equation\footnote{In this case, the coefficients of the Baxter equation are not simple rational functions of $u$, but are explicitly constructed out of other Q-functions usually denoted as $\bP_i$, $\bP^i$. }, which is a consequence of the algebraic $PSU(2,2|4)$ Q-system relations \cite{Gromov:2014caa,Gromov:2017blm}. 
 In the same way that we explained in section \ref{sec:quantize}, one can consider two independent solutions of these difference equations, one solution being analytic in the upper half $u$-plane, and the other in the lower half $u$-plane. In this way, we introduce  $\bQ_i^{\downarrow}(u)$ and $\bQ_i^{\uparrow}(u)$, and similarly $\bQ^{i \downarrow}(u)$, $\bQ^{i \uparrow}(u)$.\footnote{ In the conventional notations used in the QSC literature, $\bQ_i(u) \equiv \bQ_i^{\downarrow}(u)$ and similarly $\bQ^i(u) \equiv \bQ^{i \downarrow}(u)$.} 
 For real values of the parameters these two sets of solutions are related by complex conjugation $\bQ^{\uparrow}_i(u) = \overline{\bQ_i^{\downarrow}}(u)$ in $\mathcal{N}=4$ SYM. However, this direct relation is lost when considering generic complex points in parameter space, and for this reason we will not use it in the following. 
 
 Since they satisfy the same Baxter equation, the two sets of Q-functions analytic in the upper/lower half plane are related by an $i$-periodic matrix,
\beq\label{eq:QbarN=4}
\bQ^{\uparrow}_i(u) = \Omega_i^j(u) \, \bQ_j^{\downarrow}(u)\ , \,\,\,\, \Omega_i^j(u+i) = \Omega_i^j(u)\ .
\eeq 
Importantly,  this matrix $\Omega$ is directly related to the corresponding matrix in the fishnet model (see eq. (\ref{eq:Qbar})) after taking the double scaling limit. 

While the Q-functions in the fishnet model are meromorphic, their ``parent functions'' in $\mathcal{N}=4$ SYM have square-root branch cuts rather than poles. The branch cuts are at $u \in (-2 g, 2g) - i n$, $n \in \mathbb{N}$, where $g^2 = \lambda/(16 \pi^2)$. 
The analytic continuation of a Q-function through the branch cut on the real axis is denoted as $\widetilde \bQ^i$. The quantisation conditions for $\mathcal{N}$=4 SYM can be stated as \cite{Gromov:2014caa} 
\beq\label{eq:Qomega}
\widetilde \bQ^{i \downarrow}(u) = \omega^{ij}(u) \bQ_j^{\downarrow}(u)\ ,
\eeq
where $\omega^{ij}(u)$ is an  $i$-periodic function of $u$, satisfying the algebraic conditions
\beq\label{eq:Pfaff}
\omega^{ij}(u) = -\omega^{ji}(u)\ , \,\,\,\, 1 = \omega^{12}(u) \omega^{34} (u) - \omega^{13}(u) \omega^{24}(u) + \omega^{14}(u) \omega^{23}(u)\ .
\eeq
Combining (\ref{eq:QbarN=4}) and (\ref{eq:Qomega}), we obtain 
\beq
\label{eq:DefG}
\widetilde{\bQ}^{i \downarrow}(u)=G^{ij} {\bQ}_j^{\uparrow}(u)\ ,
\eeq
where $G^{ij}$ is defined as
\beq
G^{ik} \Omega_k^j(u) = \omega^{ik}(u)\ , 
\eeq
and is called the gluing matrix  \cite{Gromov:2015vua}. The crucial property of the gluing matrix is that it has no cuts. For the case of integer spins that we consider, it is  a constant matrix. The form of this matrix was determined for the theory without twists in \cite{Gromov:2015vua}. The case of  the theory with twists in AdS was never explicitly considered; however, our results are consistent with the assumption that the gluing matrix takes the form: 
\beq\label{eq:Gmat}
G^{ij}=\begin{pmatrix}
0 & \alpha & 0 & 0 \\
- \alpha & 0 & 0 & 0\\
0 & 0 & 0 & \beta\\
0 & 0 & -\beta & 0
\end{pmatrix},
\eeq
where $\alpha$, $\beta$ are two constants that may depend on the state and are in general non-vanishing.  
The antisymmetry of $\omega^{ij}$, combined with (\ref{eq:Gmat}), implies precisely that the four components $\Omega_1^2$, $\Omega_2^1$, $\Omega_3^4$,  $\Omega_4^3$    must vanish. This justifies the quantisation condition (\ref{eq:NumConstr}). 

Notice that the ansatz (\ref{eq:Gmat}), together with (\ref{eq:Pfaff}), imposes several further constraints on the form of the matrix $\Omega_i^j(u)$. In particular, it must take the form 
\begin{equation}\label{eq:Omform}
\Omega = \left(\begin{array}{cccc}
\Omega_1^1 & 0 & \Omega_1^3& \Omega_1^4 \\
 0 & \Omega_1^1 & \Omega_2^3& \Omega_2^4\\
 A \, \Omega_2^4 & - A\, \Omega_1^4 & \Omega_3^3 & 0 \\
-A \, \Omega_2^3 & A \,\Omega_1^3 & 0 & \Omega_3^3\end{array}\right) ,
\end{equation}
where $A = \alpha/\beta$. 
Indeed, our numerical study of twisted Baxter equations in the fishnet model confirms that, once we impose even just one of the conditions (\ref{eq:NumConstr}), the matrix $\Omega$ acquires the form (\ref{eq:Omform}). 
This provides a  strong cross-check of the correctness of the quantisation condition (\ref{eq:NumConstr}). It can be proved that the constant $A$ appearing in (\ref{eq:Omform}) is the same parameter guaranteeing the absence of poles in the bi-linear combination $Q_+(u)$ defined in (\ref{eq:Qplus0magnons}), see \cite{UpcomingFishnet}.

\section{Results for the one-magnon spectrum at equal angles from integrability}\label{app:onemag}
Starting from the solutions of the Baxter equation (\ref{eq:onemagqs}), we can compute the spectrum using the exact equation (\ref{eq:Qplus0magnons}) derived in \cite{UpcomingFishnet}:
\beq\label{eq:ratioapp}
\xi^2 = i\lim_{u \rightarrow 0} u \, \frac{ Q_+(u + 3/2 i ) }{Q_+(i/2)}\ .
\eeq
Let us discuss some of the formal steps involved. 
We listed the explicit form of the Q-functions analytic in the upper half plane in (\ref{eq:onemagqs}). The Q-functions analytic in the lower half plane are
 \beqa\label{eq:onemagqs2}
q_1^{\uparrow}(u) &=& e^{ i \pi ( -\Delta/2 + S_L + 1)} \mathcal{F}(S_R, S_L, \Delta, 1/t, -u + i )\ , \\
q_2^{\uparrow}(u)  &=& e^{ i \pi ( -\Delta/2 - S_L + 1)} \mathcal{F}(S_R, -S_L, \Delta,  t ,- u + i)\ , 
\\
q_4^{\uparrow}(u)&=& e^{ i \pi ( \Delta/2 - S_R)} \mathcal{G}(S_R, S_L, \Delta, 1/t, -u + i)\ , \\
q_3^{\uparrow}(u) &\propto& e^{ i \pi ( \Delta/2 - S_R)}  \left( \mathcal{G}(S_R, -S_L, \Delta, t, -u ) -  \mathcal{G}(S_R, S_L, \Delta, 1/t, -u + i) \right)\ . \nn
 \eeqa
Next, one needs to fix the form of the bi-linear combination $Q_+(u)$ by fixing the constant $A$ appearing in (\ref{eq:DefQplus}). 
This constant is fixed in such a way that $Q_+(u)$ is regular at $u \sim i/2$ and can easily be found case by case. We conjecture the general form of this constant for arbitrary spins to be
\beq\label{eq:Aonshell}
A = (-1)^{S_L + S_R} \, \frac{\Gamma^2(1 - \frac{\Delta}{2} + S_R) \Gamma(S_R + 1 - S_L) \Gamma(S_R + S_L + 1)}{\Gamma^2(\frac{\Delta}{2} - S_R ) \Gamma^2(2 S_R + 1) } \, ( 4 \sin^2\frac{\theta}{2}  )^{\Delta - 1} \, \left( \frac{\cos^2\frac{\theta}{2}}{\sin^2\frac{\theta}{2}}  \right)^{S_R}.
\eeq
Using the explicit form of the Q-functions, one can then explicitly evaluate (\ref{eq:ratioapp}). In particular, the evaluation of the limit $\lim_{u \rightarrow 0} Q_+(u + 3/2 i)$ is straightforward, as the pole comes from the $\Gamma(-i u)$ pre-factor included in the functions $\mathcal{F}$ and $\mathcal{G}$ functions, see (\ref{eq:defFG}).  Stripping off the pole, the result takes the generic form of a linear combination of Lerch functions $\Phi(t^{\pm1} , 1, b) \equiv 1/b \; {}_2 F_1(1, b ,b+1, t^{\pm 1}) $, multiplied by rational functions of $\Delta$ and $t$, with a prefactor $A^{\frac{1}{2} }$. 
Using identities between the Lerch functions, the result can be significantly simplified, and we find that the residue of $Q_+(u) $ at $u \sim 3/2 i$ can always be rewritten as $A^{\frac{1}{2} }$, multiplied by a rational function of $\Delta$  and $t$.  
To evaluate the denominator in (\ref{eq:ratioapp}), it is convenient to rewrite the function $Q_+(u)$ in the form
\beq\label{eq:Qplusother}
\begin{split}
Q_+(u) &=  -A \left(  q_1^{\downarrow}\left(u +\frac{i}{2}\right)\qc_2 \left(u-\frac{3 i}{2}  \right) -  q_2^{\downarrow}\left(u+\frac{i}{2}\right)\qc_1 \left(u-\frac{3i}{2}  \right) \right)\\
&- q_3^{\downarrow}\left(u+\frac{i}{2}\right)\qc_4 \left(u-\frac{3i}{2}  \right) + q_4^{\downarrow}\left(u+\frac{i}{2}\right)\qc_3 \left(u-\frac{3i}{2}  \right) \ .
\end{split} 
\eeq
This alternative expression is completely equivalent to (\ref{eq:DefQplus}) once we choose the value of $A$ canceling the poles, namely (\ref{eq:Aonshell}). This can be proved using the form of $\Omega$ in (\ref{eq:Omform}). The expression  (\ref{eq:Qplusother}) has the advantage that it manifestly has no pole at $u = i/2$. The direct evaluation of $Q_+(i/2)$ then yields a quadratic combination of Lerch functions. Again, this can be further simplified and leads in the end to a simple linear combination involving a basis of four Lerch functions. For instance, for integer spins we can always bring the result to the form
\beqa
\frac{(1 + t )^{2 S_R} \, \prod_{n =1}^{S_R
} (\Delta^2 - (2 n)^2 )  }{t^{S_R} \, \xi^2} &=&   t^{S_L} \frac{(1 + t)^2}{t} \,  \, R^{(0)}(t, \Delta, S_R, S_L)  \\
&&+ t^{S_L} \, {  R^{(1)}(t, \Delta, S_R, S_L)  } \, \frac{2 t \left( \Phi(t,1,-\frac{\Delta}{2} ) + \Phi(1/t ,1,\frac{\Delta}{2} )\right)}{ \Delta \, (-1 + t^2)  
}\nn \\&&
 - t^{S_L} \, {  R^{(1)}(t, -\Delta, S_R, S_L)  } \, \frac{2 t \left( \Phi(t,1, \frac{\Delta}{2} ) + \Phi(1/t ,1,-\frac{\Delta}{2} )\right)}{ \Delta \, (-1 + t^2) }\  \nn   , 
\eeqa
where $R^{(i)}(t, \Delta, S_R, S_L)$ are polynomials in $t^{\pm 1} $ and $\Delta$. 
A small sample of results is listed in the table below. 
\beq\la{tableR}
\footnotesize
\begin{array}{|c|c| l|l|}\hline
S_{R} & S_L & R^{(0)}(\theta, \Delta, S_R, S_L) & R^{(1)}(\theta, \Delta, S_R, S_L ) \\ \hline
 0 & 0 & 0 & 1 \\ \hline
 1 &  \left\{-1,0,1\right\} & 4 S_L^2 & \begin{array}{l} S_L^2 \left(2 (\Delta -2) t-2 (\Delta +2) t^{-1} \right)\\+( (\Delta + 2)  t^{-\frac{1}{2}}+(\Delta -2) t^{\frac{1}{2} } )^2
 \end{array}\\  \hline
 2 & 0 & -384 
 &  \left(6 \Delta (\frac{1}{t}-t) +\Delta ^2 \frac{(t+1)^2}{t} +8 ( t-4 +\frac{1}{t} )\right)^2 \\
\hline
 2 & \left\{-1,1\right\}  & 4 \left(\Delta ^2 t^{-1} +\Delta ^2 t+2 \Delta ^2 -64 \right) 
 &  \begin{array}{l}
     (\Delta +(\Delta -4) t +4) {\Big ( }\Delta ^3 \frac{(t+1)^3}{t^2}-96 (\frac{1}{t}-1) \\
     +8 \Delta  (\frac{1}{t}+1) (t-7 +\frac{1}{t} ) -6 \Delta ^2 (t-1) (\frac{1}{t}+1)^2 { \Big ) }
 \end{array} \\
\hline
 2 & \left\{-2,2\right\}  & \begin{array}{l}  16 \Delta ^2 t^{-1}+4 \Delta ^2 t^2-16 t^2+16 \Delta ^2 t 
 -96 t\\+24 \Delta ^2+ 4 \Delta ^2 t^{-2}-224 -{16}{t^{-2}}-96 t^{-1}
 \end{array}
 &  \begin{array}{l}
    384 +16 \Delta  \left(t^2+8 t-\frac{8}{ t}-\frac{1}{t^2}\right)+\Delta ^4 \frac{(t+1)^4}{t^2}\\
   -4 \Delta ^3 \frac{(t-1) (t+1)^3}{t^2}-4 \Delta ^2 (1+\frac{1}{t} )^2 (t (t+14)+1)
 \end{array} \\
\hline
\end{array}
\nn
\eeq

\end{document}